\documentclass[PAPER, atlasdraft=false, texlive=2023, cernpreprint, UKenglish, texmf, orcidlogo]{atlasdoc}
\usepackage[subfigure]{atlaspackage}
\usepackage{siunitx}
\usepackage{atlasbiblatex}

\usepackage{atlasphysics}

\graphicspath{{logos/}{figures/}}

\usepackage{ANA-HIGG-2023-11-PAPER-defs}

\AtlasTitle{Characterising the Higgs boson with ATLAS data from Run~2 of the LHC}

\AtlasAbstract{%
The Higgs boson was discovered by the ATLAS and CMS Collaborations in 2012 using data from Run~1 of the Large Hadron Collider (2010$-$2012). In Run~2 (2015$-$2018), about 140~fb$^{-1}$ of proton--proton collisions at a centre-of-mass energy of 13~\TeV\ were collected by the ATLAS experiment. This review presents the most important Run~2 results obtained by the ATLAS Collaboration regarding the properties of the Higgs boson and its interactions with other particles. The performed studies significantly enhance the understanding of the Higgs boson, while hunting for deviations from the predictions of the Standard Model of particle physics.
\newline \newline \textit{This work is dedicated to the memory of Prof. Peter Ware Higgs, whose groundbreaking research on broken symmetries and mass generation predicted the boson that is the focus of this Report.}
}

\AtlasRefCode{HIGG-2023-11}

\PreprintIdNumber{CERN-EP-2024-087}

\AtlasJournalRef{Phys. Rept. 1116 (2025) 4 - 56}
\AtlasDOI{DOI:10.1016/j.physrep.2024.11.001}

\hypersetup{pdftitle={ATLAS document},pdfauthor={The ATLAS Collaboration}}

\begin{document}

\maketitle

\tableofcontents

%

\section{Introduction}
\label{sec:intro}


%

One of the main motivations for building the Large Hadron Collider (LHC) was to investigate the nature of electroweak symmetry breaking. After the discovery of the Higgs boson~\cite{HIGG-2012-27,CMS-HIG-12-028} by the ATLAS and CMS Collaborations~\cite{PERF-2007-01,CMS-CMS-00-001}, the focus has shifted to the measurement of its properties. Ever since, an important goal has been to determine whether the Higgs boson can be a portal to so-far undiscovered phenomena.

This report summarises the expansive Higgs boson analysis programme performed by the ATLAS Collaboration with the data recorded during Run~2 of the LHC (2015$-$2018) and discusses the wealth of exciting results available at the time of its publication. Unless otherwise specified, all results are based on the full Run~2 data sample, corresponding to 140 fb$^{-1}$ of proton--proton ($pp$) collisions  recorded with the ATLAS detector~\cite{DAPR-2021-01}.
For detailed discussions of technical aspects of the conducted analyses, the reader is referred to the original publications, which are cited throughout the report. Searches for exotic Higgs boson decays and additional Higgs bosons are covered elsewhere~\cite{HDBSreport}.


\subsection{The Higgs mechanism}
\label{sec:introduction_theory}

The Standard Model of particle physics (SM) is a relativistic quantum field theory  based on gauge symmetries that describes the elementary particles and their interactions~\cite{Weinberg:1967tq,Glashow:1961tr,Salam:1968rm,tHooft:1972tcz}.  Seemingly forbidden by gauge invariance, massive  $W$ and $Z$ bosons are enabled by electroweak symmetry breaking. In its simplest form, electroweak symmetry breaking is realised through the introduction of a complex doublet scalar field, which manifests itself in a neutral scalar boson, the Higgs boson~\cite{Higgs:1964ia,Higgs:1964pj,Englert:1964et,Guralnik:1964eu,Kibble:1967sv}.
While the Higgs field potential itself is invariant under the SU(2) transformation, its shape allows for spontaneous symmetry breaking, which keeps the gauge symmetries intact but leads to a vacuum expectation value that is non-zero and non-invariant under SU(2).
Interactions between the Higgs field and the fermions, the Yukawa interactions, give rise to the masses of the fermions. The Yukawa couplings and the Higgs boson mass  are free parameters in the SM.

The LHC offers a unique opportunity to study the Higgs sector and therefore the nature of electroweak symmetry breaking. With the measurement of the Higgs boson mass, the SM Higgs sector is fully determined by experimentally measured parameters.
Measurements of the Higgs boson couplings to other particles and the Higgs boson self-coupling are of particular importance. The Yukawa sector describes a
completely new boson--fermion interaction that is the only interaction in the SM that distinguishes between particles of different masses and therefore flavours. The Higgs boson self-coupling is mainly probed through double Higgs boson production and is an important parameter of the Higgs field potential, which is related to the structure of the vacuum and governs the properties of the electroweak phase transition in the early universe~\cite{Degrassi:2012ry,Reichert:2017puo}.

\begin{figure}[!htb]
\centering
\subfigure[]{\includegraphics[height=2cm]{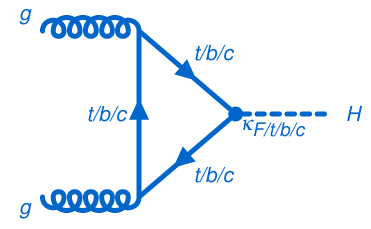}}\hfill
\subfigure[]{\includegraphics[height=2cm]{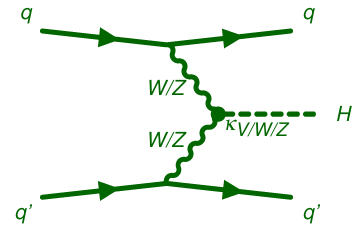}}\hfill
\subfigure[]{\includegraphics[height=2cm]{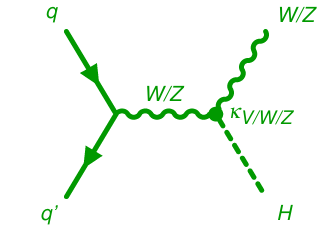}}\hfill
\subfigure[]{\includegraphics[height=2cm]{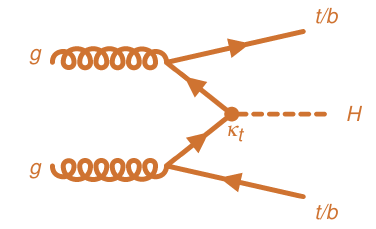}}\hfill
\subfigure[]{\includegraphics[height=2cm]{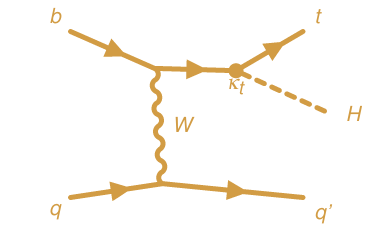}}\\

\subfigure[]{\includegraphics[height=1.93cm]{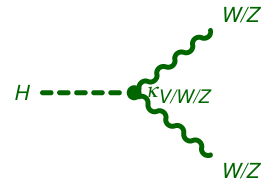}}\hfill
\subfigure[]{\includegraphics[height=1.93cm]{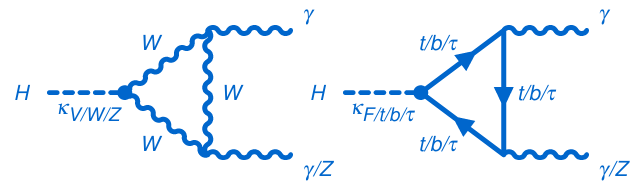}}\hfill
\subfigure[]{\includegraphics[height=1.93cm]{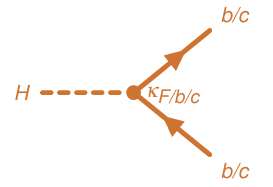}}\hfill
\subfigure[]{\includegraphics[height=1.93cm]{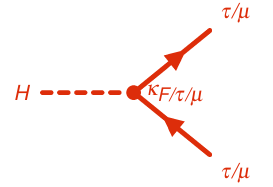}}\\

\caption{Top: Higgs boson production via (a) gluon--gluon fusion, (b) vector-boson fusion, and (c) associated production with vector bosons, (d) top- or bottom-quark pairs, or (e) a single top quark. Bottom: Higgs boson decays into (f) a pair of vector bosons, (g) a pair of photons or a $Z$ boson and a photon, (h) a pair of quarks, and (i) a pair of charged leptons~\cite{ATLAS:2022vkf}.  \label{fig:prod_decays}}
\end{figure}

\begin{figure}[tbp]
\centering
\subfigure[]{\includegraphics[height=2.5cm]{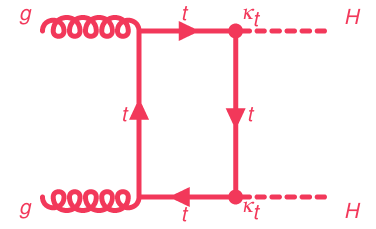}}\hskip1.cm
\subfigure[]{\includegraphics[height=2.5cm]{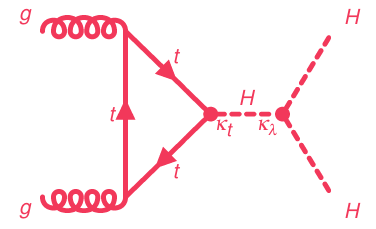}}
\\
\subfigure[]{\includegraphics[height=2.5cm]{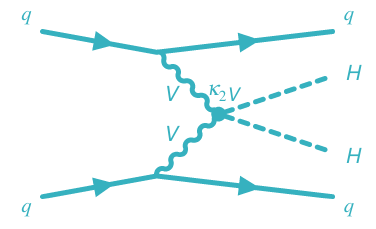}}
\subfigure[]{\includegraphics[height=2.5cm]{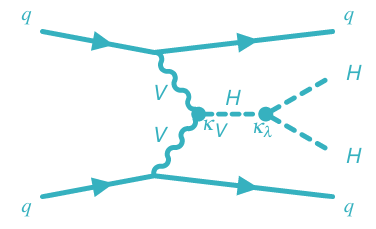}}
\subfigure[]{\includegraphics[height=2.5cm]{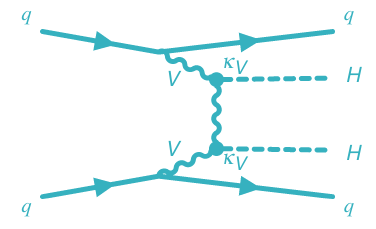}}\\
\caption{Examples of leading-order Feynman
diagrams for Higgs boson pair production in the  (a and b) \ggFHH\ and  (c, d and e) \VBFHH\ modes.}
\label{fig:hh-diagrams}
\end{figure}

The SM Higgs boson is a scalar boson: spin zero and its interactions are even under charge--parity (CP) conjugation.
In $pp$ collisions, the main Higgs boson production mechanisms predicted by the SM are gluon--gluon fusion (ggF), vector-boson fusion (VBF), and associated production with vector bosons ($VH$, where $V$ represents a $W$ or $Z$ boson), top- or bottom-quark pairs (\ttH, \bbH), or a single top quark (\tH). The different production processes  can be tagged through dedicated selection criteria, taking advantage of the particular signatures of the different final states. They are shown, together with the main predicted decay processes, in Figure~\ref{fig:prod_decays}. The corresponding diagrams for double Higgs boson production through the ggF and VBF mechanisms are shown in Figure~\ref{fig:hh-diagrams}.
The expected width of the SM Higgs boson for a mass of 125~\GeV\ is 4.1~\MeV~\cite{deFlorian:2016spz}.

\subsection{SM predictions for Higgs boson production and decay}
\label{subsec:simulation}
Complementing the measurement programme at the LHC, there is an intense effort ongoing to calculate Higgs boson production cross-sections and branching ratios, as well as their associated uncertainties, to the best possible accuracy in various phase space regions. Furthermore, Monte Carlo (MC) simulations are developed to optimise the experimental data analysis and to compare observed and expected results. These differential predictions are more challenging and therefore only available at lower precision than the total production cross-sections.

SM Higgs boson production cross-sections are calculated for the different production modes shown in Figure~\ref{fig:prod_decays} (a)-(e). The total cross-sections are listed in Table~\ref{tab:XS}, including the achieved accuracy in the electroweak (EW) and strong (QCD) coupling constant, i.e., tree-level or leading order (LO), next-to-leading order (NLO), next-to-next-to leading order (NNLO), etc. They are calculated assuming a Higgs boson mass of 125.09~\GeV. This value is based on the combined Higgs boson mass measurement by the ATLAS and CMS Collaborations in the LHC Run 1~\cite{HIGG-2014-14}.

\begin{table}[htbp]
\centering
\caption{\label{tab:XS} SM predictions of Higgs boson production cross-sections for a Higgs boson mass of 125.09~\GeV\ in $pp$ collisions at a centre-of-mass energy of $\sqrt{s} = 13$~\TeV. The values are based on
Refs.~\cite{Anastasiou:2016cez,Anastasiou:2015ema,Dulat:2018rbf,Harlander:2009mq,Harlander:2009bw,Harlander:2009my,Pak:2009dg,Actis:2008ug,Actis:2008ts,Bonetti:2018ukf,Aglietti:2004nj,Bagnaschi:2011tu,Hamilton:2015nsa} for ggF, Refs.~\cite{Ciccolini:2007jr,Ciccolini:2007ec,Bolzoni:2010xr} for VBF, Refs.~\cite{Brein:2012ne,Harlander:2018yio,Brein:2003wg,Brein:2011vx,Altenkamp:2012sx,Harlander:2014wda,Denner:2014cla,Ciccolini:2003jy,Denner:2011id} for $WH$ and $ZH$, Refs.~\cite{Beenakker:2002nc,Dawson:2003zu,Yu:2014cka,Frixione:2015zaa} for \ttH, Refs.~\cite{Dawson:2003kb,Dittmaier:2003ej,Harlander:2011aa} for \bbH, and Refs.~\cite{Demartin:2015uha,Demartin:2016axk} for \tH. They are compiled in Ref.~\cite{deFlorian:2016spz}.
}
{\def\arraystretch{1.2}
\begin{tabular}{l l l S[table-format=2.3]
S[table-format=2.2] }
\hline\hline
Process  & Precision in QCD & Precision in EW & \hbox{Cross-section [pb]} & \hbox{Fraction [\%]}\\\hline
ggF & N$^3$LO &  NLO & 48.5$^{+2.7}_{-3.6}$ &  87.2 \\
VBF & approx. NNLO & NLO & 3.78$\pm 0.08$ &  6.8 \\
$qq \rightarrow \WH$ & NNLO & NLO &  1.37$\pm 0.03 $ & 2.44 \\
$qq \rightarrow \ZH$ & NNLO & NLO &  0.76$^{+0.01}_{-0.02}$  & 1.35 \\
$gg \rightarrow \ZH$ & NLO & LO & 0.12$^{+0.03}_{-0.02}$ & 0.22\\
\ttH & NLO & NLO & 0.51$^{+0.03}_{-0.05}$ &  0.92\\
\bbH & NNLO & LO & 0.49$^{+0.10}_{-0.12}$ & 0.88\\
\tH & NLO & LO & 0.092$^{+0.006}_{-0.011}$ &  0.16\\
\hline\hline
\end{tabular}}
\end{table}

The SM predicts the double Higgs boson production cross-section to be more than a thousand times smaller than single Higgs boson production: $\sigma_{\ggFHH} = 31.0^{+2.1}_{-7.2}$~fb~\cite{Grazzini:2018bsd,Heinrich:2019bkc,deFlorian:2016spz,Dawson:1998py,Borowka:2016ehy,Baglio:2018lrj,Bonciani:2018omm,deFlorian:2013jea,Shao:2013bz,deFlorian:2015moa,Baglio:2021}, computed at NNLO in QCD including approximated mass top effects,
and $\sigma_{\VBFHH}=1.73\pm 0.04$~fb, at N$^3$LO in QCD, for  a Higgs boson mass of 125~\GeV~\cite{Dreyer:2018,Baglio:2013,Ling:2014}.

Branching fractions, or decay probabilities for the different Higgs boson decays, as depicted in Figure~\ref{fig:prod_decays} (f)-(i), also depend on the Higgs boson mass. For a mass of 125.09~\GeV, the SM predicts that decays into $W$ bosons have a branching fraction of 22\%, $Z$ bosons of 3\%, photons ($\gamma$) of 0.2\%, and $Z$ boson and photon of 0.2\%. Decays into fermion pairs have the following branching ratios: bottom quarks 58\%, charm quarks 3\%, $\tau$ leptons 6\%, and muons 0.02\%~\cite{Djouadi:1997yw,Spira:1997dg,Bredenstein:2006rh,Bredenstein:2006ha,Mihaila:2015lwa,}.

For most of the analyses discussed in this report, samples were produced with MC generators as follows: ggF events were simulated using the \POWHEG NNLOPS program~\cite{Nason:2004rx,Frixione:2007vw,Alioli:2010xd,Hamilton:2013fea,Hamilton:2015nsa,Catani:2007vq}, which provides NNLO accuracy in QCD\@.  The other production modes were simulated with the \POWHEG generator~\cite{Nason:2004rx,Frixione:2007vw,Alioli:2010xd,Nason:2009ai,Cullen:2011ac,Hartanto:2015uka} at an accuracy of NLO in QCD,  except for \bbH and \tH, which were simulated using the \MGNLO generator, also at NLO accuracy in QCD~\cite{Alwall:2014hca}.
In the majority of cases, the \PYTHIA program is used to simulate decays and to model the effects of parton showering, hadronisation and the underlying event. To evaluate modeling uncertainties, alternative samples showered with the \HERWIG program are used.
The effect of multiple interactions in the same and neighbouring LHC bunch crossings (pile-up) was modelled by overlaying the original hard-scattering event with simulated inelastic $pp$ events. The generated events were passed through a simulation of the ATLAS detector~\cite{SOFT-2010-01} using \GEANT~\cite{Geant4}.


\subsection{The LHC Run~1 legacy: From the discovery of the Higgs boson to the first property measurements}

\subsubsection{The Higgs boson discovery}

By 2012, both the LEP collider at CERN and the Tevatron collider at Fermilab had searched for the Higgs boson and excluded the existence of a SM Higgs boson with a mass below 114.4~\GeV~\cite{LEPWorkingGroupforHiggsbosonsearches:2003ing} or between 149 and 182~\GeV~\cite{CDF:2013kiv} at 95\% confidence level (CL). SM consistency fits based on precision electroweak measurements hinted at a Higgs boson with a mass of $\sim$90 $\pm$ 30~\GeV~\cite{ALEPH:2010aa,Baak:2011ze}.

Due to record centre-of-mass energies and a steady increase in collected $pp$ data, Run~1 of the LHC saw a continuous widening of the excluded mass ranges for a SM Higgs boson, except for masses around 125~\GeV, where, during the 2011 and 2012 data taking, the data recorded by the ATLAS and CMS detectors showed an excess over the background-only hypothesis.

On July 4th, 2012, based on approximately 10~\ifb\ of recorded LHC data each, the ATLAS and CMS Collaborations individually announced the discovery of a neutral  boson with a mass of about 125~\GeV~\cite{HIGG-2012-27,CMS-HIG-12-028}. As shown in Figure~\ref{fig:pvalue_discovery}, multiple decay channels contributed to the combined significance of the excess, which exceeded the five standard deviation ($\sigma$) discovery threshold. The ones with the highest significances were the \yy and four-lepton (\ZZfourl, $\ell=e, \mu$) final states, which provide an excellent invariant mass resolution and favourable signal-to-background ratios, as well as the \WWevmuv final state.
Figure~\ref{fig:peakplots_discovery} shows the observed excess in the \yy and \fourl invariant mass distributions, and in the transverse mass distribution for the \HWWenumunu decay channel, where the transverse momentum of the neutrinos is estimated through the missing transverse momentum in the detector. This results in the wider distribution seen in Figure~\ref{fig:peakplots_discovery} (c). The transverse mass is defined as $m_\mathrm{T} = \sqrt{(E_\mathrm{T}^{\ell \ell} + E_\mathrm{T}^\mathrm{miss})^2 - |\vec{p}_\mathrm{T}^{~\ell \ell} + \vec{p}_\mathrm{T}^\mathrm{~miss}|^2}$, where $\vec{p}_\mathrm{T}^{~\ell \ell}$ is the transverse momentum vector of the dilepton
system, $\vec{p}_\mathrm{T}^\mathrm{~miss}$ is the missing transverse momentum vector with magnitude $E_\mathrm{T}^\mathrm{miss}$, and $E_\mathrm{T}^{\ell \ell}$ is the square root of sum in quadrature of the magnitude of $\vec{p}_\mathrm{T}^{~\ell \ell}$ and the invariant mass of the dilepton system.

\begin{figure}[!tbp]
\centering
\includegraphics[width=0.6\textwidth]{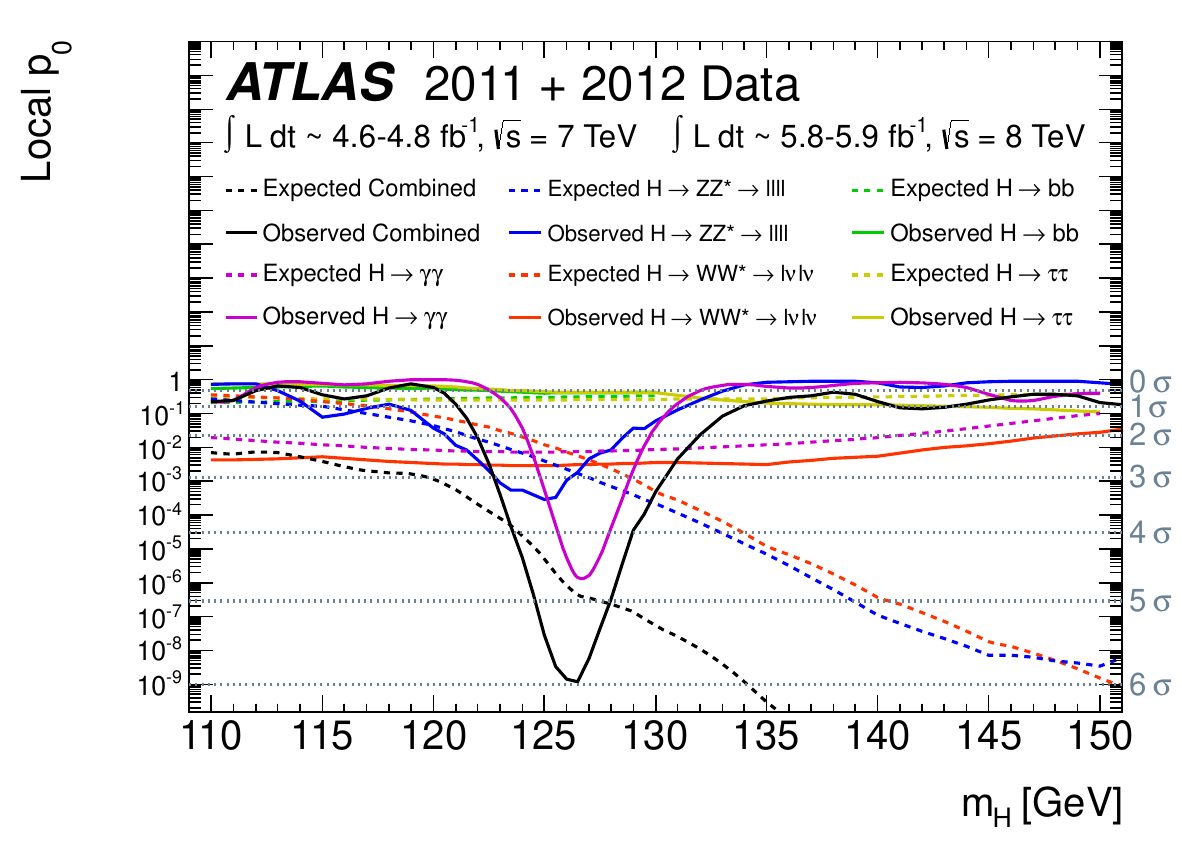}
\caption{Left vertical axis: The $p$-value, i.e., the probability for a background-only experiment to be more like the signal than the observation, for individual decay channels and their statistical  combination. Right vertical axis: The translation to standard deviations. The horizontal axis shows the Higgs boson mass hypothesis~\cite{HIGG-2012-27}.
\label{fig:pvalue_discovery}}
\end{figure}

\begin{figure}
\centering
\subfigure[]{\includegraphics[width=0.4\textwidth]{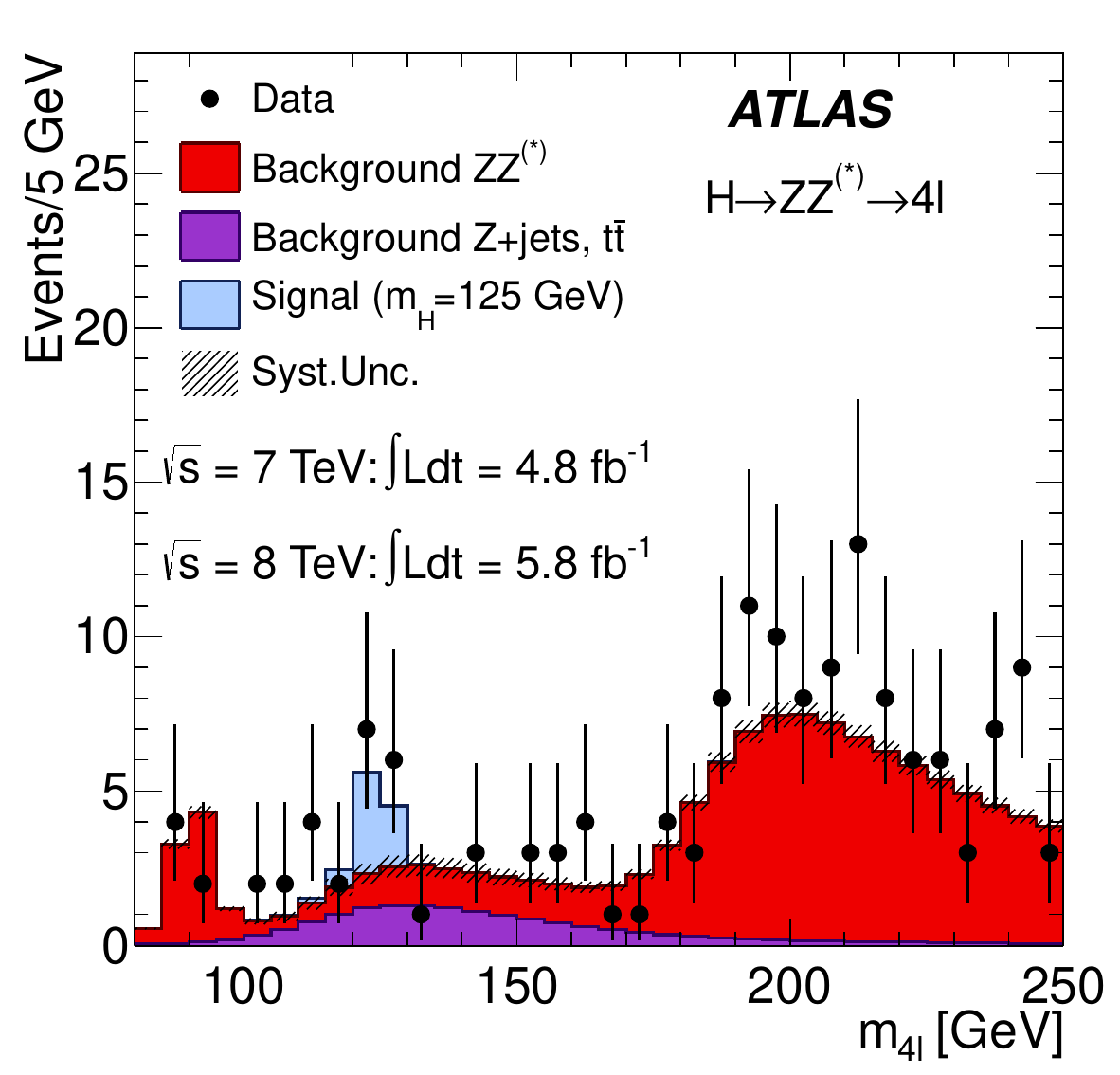}}\\
\subfigure[]{ \includegraphics[width=0.43\textwidth]{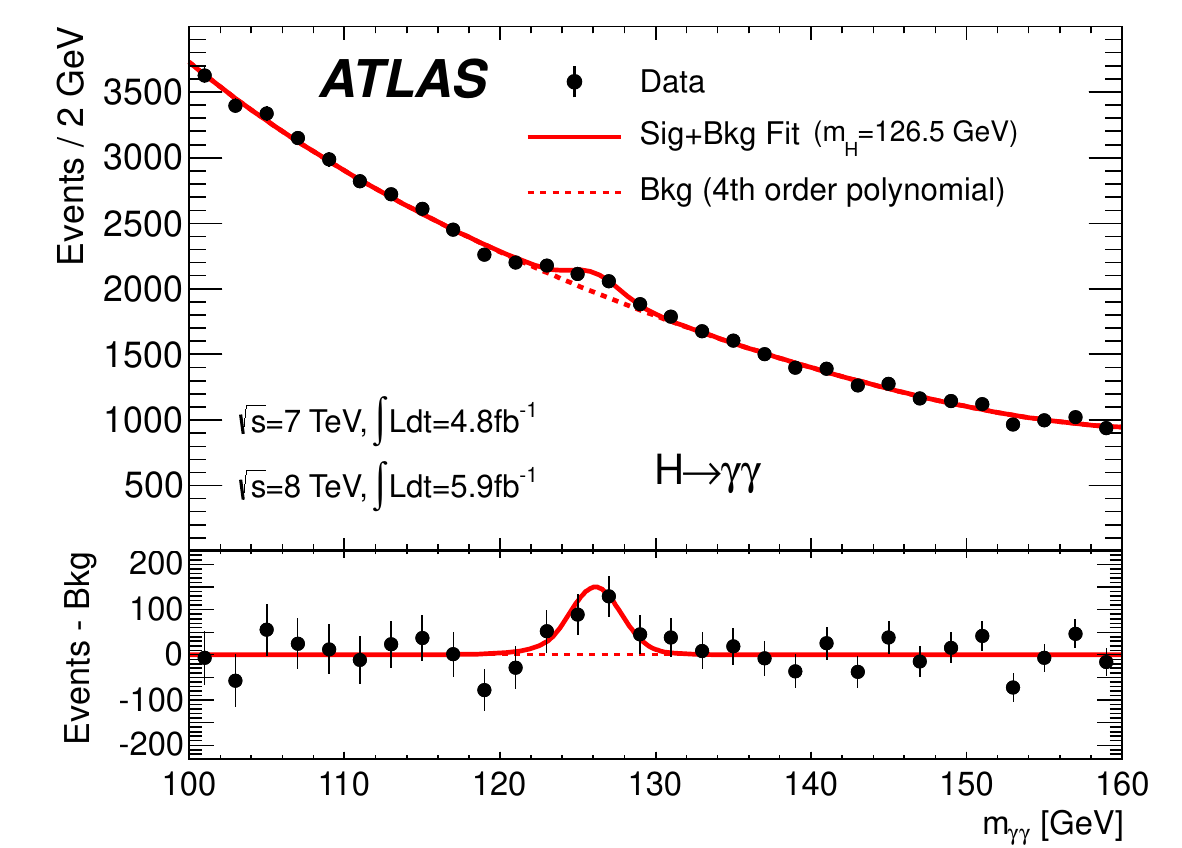}}
\subfigure[]{ \includegraphics[width=0.43\textwidth]{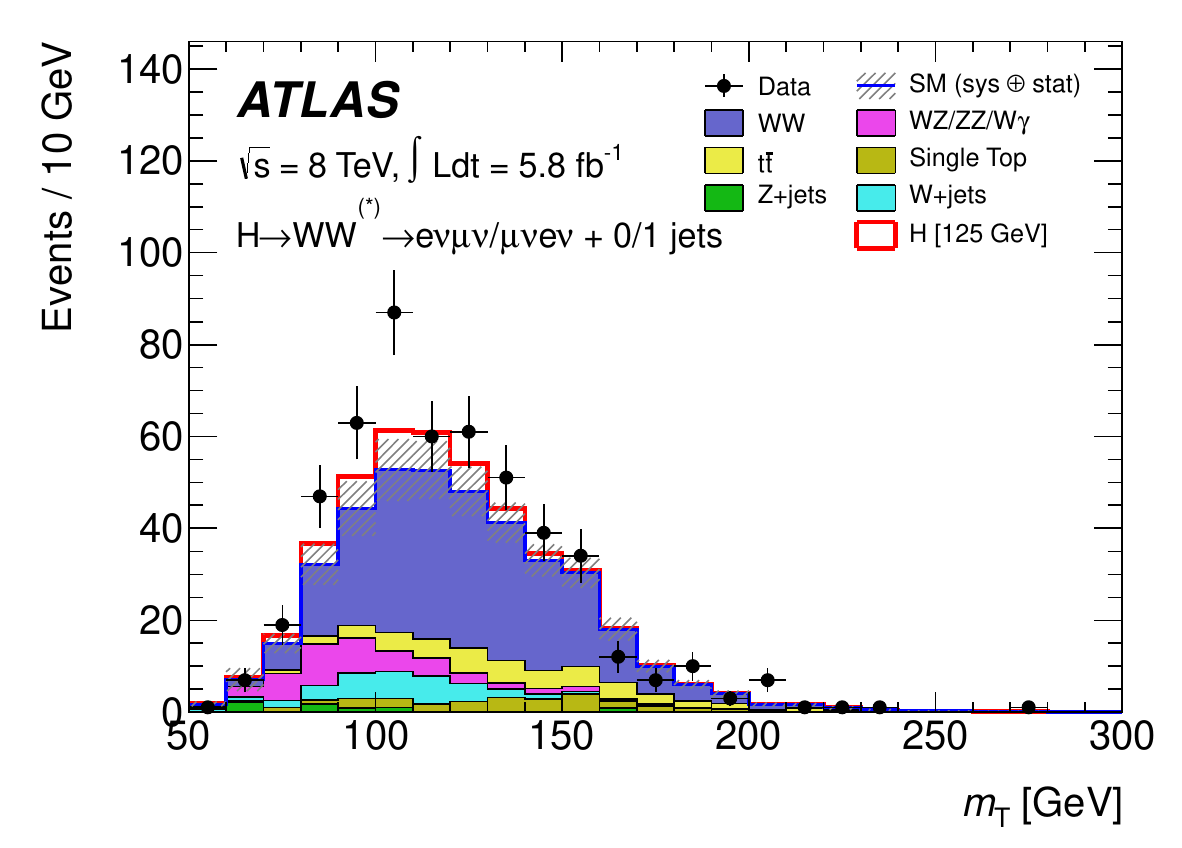}}
\caption{ Invariant mass distributions in the (a) \ZZfourl and (b) \yy final states, and (c) the transverse mass in the \WWevmuv final state, showing the excesses in data events over the background-only hypotheses that led to the discovery of the Higgs boson~\cite{HIGG-2012-27}.
\label{fig:peakplots_discovery}}
\end{figure}

\subsubsection{First measurements of the Higgs boson properties}

The data produced by the LHC and recorded by the ATLAS and CMS detectors in Run~1, in particular the 20~\ifb\ of data at 8~\TeV\ centre-of-mass energy recorded in 2012, allowed the first measurements of the properties of the newly discovered particle.

A Higgs boson mass measurement in the \HZZfourl\ and \Hyy\ decay channels, combining ATLAS and CMS data, resulted in a value of \mH = $125.09 \pm 0.24$~\GeV~\cite{HIGG-2014-14}.
This value is used as input to almost all measurements performed by the ATLAS Collaboration based on the LHC Run~2 data. In most measurements, the effect of the associated uncertainty is negligible.

Combining ATLAS and CMS Higgs boson measurements from Run~1 in all accessible production and decay channels, the Higgs boson signal yield relative to the SM prediction, also called signal strength $\mu$, was measured to be 1.09 $\pm$ 0.11~\cite{HIGG-2015-07}. Each experiment observed, with more than 5$\sigma$ significance, the ggF production process and the decays into two $Z$ bosons, two $W$ bosons, and two photons. The combination also allowed observation of VBF production and decays into two \tauls.
Assuming SM coupling structures, the couplings of the Higgs boson to the gauge bosons were measured with an accuracy of $\sim$10\%, to top quarks and \tauls with $\sim$17\%, and to bottom quarks with $\sim$30\%.

Differential cross-section measurements were performed in the \Hyy, \HZZfourl, and \HWWenumunu decay channels~\cite{HIGG-2013-10,HIGG-2013-22,HIGG-2014-11,HIGG-2015-04}, analyzing Higgs boson kinematics and properties of associated hadronic jet production.

First statements could also be made in Run~1 regarding the spin and CP properties of the newly discovered particle. The fact that decays into two bosons were observed clearly characterises the particle as a boson, with the diphoton decay excluding the spin-1 possibility~\cite{Landau:1948kw,Yang:1950rg}. Spin and CP studies of diboson decays~\cite{HIGG-2013-17} confronted the scalar hypothesis with several alternative spin scenarios, including non-SM spin-0 and spin-2 models with universal and non-universal couplings to fermions and vector bosons. All tested alternative models were excluded in favour of the SM Higgs boson hypothesis at a CL of more than 99.9\%. No sign of CP violation was found in VBF production with \taul decays~\cite{HIGG-2015-06}.

An upper limit was set on the width of the Higgs boson by the ATLAS Collaboration of 22.7~\MeV~\cite{HIGG-2014-10}, which corresponds to roughly six times the SM expectation. This assumed that the signal strength on the Higgs boson peak and off-shell is the same~\cite{Kauer:2012hd,Caola:2013yja,Campbell:2013una,Campbell:2013wga}.

Within the achieved statistical and systematic uncertainties, all measurements in Run 1 agreed with the predictions for a SM Higgs boson.

In Run~2 of the LHC, the ATLAS detector recorded 140 fb$^{-1}$ of $pp$ collision data. For this data sample, the SM predicts a total of $\sim$9 million Higgs bosons produced at the ATLAS interaction point, of which 0.3\% are experimentally accessible~\cite{ATLAS:2022vkf}. Making use of this unprecedented amount of data, the goals for Run~2 included a more precise determination of the Higgs boson mass and width (see Section~\ref{sec:mass_width}), and of fiducial and differential cross-sections (see Section~\ref{sec:higgs_diffxs}). Measurements of the Higgs boson couplings to fermions, in particular the direct observation of the Yukawa couplings to third generation quarks (see Section~\ref{sec:higgs_fermions}) were eagerly anticipated, as were the searches for rare decays (see Sections~\ref{sec:higgs_fermions} and \ref{sec:higgs_rare}). The larger data sample also allowed increased precision on the measured Higgs boson production cross-sections and couplings to other particles (see Section~\ref{sec:higgs_combination}) and the study of the CP properties of both the boson and fermion couplings (see Section~\ref{sec:higgs_cp}). Finally, enormous progress was made regarding the search for Higgs boson pair production and the constraints on the Higgs boson self-coupling (see Section~\ref{sec:higgs_self}).

\FloatBarrier


%
%
%

\section{Higgs boson mass and width}
\label{sec:mass_width}

\subsection{Higgs boson mass }

The SM predicts the existence of a Higgs boson but not its mass, which therefore has to be measured experimentally. Once the mass of the Higgs boson is known, together with the masses of fermions, the SM production cross-section and branching ratios can be calculated. They can then be checked experimentally to search for deviations from SM predictions due to potential new phenomena. The value of the Higgs boson mass also affects the SM predictions of the effective weak mixing angle and, logarithmically, the $W$ boson mass (see e.g., Ref.~\cite{Haller:2018nnx}). Moreover, the shape and energy evolution of the Higgs potential are directly related to the Higgs boson mass, which therefore determines the stability of the electroweak vacuum~\cite{Sher:1988mj,Degrassi:2012ry}.

Among the various decay modes, \Hyy\  and \HZZfourl\ play a special role in  the characterisation of the Higgs boson and in particular the measurement of its mass. Both final states are fully reconstructable, provide an excellent mass resolution, and are accompanied by well-understood backgrounds.  Among the two, \Hyy yields more signal events, while \HZZfourl benefits from a higher signal-to-background ratio.

\begin{figure}
\centering
\subfigure[]{\includegraphics[height=6.1cm]{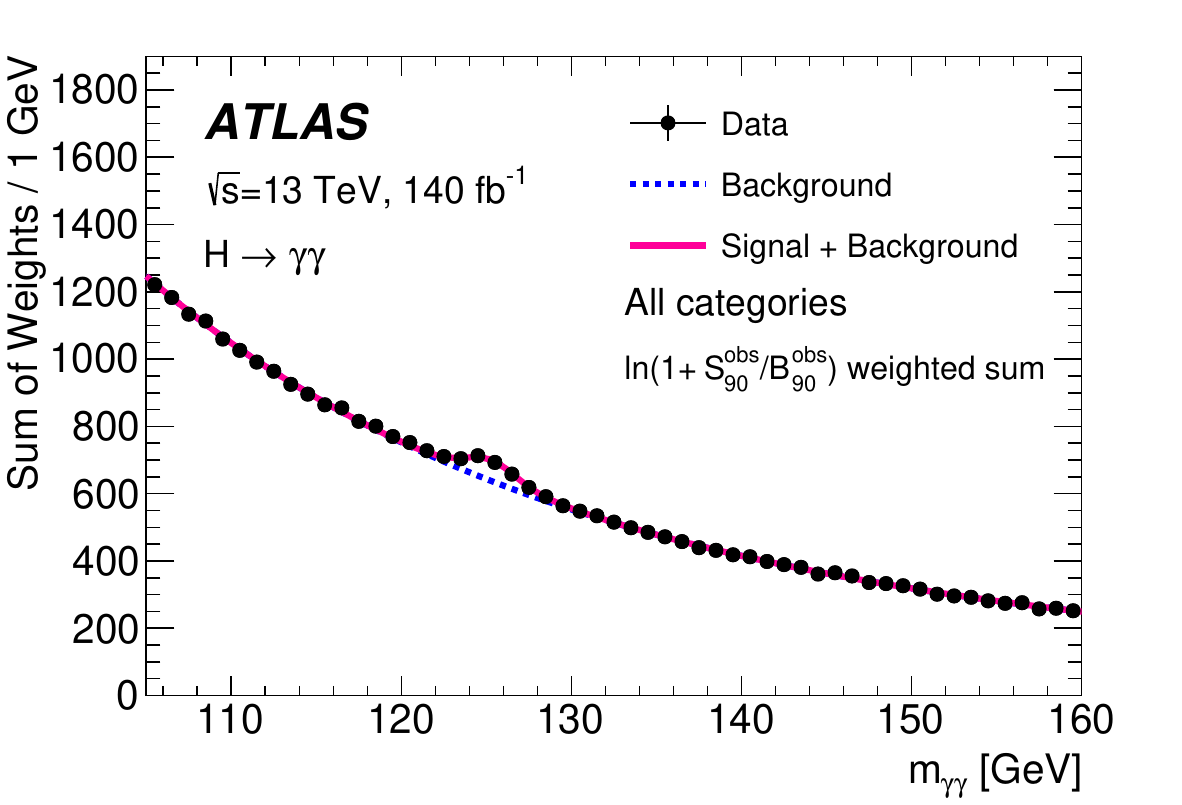}}
\subfigure[]{\includegraphics[height=6.1cm]{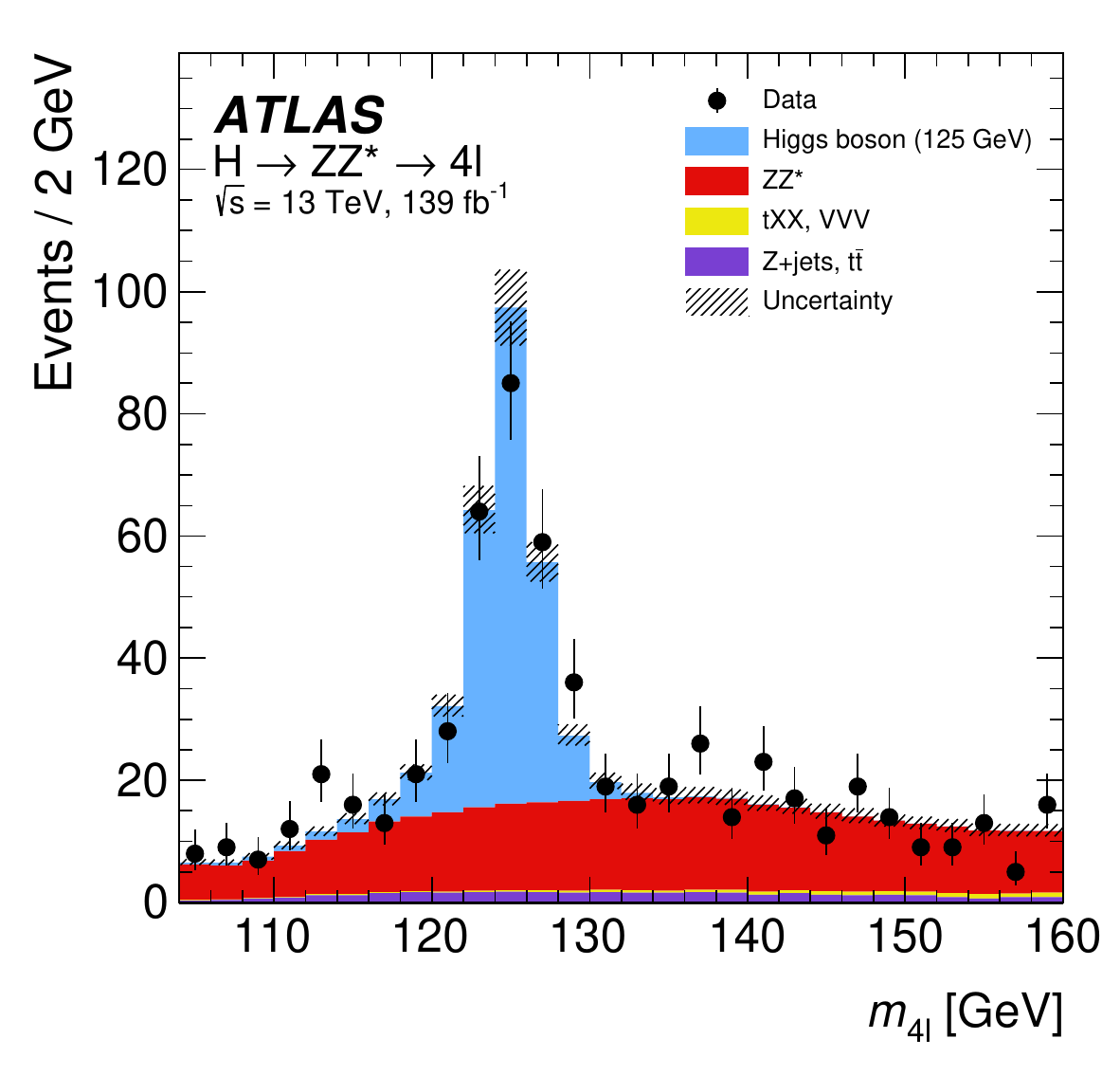}}
\caption{(a)  Weighted diphoton invariant mass distribution of all selected data events, overlaid with the result of the  fit~\cite{HIGG-2019-16}. (b) Four-lepton invariant mass distribution of the selected events in data compared with the simulation~\cite{ATLAS:2022zjg}.
\label{fig:mass_zz_yy}}
\end{figure}

\textbf{The \Hyy decay channel}~\cite{HIGG-2019-16} allows the direct observation of a narrow mass peak over a smooth background that can be determined directly from the collected data. The typical mass resolution in the ATLAS detector is 1.7~\GeV\ for a Higgs boson mass of 125~\GeV. The main background process is non-resonant $\gamma\gamma$ production with a smaller contribution of about 25\% from $\gamma$+jet and dijet processes where one or two of the hadronic jets are misidentified as photons.
In the \Hyy\  analysis,  the Higgs boson properties (e.g., signal yield, mass) are  measured in events with
at least two isolated and well-identified photon candidates.
The selection efficiency for events with a SM Higgs boson decaying into two photons is about 36\%. Depending on the measurement being performed,  events are classified into different categories to improve the specific sensitivity to a given observable. In the case of the Higgs boson mass measurement, the selected events are classified into 14 categories with different signal-to-background ratios, invariant mass resolutions and photon energy scale uncertainties.

The invariant diphoton mass distribution  (\myy)
of the simulated signal in each category is found to be very well described by a double-sided Crystal Ball~\cite{Oreglia:1980cs} probability
density function.  The normalisation factors of the signal, one for each category,  are treated as free parameters in the fit to reduce the  model-dependence of the measurement.
In each category, the  \myy\ distribution of the background processes is represented by either an exponential function, a power-law function, or the exponential of a second-order polynomial.

One key aspect of the mass measurement in the \Hyy\ channel is the careful control of the systematic uncertainties, mainly arising from the photon energy scale. A significant effort was devoted during Run~2 to substantially reduce these uncertainties, resulting in a decrease of a factor close to four compared to the mass measurement performed with the initial 36~\ifb\ of LHC Run~2 data~\cite{HIGG-2016-33}. This improvement in the photon energy scale calibration was achieved through a better  understanding of the energy response across the longitudinal layers of the ATLAS electromagnetic calorimeter and a new correction implemented in the extrapolation of the electron energy scale measured in $Z\to ee$ events to photons~\cite{ATLAS:2023mnw}. The impact of the photon energy scale uncertainty on the mass measurement was reduced to 83~\MeV. Systematic uncertainties related to the choice of the function used to describe the background are also taken into account, with a minor impact on the mass measurement, but more relevant for analyses targeting the measured signal yield. The measured mass of the Higgs boson in the \Hyy\ final state using the full Run~2 data sample is
$m_H=125.17 \pm 0.11 (\mathrm{stat.})\pm 0.09 (\mathrm{syst.})$~\GeV. The measurement is statistically limited,  hence opening the possibility  for substantial improvements with the  future LHC runs. The diphoton invariant mass distribution of all selected data events, weighted according to the signal-to-background ratio in the respective category, and the fitted functions are shown in Figure~\ref{fig:mass_zz_yy} (a).

\textbf{\HZZfourl decays}~\cite{ATLAS:2022zjg} are reconstructed by requiring two pairs of oppositely charged same-flavour isolated electrons or muons  in the final state. One of the two pairs, the leading lepton pair is required to have an invariant mass compatible with that of the $Z$ boson. The events are split into four sub-channels according to the flavour of the leading  and subleading lepton pair ($4\mu, 2e2\mu, 2\mu 2e, 4e$). The selection efficiency ranges between 31\% and 16\%, depending on the sub-channel. The dominant background process is non-resonant $ZZ\to 4\ell$ production, which is suppressed in this specific measurement by using a neural network (NN) based classifier. The measurement of the Higgs boson mass is carried out by simultaneously fitting the reconstructed invariant mass of the four-lepton system ($m_{4\ell}$) in the four sub-channels using an unbinned maximum-likelihood approach. The resolution of $m_{4\ell}$ varies from approximately 1.5~\GeV\ to 2.1~\GeV. The signal model comprises a double-sided Crystal Ball probability density function, with the mean of its Gaussian core determined as a function of $m_H$ and the standard deviation expressed as a function of the predicted event-level resolution. In the fitting procedure,  the normalisations of the signal and background components are treated as free parameters for each of the four sub-channels.

Figure~\ref{fig:mass_zz_yy} (b) shows the four-lepton invariant mass distribution of the selected events in data compared with the simulation.
The Higgs boson mass measured in the \HZZfourl\ channel using the Run~2 data sample is $m_H=124.99 \pm 0.18 (\mathrm{stat.})\pm 0.04 (\mathrm{syst.})$~\GeV, with the dominant source of systematic uncertainty arising from the uncertainties in the muon momentum scale, resolution and sagitta
bias correction (28 MeV) and the electron energy scale (19 MeV).

\textbf{The combination of the  mass measurements} performed with the
\Hyy\ and \HZZfourl\  channels  using Run~1 and  Run~2~\cite{HIGG-2014-14} data results in a  Higgs boson mass of $m_H=125.11 \pm 0.09 (\mathrm{stat.})\pm 0.06 (\mathrm{syst.})$~\GeV~\cite{HIGG-2022-20}, which corresponds to an experimental precision of 0.09\%. With the current precision achieved in the Higgs boson mass, the impact of its uncertainty on the Higgs boson production cross-sections and branching ratios is below 1\%.
A summary plot of the Higgs boson mass  measurements performed in Run~1 and Run~2 and their combination is shown in Figure~\ref{fig:masscomb}.

\begin{figure}
\centering
\subfigure{\includegraphics[width=0.70\textwidth]{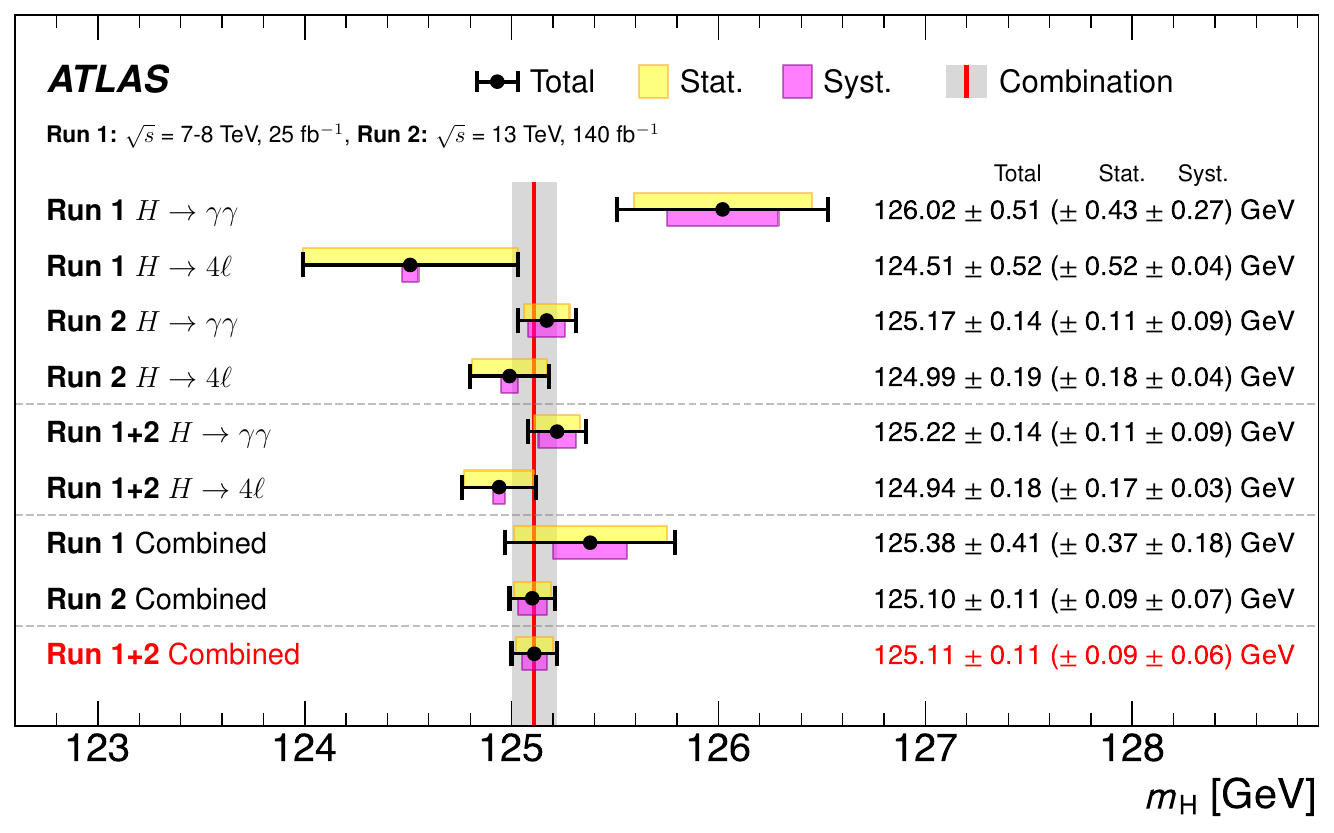}}
\caption{ATLAS Higgs boson mass measurement combining LHC Run~1 and Run~2 results in the \Hyy\ and \HZZfourl\ final states~\cite{HIGG-2022-20}.
\label{fig:masscomb}}
\end{figure}

\subsection{Higgs boson width}
Once the mass of the Higgs boson is known, the SM allows the width to be computed. For a Higgs boson with a mass of 125~\GeV, the expected width is  only $\Gamma_H=4.1$~\MeV~\cite{deFlorian:2016spz}.
Any deviation from this value would indicate the presence of phenomena beyond the SM, e.g., Higgs boson decays into so-far unknown particles.

Due to the  experimental mass resolution in the most precise channels (\Hyy\ and \HZZfourl) of the order of 1$-$2~\GeV, a direct measurement of the SM width of the Higgs boson is not achievable from the detected lineshapes. Similarly, the corresponding lifetime is so short that  the  decay length is extremely small, well beyond the experimental resolution on the reconstruction of the displaced vertex from the beam spot. In addition, the width cannot be extracted from the rate measurement in the different decay channels because only cross-section times branching-ratio measurements are possible at the LHC\@.
However, in 2012, a different method was proposed, which relies on measuring on-shell and off-shell Higgs boson production~\cite{Kauer:2012hd,Caola:2013yja,Campbell:2013una,Campbell:2013wga}. In the SM, despite the small total width, about 10\% of the $gg\to H^*\to VV$ ($V=W,Z$) events are produced with an invariant mass larger than twice the vector bosons masses.
In particular, the vector bosons from the Higgs boson decay and the top quarks running in the ggF production loop can be on-shell
above the thresholds of 2$m_{V}$ and $2m_{t}$, respectively. On-shell Higgs boson production is inversely  proportional to the width,  e.g., for ggF production with the Higgs boson decaying into two $V$ bosons
\begin{align}
\sigma^{\mathrm{on-shell}}_{gg\to H\to VV}\propto
\frac{g^2_{ggF, \mathrm{on-shell}}g^2_{HVV,\mathrm{on-shell}}}{\Gamma_{H}m_H},
\end{align}
where $g_{ggF, \mathrm{on-shell }}$ and $g_{HVV, \mathrm{on-shell }}$ are the on-shell couplings associated with the Higgs boson production and decay. On the other hand, off-shell Higgs boson production has negligible dependence on the width, e.g.,
\begin{align}
\sigma^{\mathrm{off-shell}}_{gg\to H\to VV}\propto \frac{g^2_{ggF,\mathrm{off-shell}} g^2_{HVV,\mathrm{off-shell}}}{m^2_{VV}}.
\end{align}
Similar expressions apply to electroweak production, where the $g_{ggF}$ coupling is replaced by the $g_{HVV}$ coupling. If the effective couplings have a known relationship (e.g., the one predicted in the SM)  in the on-shell and off-shell regimes, the ratio of the off-shell to the on-shell production allows  the determination of $\Gamma_H$.

The off-shell Higgs boson signal cannot be treated independently of the $gg\to\ VV$ background due to  significant negative interference between the two processes~\cite{Kauer:2012hd}. For a given off-shell signal strength $\mu_{\mathrm{off-shell}}$, the expected number of events as a function of the invariant mass of the diboson system, $m_{VV}$, is positive and corresponds to the sum of the $gg\to\ VV$ background, the off-shell signal, which is proportional to $\mu_{\mathrm{off-shell}}$, and the (negative) interference term, which is proportional to $\sqrt{\mu_{\mathrm{off-shell}}}$. Additionally, when extracting the width of the Higgs boson, it must also be assumed  that any new phenomena affecting the off-shell signal strength and the off-shell couplings do not modify the relative phase of the interfering signal and background processes. The full Run~2 data sample is used to study off-shell Higgs boson production and set an upper limit on the width of the Higgs boson through the $ZZ\to 4\ell$ and $ZZ\to 2\ell 2\nu$ final states~\cite{ATLAS:2023dnm}, which were identified as the most sensitive channels in Run 1~\cite{ATLAS:2015cuo}.

\textbf{The $ZZ \rightarrow 4\ell$ analysis} selects  events with two same-flavour oppositely charged dilepton pairs and a four-lepton invariant mass above the on-shell $ZZ$ production threshold. The mass of each dilepton pair is required to be compatible with that of a $Z$ boson. The signal regions are designed to be sensitive to both the electroweak and ggF production modes, considering only events with $m_{4\ell} > 220$~\GeV, while events with $180 < m_{4\ell} < 220$~\GeV are used to constrain the normalisation of the dominant $qq\to ZZ$ process.
To enhance the signal sensitivity, a multi-class dense NN is used to distinguish among the off-shell Higgs boson signal, the interfering background and non-interfering background. Two separate NNs are trained for ggF and electroweak-induced signals. The NN's output is employed to define the final observable, serving as a discriminating variable in the corresponding signal region. Figure~\ref{fig:offshell} (a) shows the observed distribution of the NN based discriminant in the ggF signal region compared with the SM expectations.

\begin{figure}
\centering
\subfigure[]{\includegraphics[width=0.45\textwidth]{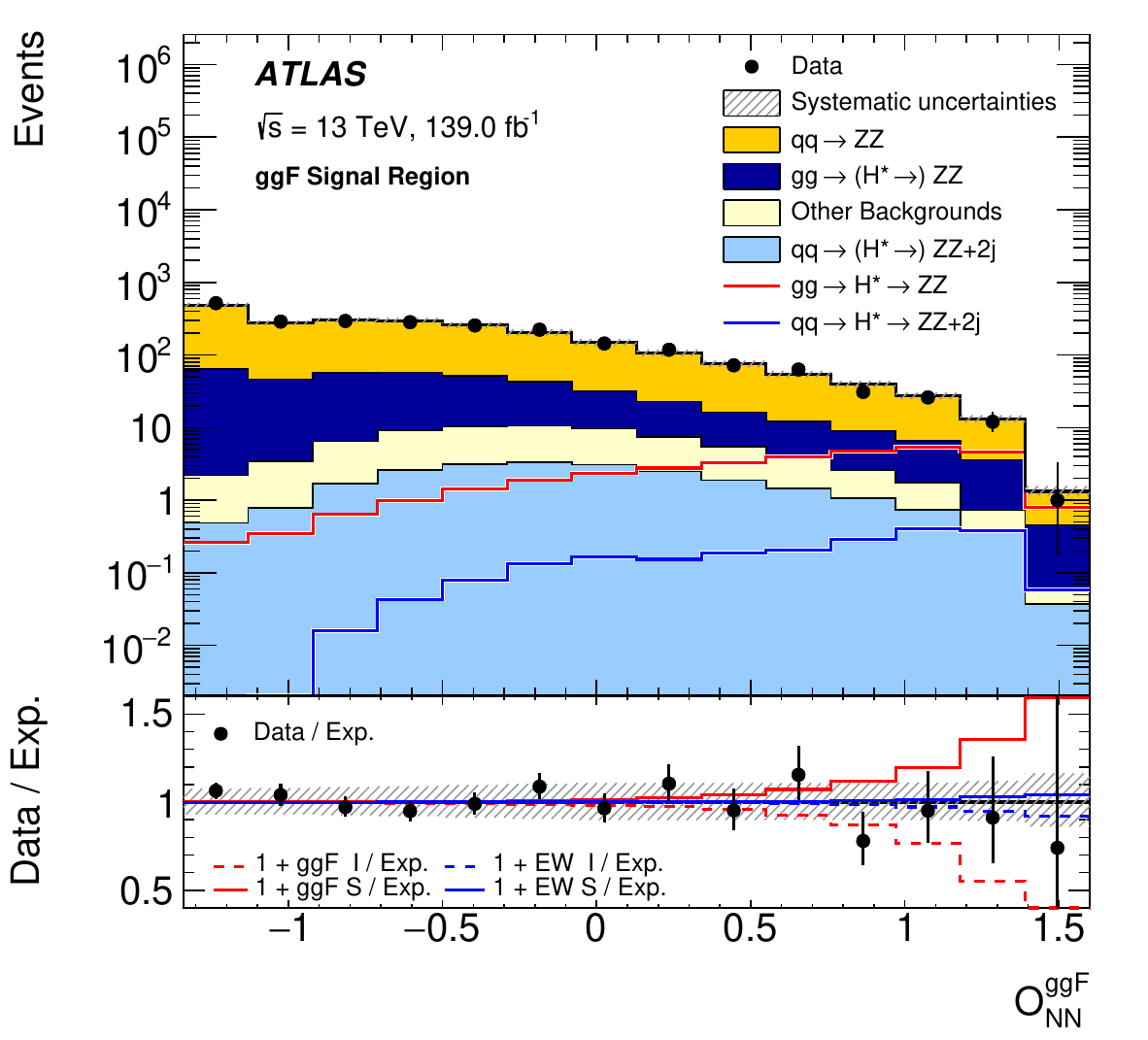}}
\subfigure[]{\includegraphics[width=0.45\textwidth]{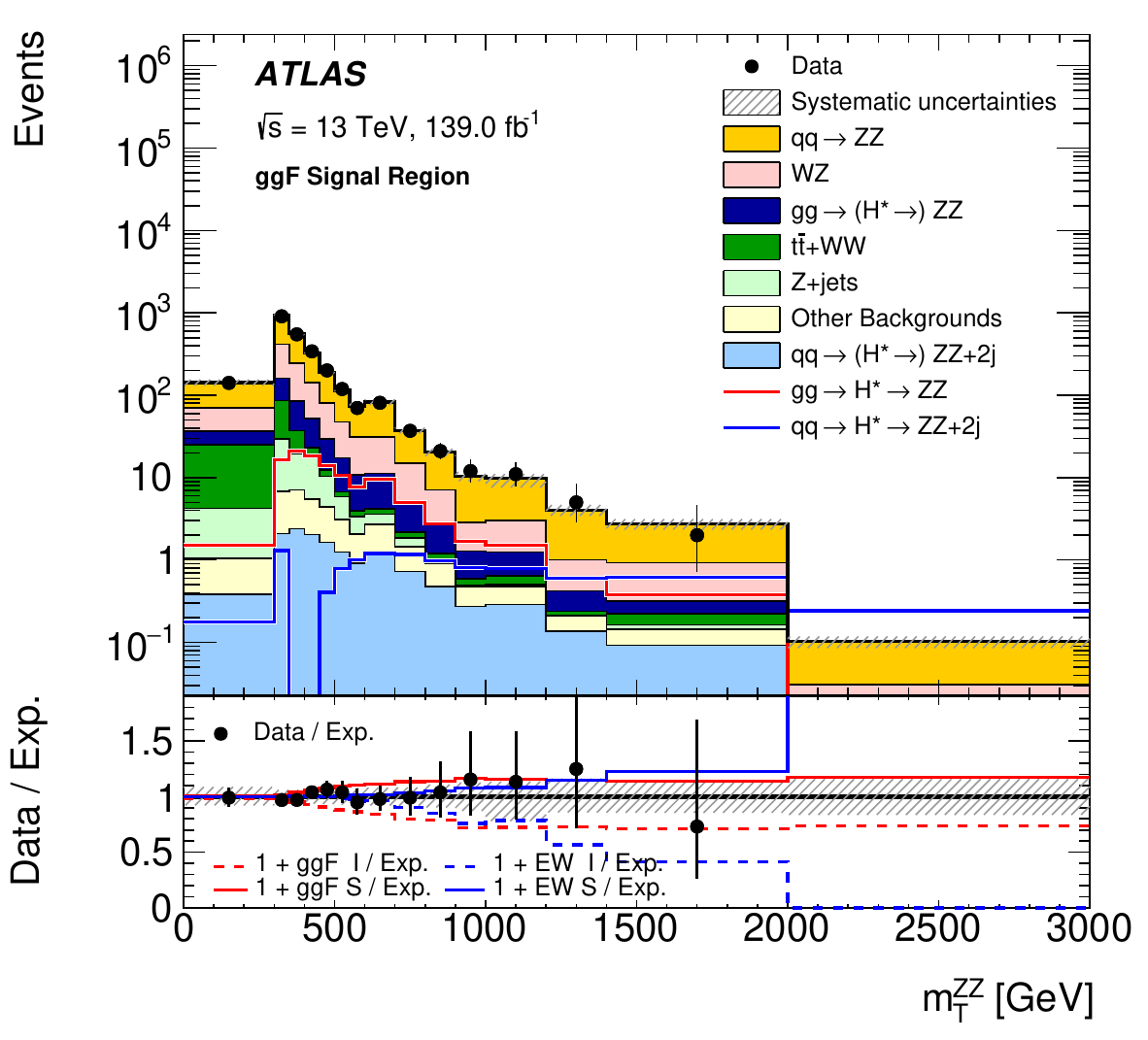}}

\caption{(a)  Observed and expected distribution of the NN-based observable sensitive to the off-shell production in the $4\ell$ channel in the ggF signal region~\cite{ATLAS:2023dnm}. (b)  Observed and expected transverse mass distribution in the $2\ell 2\nu$ channel in the ggF signal region~\cite{ATLAS:2023dnm}.
\label{fig:offshell}}
\end{figure}
\textbf{In the $ZZ\to 2\ell 2\nu$ final state}, the presence of neutrinos prevents the kinematic reconstruction of $m_{ZZ}$. The discriminating variable used to enhance sensitivity to the off-shell production of the Higgs boson is therefore the transverse mass, computed from the transverse components of the momenta, defined as
\begin{align}
m_{\mathrm{T}}^{ZZ} \equiv \sqrt{ \left( \sqrt{m^2_Z+(p_{\mathrm{T}}^{\ell\ell})^2}+\sqrt{m^2_Z+(E_{\mathrm{T}}^{\mathrm{miss}})^2}\right)^2 - \left| \vec{p}_{\mathrm{T}}^{\ell\ell}+\vec{p}_{\mathrm{T}}^{\mathrm{~miss}}\right|^{2}},
\end{align}
which is reconstructed from the mass of the $Z$ boson $m_Z$, the transverse momentum vector of the dilepton system  $\vec{p}_{\mathrm{T}}^{\ell\ell}$ and  the missing transverse momentum vector $\vec{p}_{\mathrm{T}}^{\mathrm{~miss}}$, along with their respective magnitudes.
Signal events are selected by requiring two oppositely charged electron or muon candidates with a dilepton invariant mass compatible with the $Z$ boson mass and high \MET, vetoing events with a third lepton.
Additional selection criteria are optimised to maximize the signal significance relative to the main backgrounds, which include $ZZ$, $WZ$, $WW$, top-quark, and $W/Z$+jets processes. Figure~\ref{fig:offshell} (b) shows the  observed $m_\mathrm{T}^{ZZ}$ distribution compared with the SM prediction in the ggF signal region.

Combining the 4$\ell$ and $2\ell2\nu$ channels, the ratio of the off-shell Higgs boson production rate to its SM prediction ($\mu_{\mathrm{offshell}}$) is measured.  The data are found to be consistent with SM predictions, rejecting the hypothesis of no off-shell Higgs boson production ($\mu_{\mathrm{offshell}}=0$) with an observed (expected) statistical significance of 3.3 (2.2)$\sigma$, thus providing experimental evidence for this process.
In addition, these results are combined with the
on-shell Higgs boson measurements in the \HZZfourl channel ~\cite{HIGG-2018-28}, obtaining an indirect  measurement of the  width of the Higgs boson of $\Gamma_H=4.5^{+3.0}_{-2.5}$~\MeV, in agreement with the SM expectation, and observed (expected) lower and upper limits on the total width of  $0.5~(0.1) < \Gamma_{H} < 10.2~(10.6)$~\MeV  at 95\%~CL.


\section{Fiducial and differential cross-sections}
\label{sec:higgs_diffxs}

Fiducial and differential cross-section measurements allow the characterisation of the Higgs boson through production- and decay-related distributions. They are usually performed inclusively over all production modes. Compared to measurements of event yields, these cross-section measurements have the advantage that they can be easily interpreted and used by physicists outside the ATLAS Collaboration, as they include corrections for detector effects.

In each final state, a fiducial phase space is chosen to reduce the extrapolation from the measured phase space and therefore the model dependence of the measurement. The fiducial phase space is defined at particle level, i.e., based on generated stable particles, approximating the selection of reconstructed events.
After selecting the events, the extraction of the signal event yields and the correction to the fiducial phase space are performed simultaneously using a likelihood fit.

\subsection{Fiducial and total cross-sections}
\label{subsec:fidtot}

Higgs boson production cross-sections are measured in the \yy and \ZZfourl final states. The event selections are very similar to those discussed in Section~\ref{sec:mass_width}, but the events are not separated into different categories.
The Higgs boson production cross-section times branching ratio measured in the \yy fiducial volume is $67 \pm 6$~fb~\cite{HIGG-2019-13}, while the cross-section times branching ratio measured in the \ZZfourl fiducial volume is $3.28 \pm 0.32$~fb~\cite{HIGG-2018-29}. Both results are in agreement with the SM predictions ($64 \pm 4$~fb and $3.41 \pm 0.18$~fb, respectively~\cite{deFlorian:2016spz}), and achieve a precision of about 10\%.

To allow the combination of the \ZZfourl and \yy measurements~\cite{HIGG-2022-04}, total cross-sections are extracted by extrapolating from the fiducial to the total phase space, also correcting for the branching fractions. The combination improves the statistical power by roughly doubling the analysed data sample, as the lower cross-section times branching fraction of the \ZZfourl decay channel is balanced by a better signal-over-background ratio compared to the \yy measurement.
On the other hand, the combination increases the model-dependence of the result. The extrapolation factors rely on SM predictions and uncertainties related to the measured cross-sections of the different Higgs boson production modes are taken into account. The reduction in statistical uncertainties is significantly larger than the additional uncertainties due to the extrapolation, which are at the per-cent level. The cross-sections measured at different centre-of-mass energies are presented in Figure~\ref{fig:totalxs}, together with the SM prediction. For $pp$ collisions at 13~\TeV centre-of-mass energy, the measured cross-section is $55.5 \pm 3.2\mathrm{(stat.)}^{+2.4}_{-2.2}\mathrm{
(syst.)} = 55.5^{+4.0}_{-3.8}$~pb, achieving a precision of about 7\%. Good agreement with the SM prediction (55.6 $\pm$ 2.2 pb~\cite{deFlorian:2016spz})  is observed.

\begin{figure}
\centering
\includegraphics[width=0.65\textwidth]{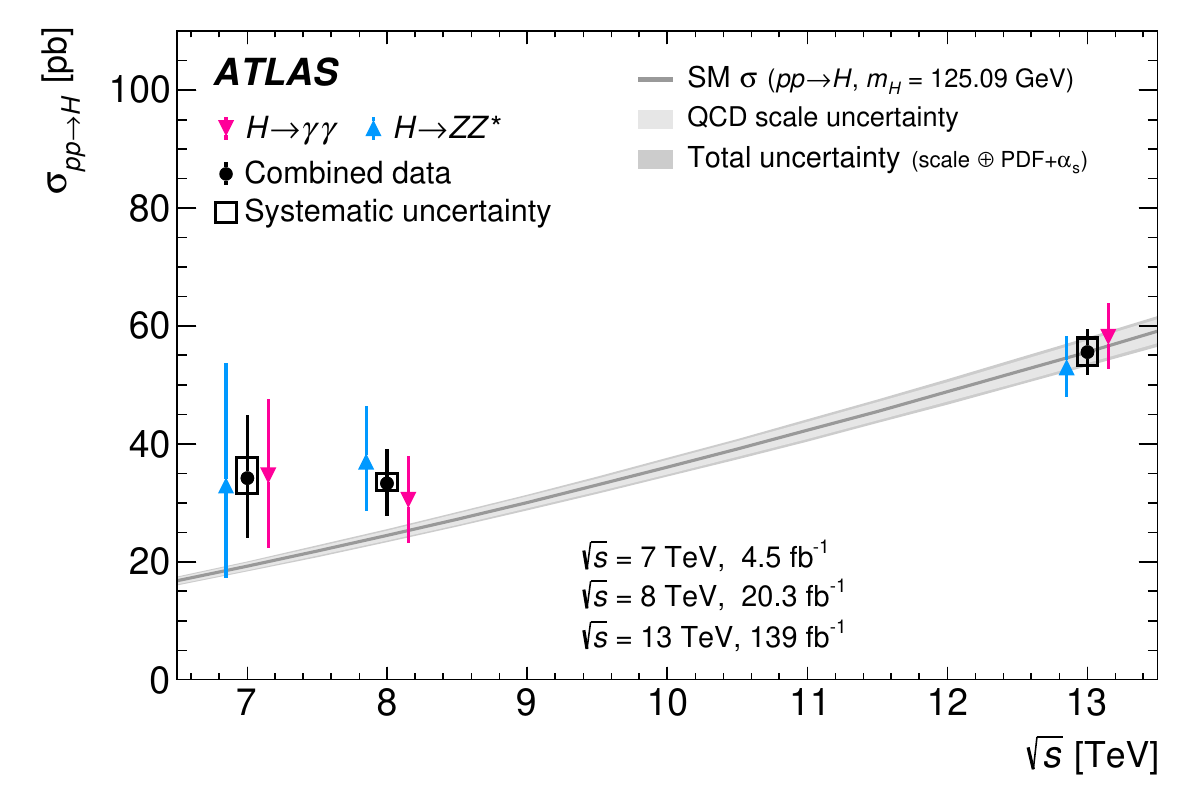}
\caption{Total $pp \rightarrow H + X$ cross-sections measured at centre-of-mass energies of 7, 8 and 13~\TeV\ in the \ZZfourl and \yy decay channels, and their combination, compared with SM predictions~\cite{HIGG-2022-04}.
\label{fig:totalxs}}
\end{figure}

\subsection{Differential cross-sections}

Measurements of differential cross-sections allow probing of Higgs boson kinematics, such as the Higgs boson transverse momentum \ptH and rapidity, and other aspects of Higgs boson production and decay, such as the number and properties of the associated hadronic jets.
The \ZZfourl and \yy differential cross-section measurements are performed individually and combined, to benefit from the reduced statistical uncertainties. As for the total cross-section, the necessary extrapolation introduces additional model-dependent uncertainties.

For each observable, a likelihood fit performs the extraction of the signal event yields in each bin. It uses dedicated factors for the correction of measurement inefficiencies, a migration matrix for the unfolding of resolution effects, and the measured luminosity to determine the cross sections. No regularisation is applied.
In addition to the experimental and background uncertainties, uncertainties related to the model dependence of the unfolding are included.

The \ptH distribution of the Higgs boson is one of the most interesting observables to measure as it could be sensitive to different phenomena beyond the SM (BSM). It could differ from the SM prediction  if the Higgs boson were produced together with an invisible particle or if so-far unknown particles contributed to the ggF loop.
In addition, the \ptH distribution, the overall production rate and the \Hyy branching fraction are sensitive to the Yukawa couplings of the Higgs
boson to the bottom and charm quarks. Both the contributions of these quarks to the loop-induced ggF production and $b\bar{b}$ and $c\bar{c}$ initiated Higgs boson production play an important role.
The \ptH would also be modified by the presence of new dynamics. Constraints on non-SM interactions can be set within an effective field theory model, as shown in Section~\ref{sec:higgs_combination}.   A combined measurement of \ptH, extrapolated to the total phase space, is shown in Figure~\ref{fig:diffxspt} (a). The binning was chosen to adequately capture the shape of the distribution for the subsequent interpretations.  A precision of 20\% is achieved in some regions.
Figure~\ref{fig:diffxspt} (b) shows the observed limits at 95\% CL on the Yukawa coupling-strength modifiers of the bottom and charm quarks,  defined as the ratios of the measured
Yukawa couplings to their SM values~\cite{HIGG-2021-12}, based on the shape and normalisation of the \ptH distribution. The  combination of these constraints with those from the \Hbb and \Hcc analyses is discussed in Section~\ref{sec:higgs_fermions}.

\begin{figure}
\centering
\subfigure[]{\includegraphics[width=0.45\textwidth]{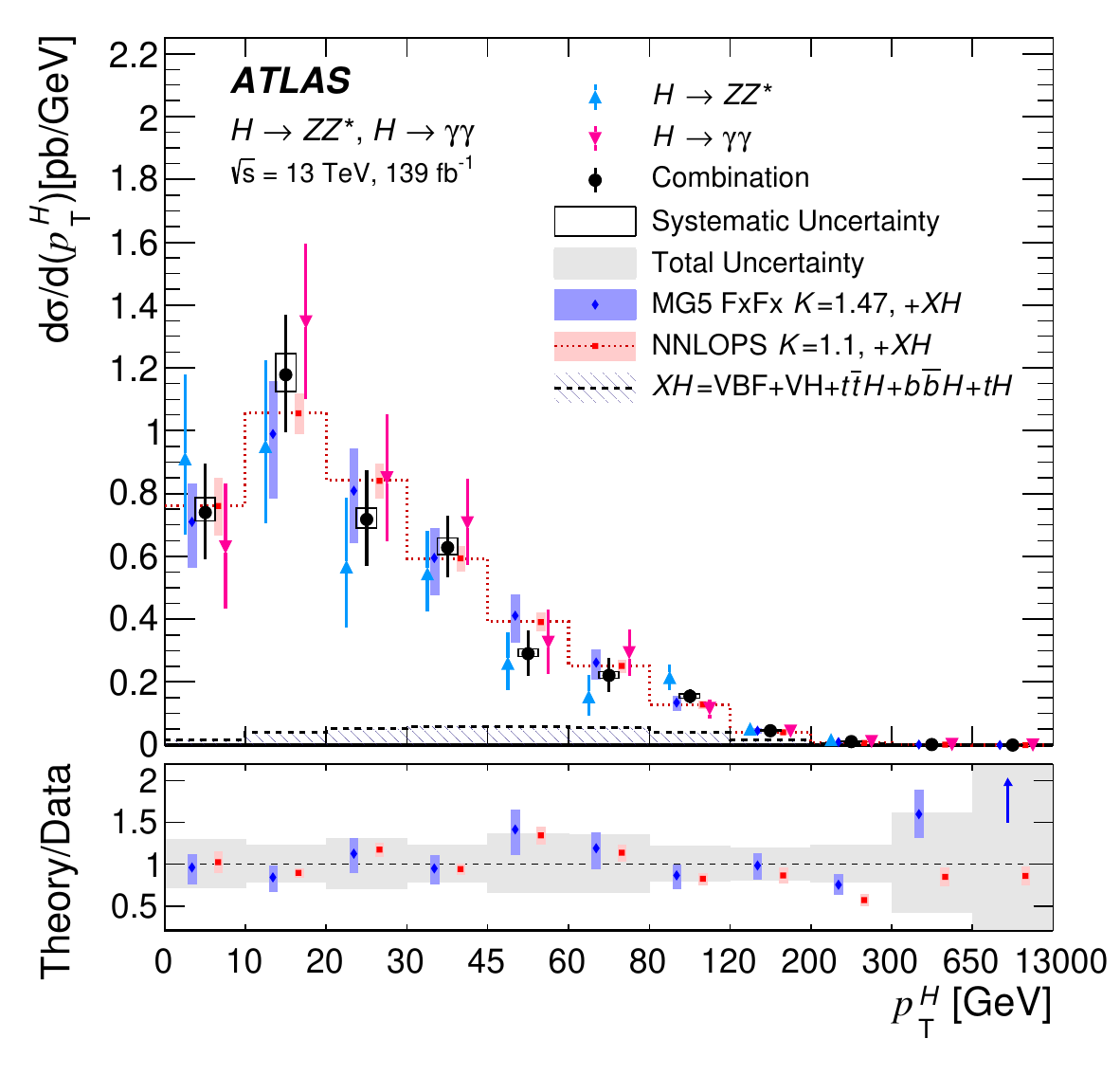}}
\subfigure[]{\includegraphics[width=0.45\textwidth]{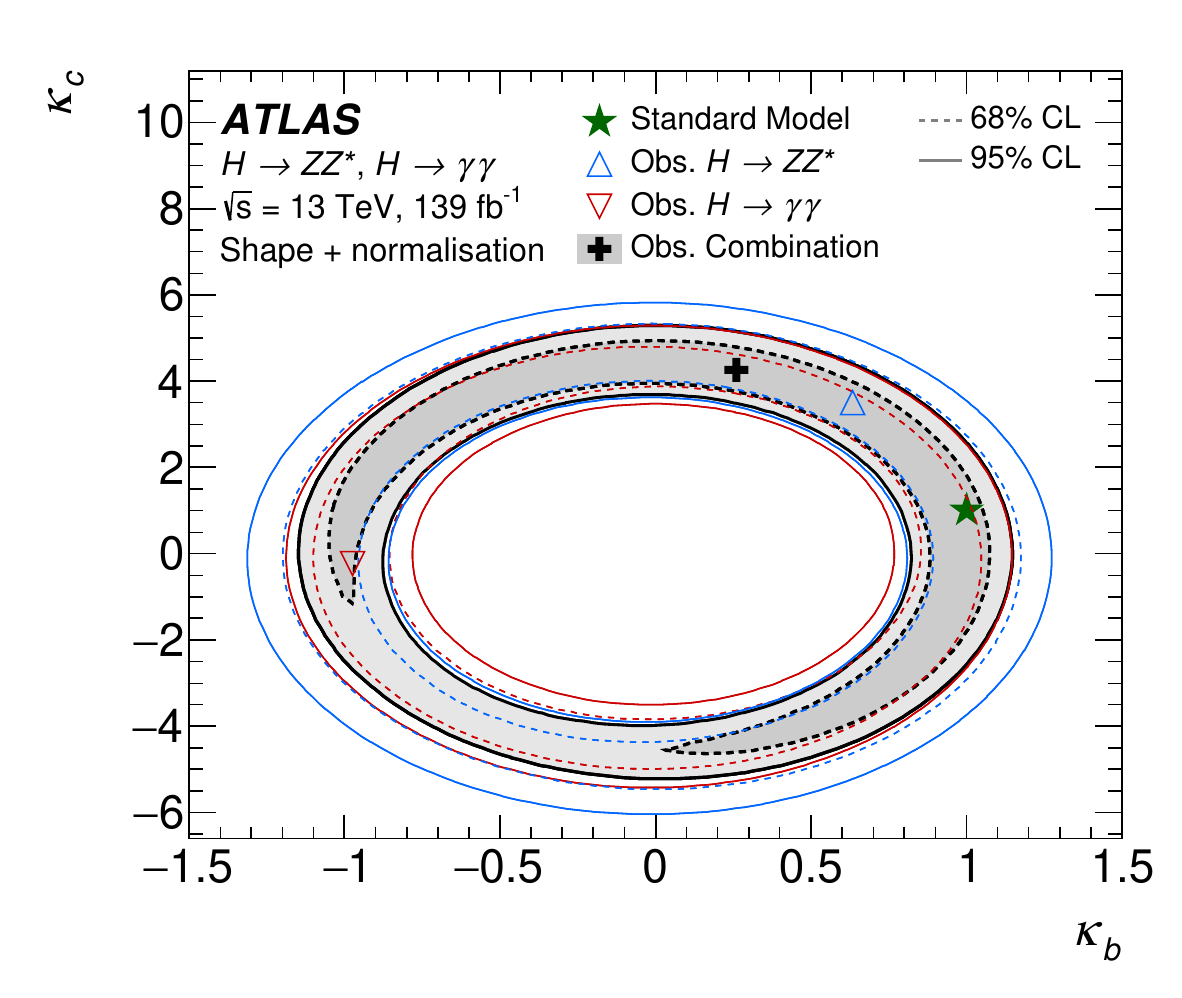}}
\caption{(a) Higgs boson transverse momentum measured in the \ZZfourl and \yy decay channels, and their combination,  compared with two SM ggF predictions, which are scaled to the total expected N$^3$LO cross-section and added to the predictions for the other production modes.  (b) Observed limits at 95\% CL on the Yukawa coupling-strength modifiers based on both the shape and normalisation of the Higgs boson transverse momentum distribution. The SM prediction and the observed best-fit values are also indicated~\cite{HIGG-2022-04}.
\label{fig:diffxspt}}
\end{figure}

Figure~\ref{fig:diffxs_obs} shows the rapidity of the Higgs boson measured in the \Hyy decay channel and the invariant mass of the off-shell $Z$ boson in the \HZZfourl decay channel. The rapidity of
the Higgs boson is sensitive to the parton distribution functions of the colliding protons, and is also influenced by QCD radiative corrections. The invariant mass of the off-shell $Z$ boson would be modified by the presence of a `dark' $Z$ boson mixing with the SM~\cite{Curtin:2013fra,Curtin:2014cca}. It is also sensitive to anomalous couplings and can be interpreted in the pseudo-observable framework~\cite{Gonzalez-Alonso:2014eva}.
The two results are a subset of the measured distributions, which cover a broad range of observables, ranging from Higgs boson kinematics to properties of the associated hadronic jets and of the Higgs boson decay products.

\begin{figure}
\centering
\subfigure[]{\includegraphics[width=0.45\textwidth]{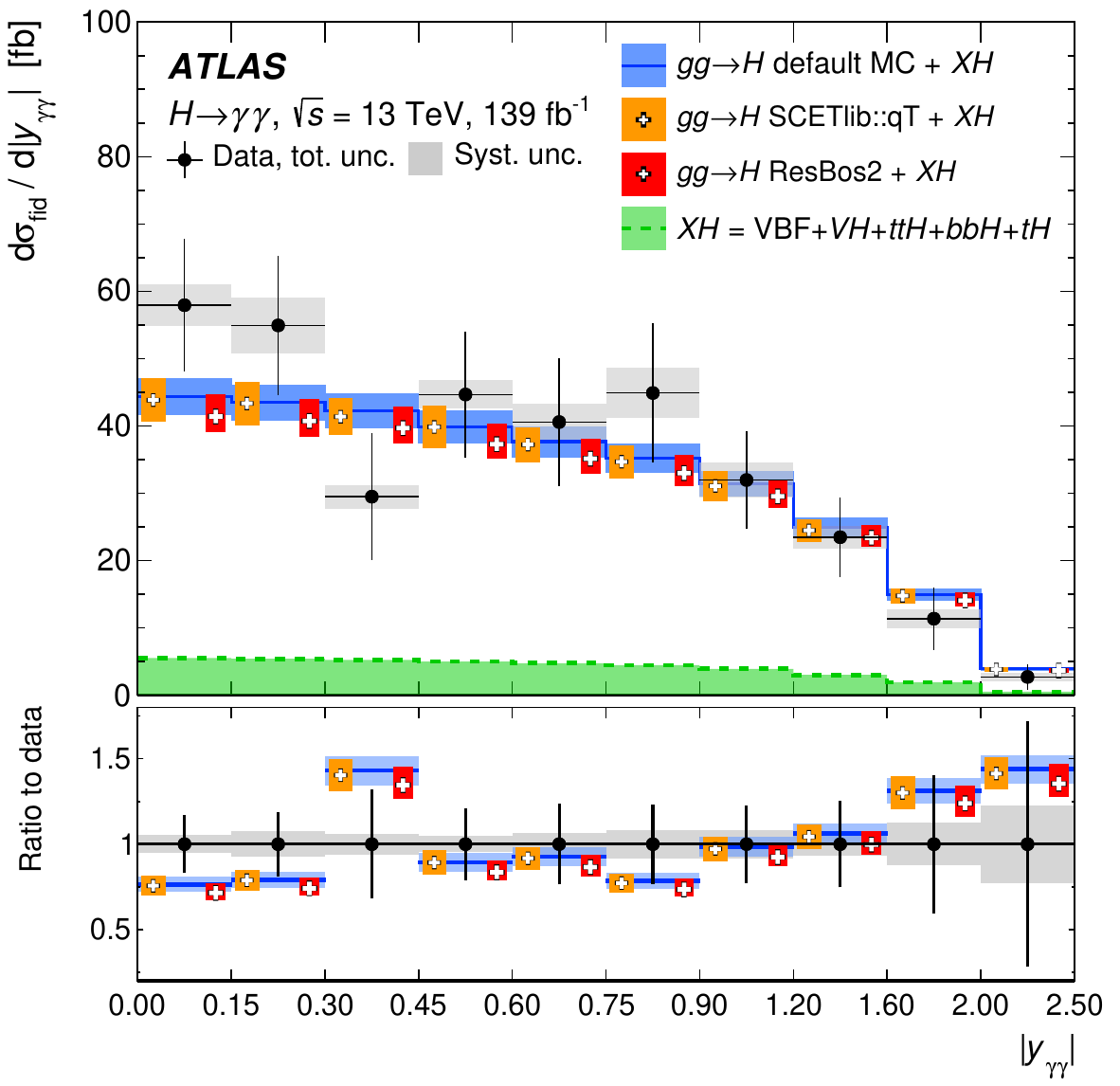}}
\subfigure[]{\includegraphics[width=0.45\textwidth]{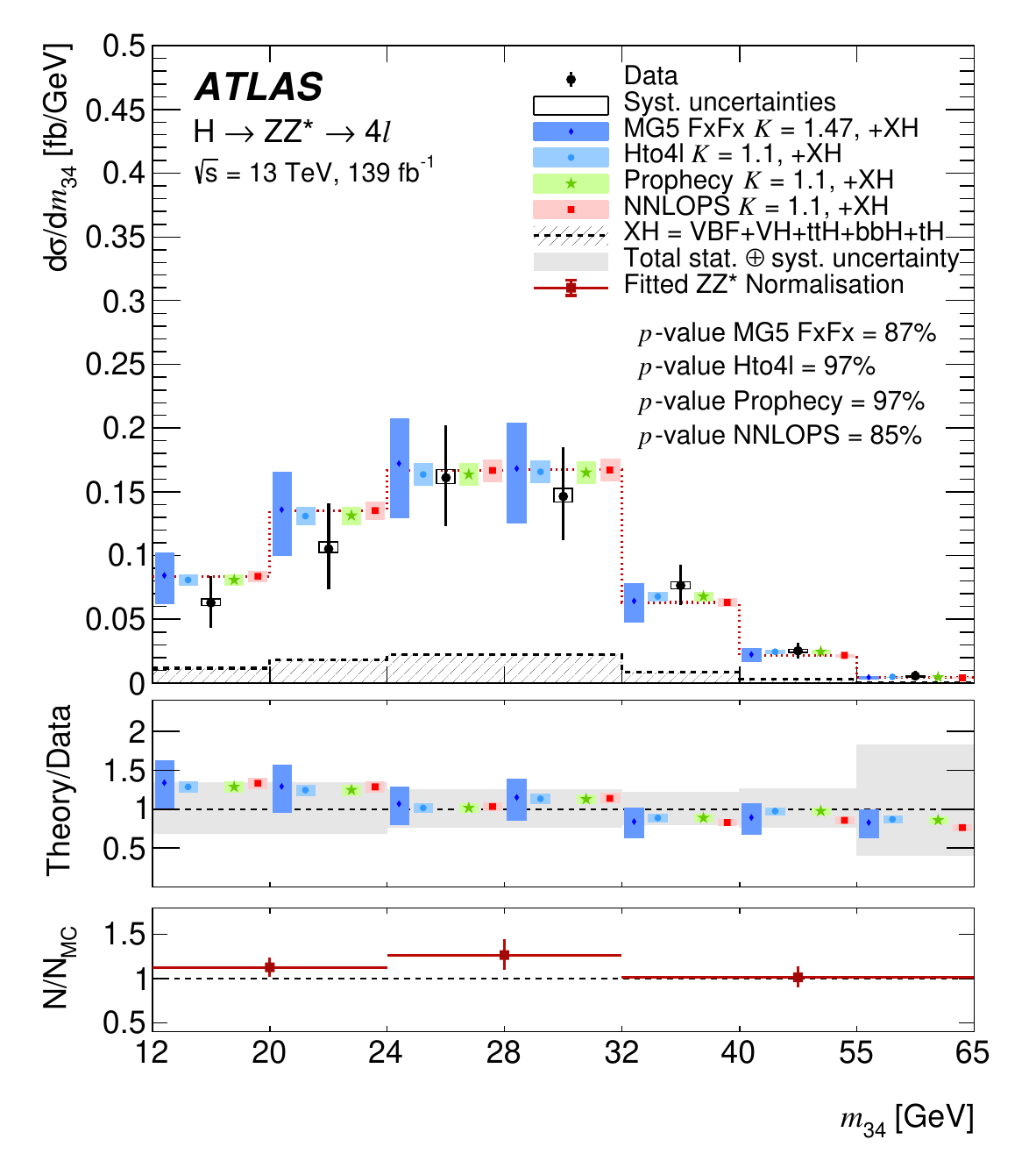}}
\caption{
(a) The rapidity of the Higgs boson measured in the \Hyy decay channel~\cite{HIGG-2019-13}, and (b) the invariant mass of the off-shell $Z$ boson in the \HZZfourl decay channel~\cite{HIGG-2018-29}. The measurements are compared with different SM ggF predictions that are added to predictions for the other production modes. For the invariant mass, the ggF predictions are scaled to the total N$^3$LO cross-sections.
\label{fig:diffxs_obs}}
\end{figure}

Despite the missing information from the neutrinos in the final state and thanks to the relatively high branching ratio, it is also possible to measure differential cross-sections in the \WWevmuv final state~\cite{HIGG-2018-49,HIGG-2020-25}. In this channel, the measurements are not performed inclusively, but separately for ggF and VBF production. Events are selected if they contain two oppositely charged, different-flavour leptons (electrons or muons). Further requirements are applied to ensure a non-negligible missing transverse momentum in the event. To suppress $\tau\tau$ background from the $\Upsilon$ meson, or from an off-shell photon or $Z$ boson (Drell--Yan) a minimum requirement on the invariant mass of the two leptons is applied. After the event selection, the dominant backgrounds to the Higgs boson signal are non-resonant $WW$ production, top pair production, the Drell--Yan process, and jets that are misidentified as leptons. Dedicated control regions are created that are enriched in these background processes. They are used in the subsequent fit to extract correction factors to the estimated yields, which are extracted from MC for most backgrounds. To estimate  the background from events with jets misidentified as leptons, correction factors are determined from data and applied to the corresponding control region. In the VBF analysis, an additional control region enriched in ggF events is added to the fit to estimate this background. In the ggF analysis, the small contributions from other Higgs boson production modes are estimated by using simulated events scaled to the expected SM cross-sections. The unfolding is performed by inverting the response matrix. In the ggF analysis, a Tikhonov regularisation term~\cite{Tikhonov} is included in the likelihood to avoid large statistical fluctuations. Figure~\ref{fig:diffxsptWW} shows \ptH measured for VBF production in the \WWevmuv decay channel.

\begin{figure}
\centering
\includegraphics[width=0.45\textwidth]{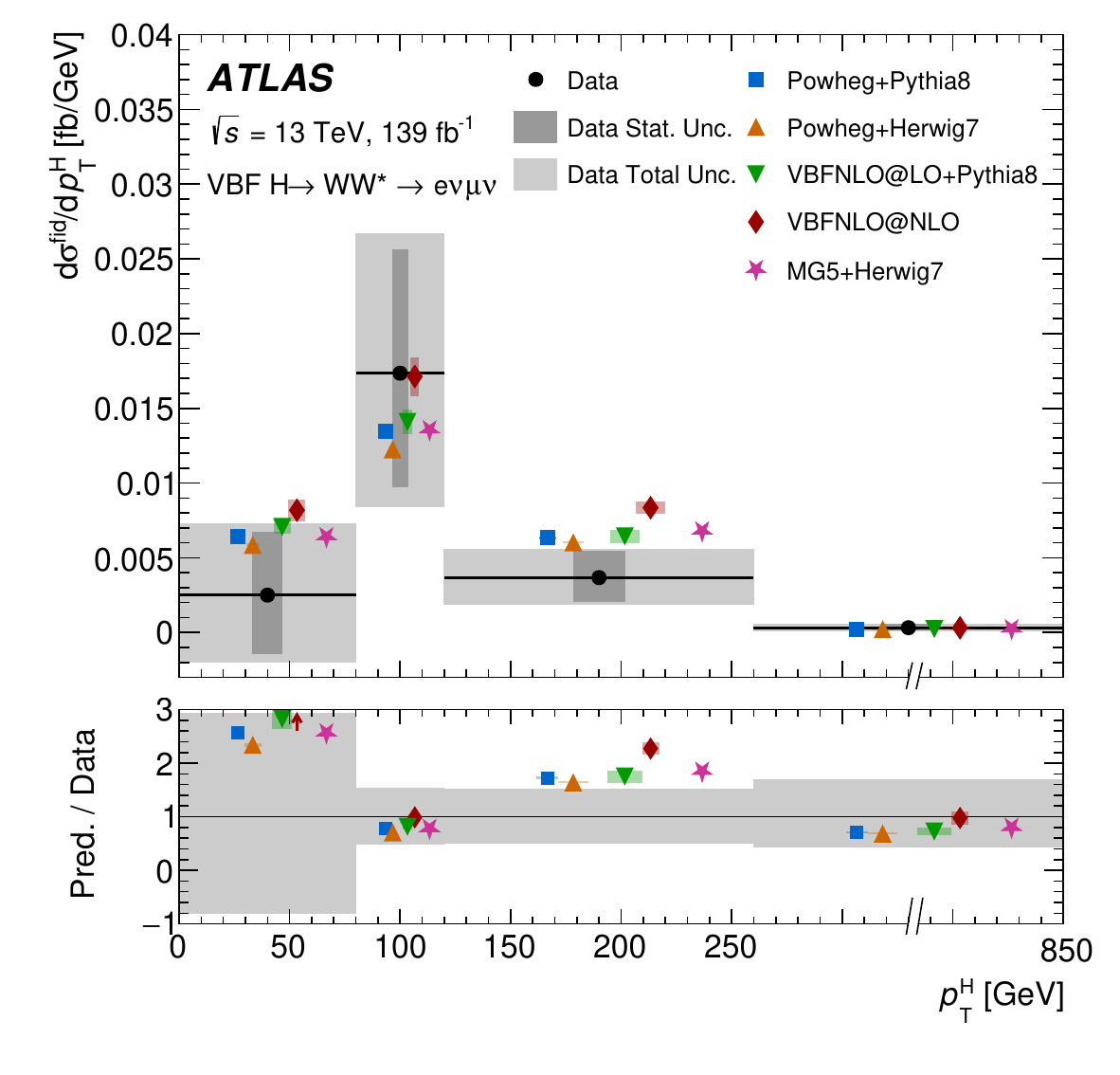}
\caption{Higgs boson transverse momentum measured for VBF production in the \WWevmuv decay channel compared with different SM predictions~\cite{HIGG-2020-25}.
\label{fig:diffxsptWW}}
\end{figure}

Good agreement is seen between all measured differential cross-sections and the state-of-the-art SM predictions.

\FloatBarrier


\section{Higgs boson couplings to fermions}
\label{sec:higgs_fermions}

\subsection{Observation of the Higgs boson couplings to third generation fermions: \tauls, top and bottom quarks }
In the SM, fermions obtain
their masses by interacting with the Higgs  field via the Yukawa coupling.
One of the major achievements during Run~2 of the LHC was the direct confirmation of the Yukawa couplings to third generation fermions. This was accomplished by studying the fermionic decays of the Higgs boson and the associated production of the Higgs boson with pairs of top quarks.

\textbf{The observation of Higgs boson decays into two \tauls} was established during Run~1 through a combination of measurements by the ATLAS and CMS Collaborations, with a significance of $5.5\sigma$~\cite{HIGG-2015-07}. ATLAS achieved a single-experiment observation using 36.1 $\text{fb}^{-1}$ of Run~2  data in combination  with Run~1 data~\cite{HIGG-2017-07}.

All combinations of hadronic and leptonic \taul decays are considered. They are treated as separate analysis channels due to differences in their background composition. In all channels, the dominant background is the $Z\to\tau\tau$ process. Therefore, the best sensitivity can be achieved for topologies in which $Z$ boson production is suppressed, such as Lorentz-boosted ggF and VBF production. The relative contributions from other backgrounds, primarily from top-quark and vector-boson decays, as well as from misidentified leptonic or hadronic \taul decays, vary considerably depending on the \taul decay mode.
The invariant mass of the $\tau\tau$ system is used to discriminate between  Higgs boson and $Z$ boson production and to extract the signal yield,  making the mass resolution  extremely important. An algorithm is employed that determines the most probable $\tau\tau$ mass for the event, assuming that the measured missing transverse momentum originates from neutrinos in a resonant $\tau\tau$ final state~\cite{Elagin:2010aw}. Several categories are defined based on the Higgs boson production mode, particularly probing VBF and  boosted Higgs boson production via ggF. The normalisations of the major backgrounds are constrained using data from different control regions. The yields of the $Z\to\tau\tau$ background are estimated directly from a fit to the final discriminating variable.

In Figure~\ref{fig:ferm_tau} (a), the reconstructed invariant mass of the $\tau\tau$ system is shown for all  signal regions using the 2015--2016 ATLAS data sample~\cite{HIGG-2017-07}. The $H\to\tau\tau$ signal
is established with an observed (expected) significance of 4.4$\sigma$ (4.1$\sigma$), which increases to 6.4$\sigma$ (5.4$\sigma$) when combined with the Run~1 results.

With the data collected by the end of Run 2, it is possible to probe the $H\to\tau\tau$ cross-section in the four major production modes (ggF, VBF, $VH$, and \ttH), and as a function of the kinematic properties of the events in ggF and electroweak production processes~\cite{HIGG-2019-09}.  In Figure~\ref{fig:ferm_tau} (b), the reconstructed invariant mass of the $\tau_{\mathrm{lep}}\tau_{\mathrm{had}}$ pair is shown, where one $\tau$-lepton decays leptonically and one hadronically.
The measured cross-sections times branching ratio for $|y_{H}|<2.5$ in the  dominant ggF and VBF production modes are $2.65  \pm 0.41 \text{(stat.)}  ^{+\,0.91}_{-\,0.67} \text{(syst.)}$~pb  and $0.197 \pm 0.028 \text{(stat.)} ^{+\,0.032}_{-\,0.026} \text{(syst.)}$~pb respectively, in agreement with the SM expectations. Cross-sections of the production of a Higgs boson decaying into \tauls are measured as a function of \ptH, the number of jets produced in association with the Higgs boson, and the invariant mass of the two leading jets m$_{jj}$, as shown in Figure~\ref{fig:ferm_tau} (c). The achieved statistical and systematic uncertainties are 24\% for the electroweak production of the Higgs boson with two jets  and m$_{jj}$ greater than 350~\GeV and about 40\% for the boosted ggF production.

\begin{figure}
\centering
\subfigure[]{\includegraphics[width=0.4\textwidth]{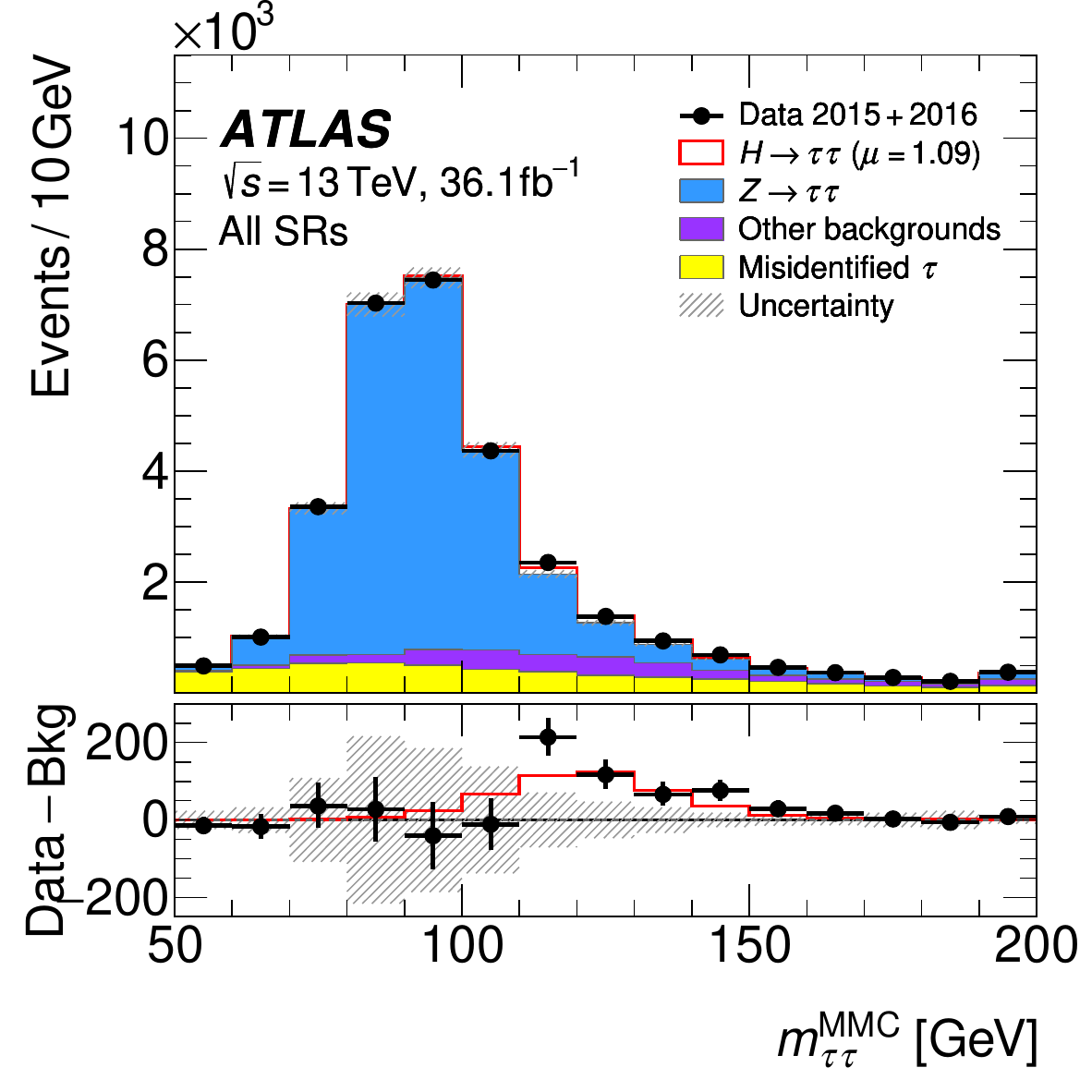}}
\subfigure[]{\includegraphics[width=0.4\textwidth]{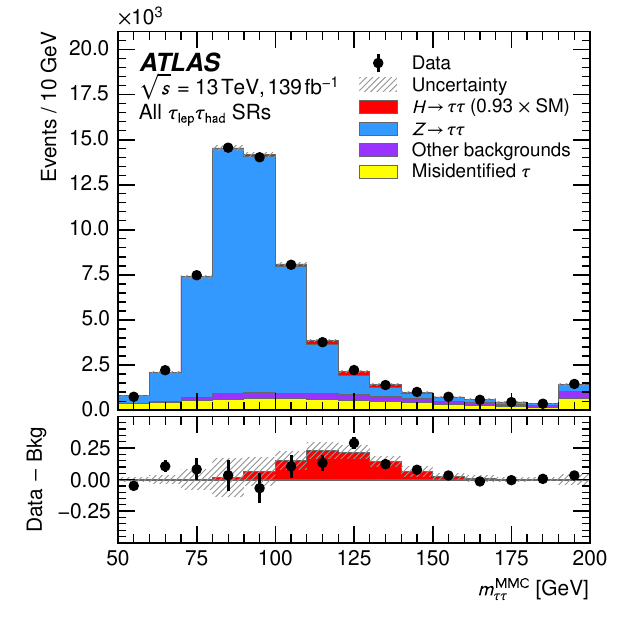}}
\subfigure[]{\includegraphics[width=0.7\textwidth]{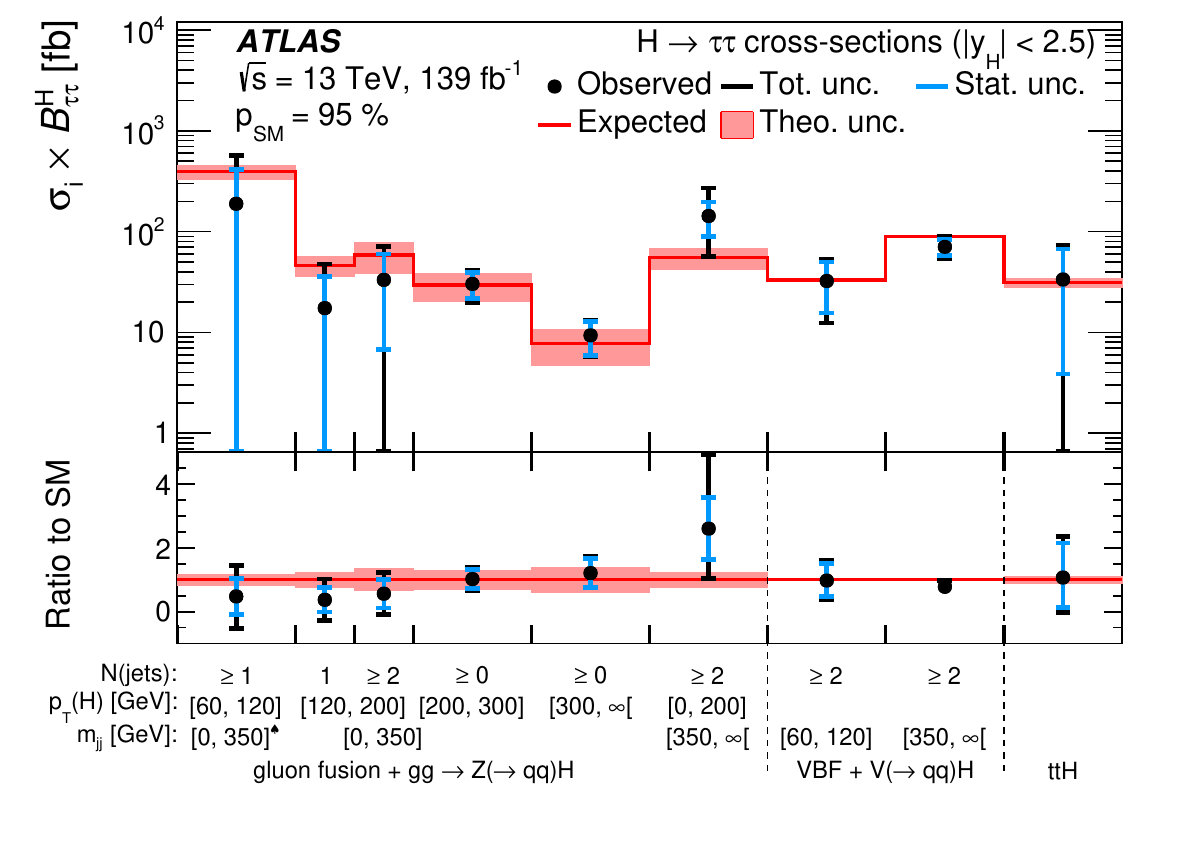}}
\caption{(a) The invariant mass of the $\tau\tau$ system for all combinations of \taul decays using 36.1 $\text{fb}^{-1}$ of data~\cite{HIGG-2017-07}. (b) The same observable considering only signal regions with one \taul decaying leptonically and one decaying hadronically for 139 $\text{fb}^{-1}$of data ~\cite{HIGG-2019-09}. (c) The measured cross-sections times branching ratio in nine different kinematic regions compared with the SM expectations~\cite{HIGG-2019-09}.
\label{fig:ferm_tau}}
\end{figure}

\textbf{The observation of \ttH production}~\cite{HIGG-2018-13} was an important breakthrough. Since the strength of the Yukawa interaction is proportional to the fermion mass, the predicted Yukawa coupling of the top quark is the largest one in the SM, with a value close to unity. Indirect constraints on this coupling were already obtained through ggF production and Higgs boson decays into photons, which occur via quantum loops (Figure~\ref{fig:prod_decays}). In the SM,  the dominant contribution to the former arises from diagrams featuring the top quark or the bottom quark in the loop, along with their interference.  Meanwhile, in the latter case, the primary contribution
is due to the $W$ boson in the loop  or the top quark~\cite{Heinemeyer:2013tqa}. However, so-far unknown phenomena could affect these loops, therefore a direct measurement through \ttH production is necessary to disentangle and probe potential BSM effects.

The presence of complex final states and large irreducible backgrounds, combined with a small predicted rate (about 1\% of the total Higgs boson production), makes the measurement of \ttH challenging. Different analyses using up to 79.8~fb$^{-1}$ of Run~2 data and targeting various Higgs boson decay channels were combined, leading to the observation of this production mode in 2018~\cite{HIGG-2018-13}.

The search for \ttH production via \Hbb decays~\cite{HIGG-2017-03} benefits from the large Higgs boson branching fraction to bottom quarks, but suffers from sizeable backgrounds, primarily from events with a top quark pair and additional heavy-flavour jets, which are also challenging to model. Events are selected in which one or both top quarks decay semileptonically. The events are further categorised using the jet multiplicity and the probability that a certain jet originates from a bottom quark ($b$-tagging). This allows nine signal-enriched regions to be defined, with a maximum purity at the level of about 5\%, and ten background-enriched control regions. Multivariate discriminants based on kinematic variables are trained for each region to further enhance the signal-to-background separation.
The signal yield is extracted through a global fit in which the normalisation of the \ttbar+heavy flavour background components is allowed to float freely. An excess over the background-only hypothesis is observed (expected) with a significance of 1.4$\sigma$ (1.6$\sigma$)  with the 2015--2016 data sample, corresponding to 36.1 \ifb of Run~2 data.

Searching for \ttH production, the \HWW, \Htautau, and the \HZZ decays are difficult to disentangle, as they lead to similar final states featuring the presence of multiple leptons. Therefore, they are treated together in a common $\ttH, H\to$ multilepton analysis~\cite{HIGG-2017-02}. The \ttH, \HZZfourl contribution is studied in a dedicated analysis and minimized to be negligible in the $\ttH, H\to$ multilepton analysis with a selection based on the four-lepton invariant mass. Events are categorised into seven orthogonal categories based on the number of light leptons and hadronically decaying \taul candidates. The \ttbar background is suppressed by requiring two leptons with the same charges or at least three leptons of any charges. The defined categories have different background compositions and a wide range of total yields and signal purities. The irreducible background is predominantly composed of SM processes with prompt leptons, such as \ttW, \ttZ, and diboson production. Control regions are used to extract the normalisation for these backgrounds and estimate the backgrounds with non-prompt light leptons and mis-identified hadronically decaying \taul
candidates. Dedicated multivariate discriminants are employed in most of the signal regions to further enhance the signal-to-background separation. In Figure \ref{fig:ferm_top} (a), the observed and expected yields in the signal regions and control regions are shown. A combined profile likelihood fit across all regions extracts the signal with an observed (expected) significance of $4.1\sigma$ ($2.8\sigma$) with the 2015--2016 data sample, corresponding to 36.1 \ifb of Run~2 data.

The \ttH, \HZZfourl final state is treated in a dedicated analysis~\cite{HIGG-2018-13}, separating events in which both top quarks decay hadronically from events in which at least one top quark decays semileptonically and employing a boosted decision tree (BDT) in each region. Due to the small Higgs boson branching ratio to four leptons, the expected rates are very low. The expected background contamination is also small. The simulation predicts 0.6 signal events, and an expected significance of $1.2\sigma$. No events are observed with 79.8~\ifb of Run~2 data.

The \ttH analysis using \Hyy decays~\cite{HIGG-2018-13} benefits from a very clean signature, with a peak in the invariant mass distribution of the diphoton system over a smooth background, but it suffers from a  low rate. The main background processes are non-resonant diphoton production and other Higgs boson production modes. Regarding the associated top quark decays, this analysis distinguishes between leptonic events (where at least one light lepton is present) and hadronic channels. BDTs are employed in various analysis categories to separate the signal from backgrounds.
As shown in Figure~\ref{fig:ferm_top}~(b), the signal is extracted from signal-plus-background fits to the $\gamma\gamma$ invariant mass distribution, with the background constrained by the mass sidebands. The resulting observed (expected) significance is 4.1$\sigma$ (3.7$\sigma$) with 79.8~\ifb of Run~2 data.

Combining the various decay modes, along with the results obtained using the Run~1 data, an observed (expected) significance of 6.3$\sigma$ (5.1$\sigma$) for \ttH production is achieved with 36.1$-$79.8~\ifb\ of Run 2 data~\cite{HIGG-2018-13}. Assuming SM branching fractions, the total \ttH production cross-section at 13~\TeV is measured to be $670 \pm 90 \mathrm{(stat.)} ^{+110}_{-100} \mathrm{(syst.)}$ fb, in agreement with the SM prediction. Figure~\ref{fig:ferm_top}~(c) shows the
combined event yields in all analysis categories as a function of $\log_{10}(S/B)$, where $S$ and $B$ are the expected signal and background yields extracted from the fit, with a clearly visible signal-like
excess over the background at high $\log_{10}(S/B)$.

\begin{figure}
\centering
\subfigure[]{\includegraphics[width=0.49\textwidth]{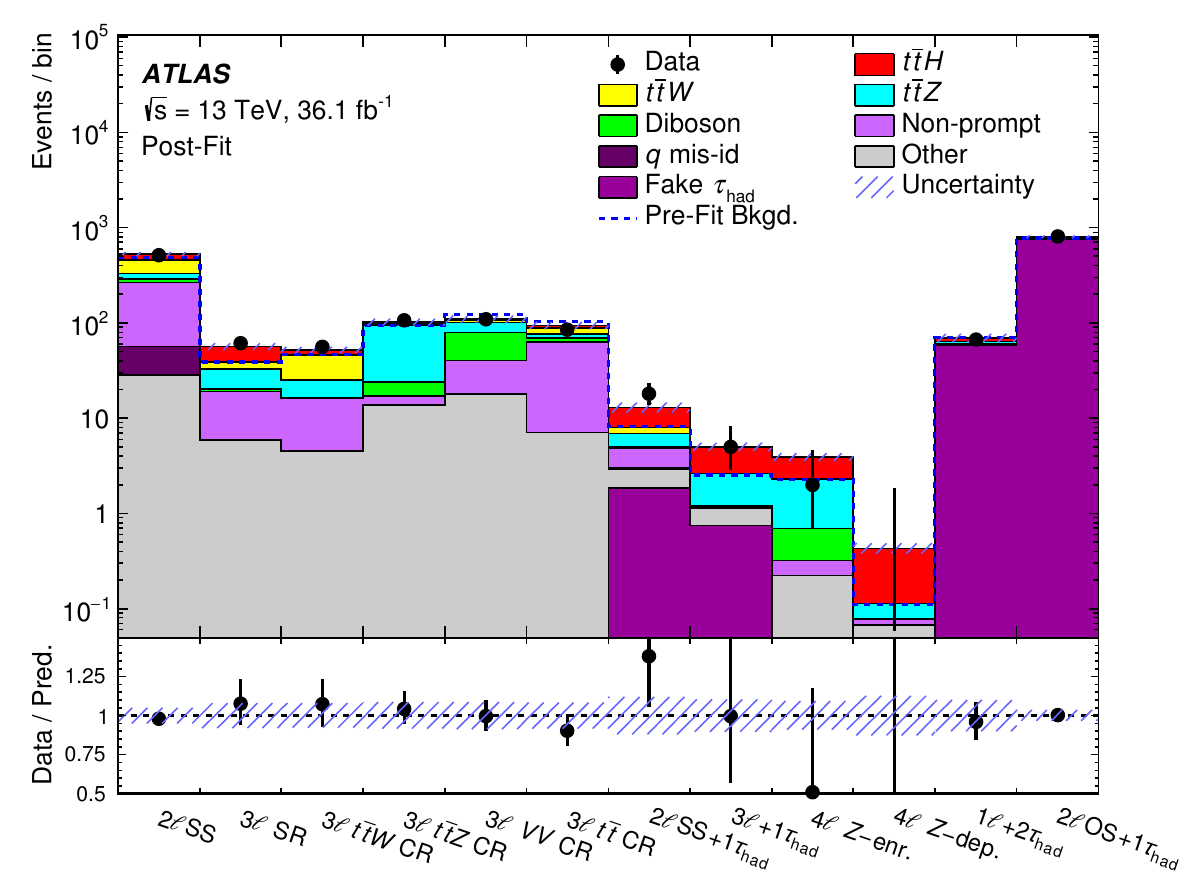}}
\subfigure[]{\includegraphics[width=0.49\textwidth]{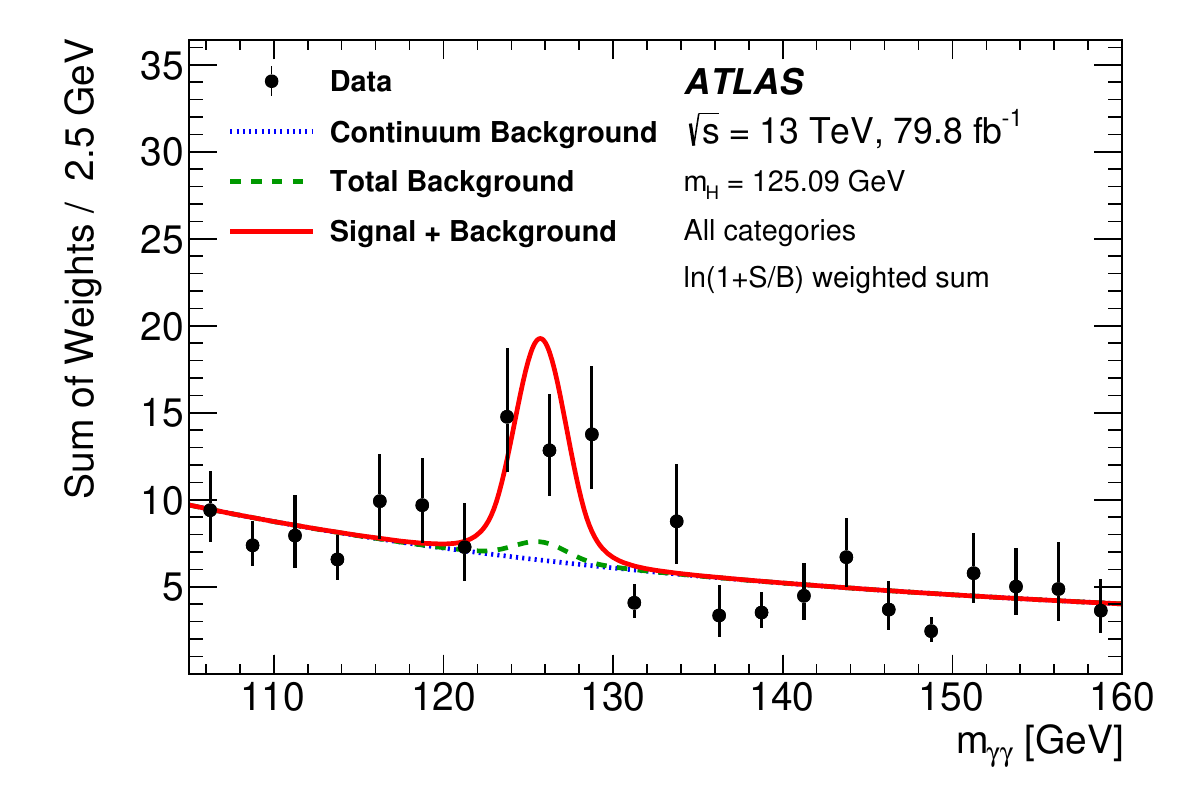}}
\subfigure[]{\includegraphics[width=0.45\textwidth]{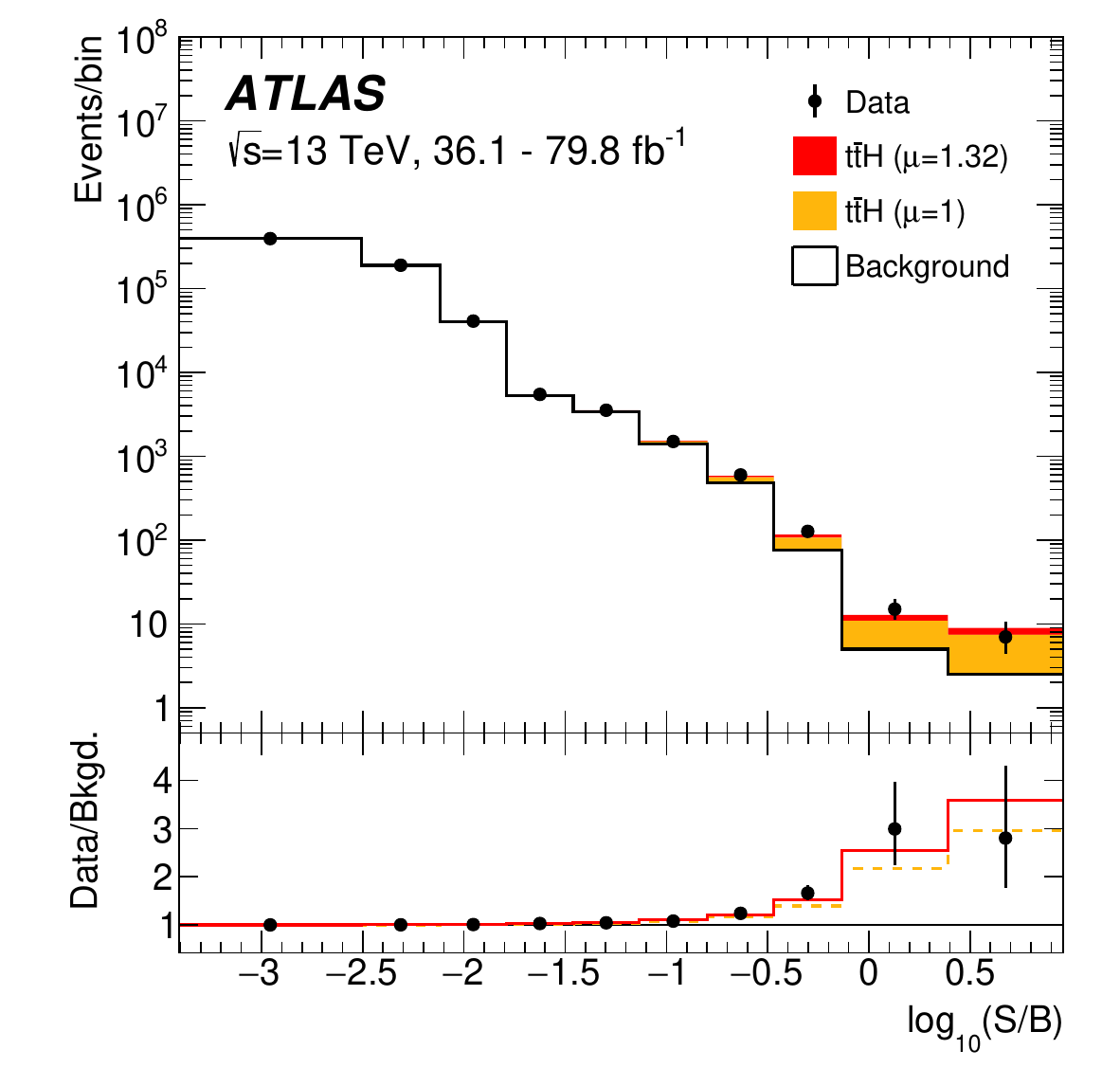}}
\subfigure[]{\includegraphics[width=0.49\textwidth]{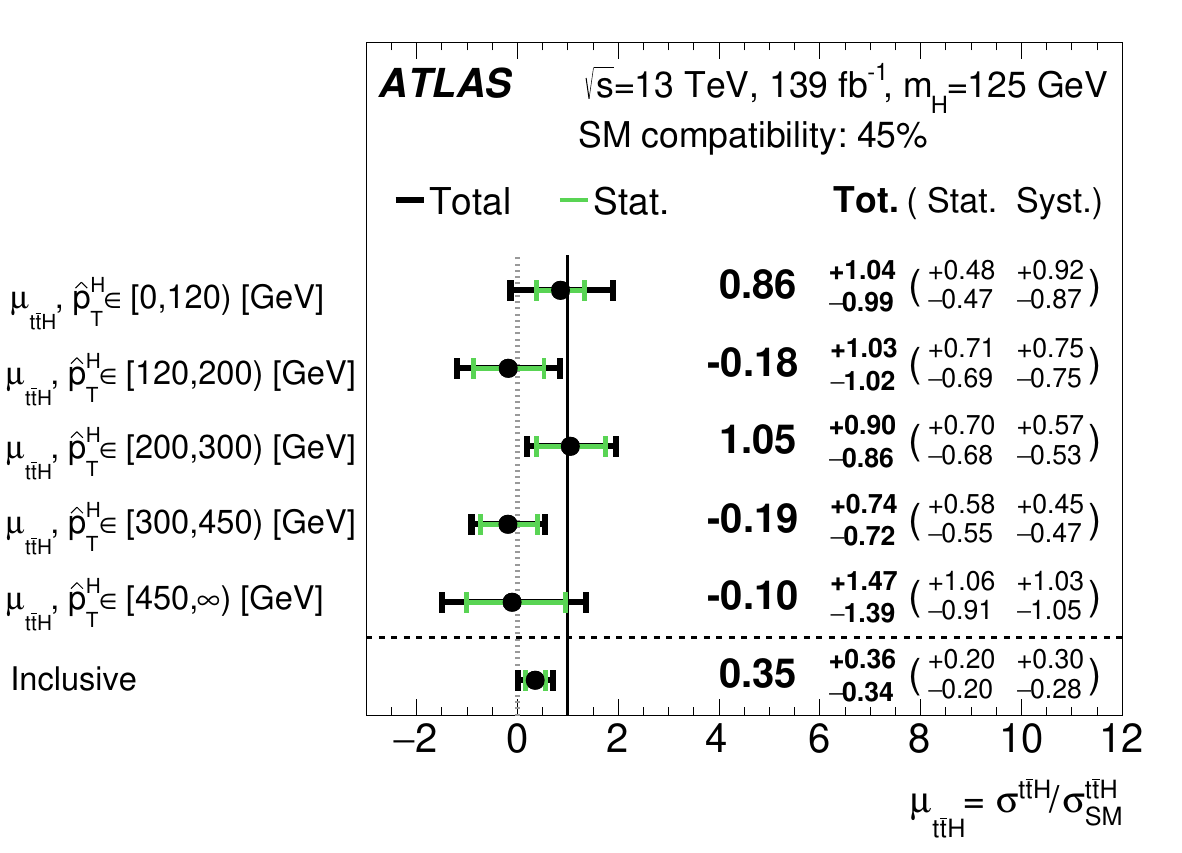}}
\caption{(a) Comparison of the data to the prediction in the different signal and control regions of the \ttH, $H \rightarrow$ multilepton analysis, after the fit is performed~\cite{HIGG-2017-02}. (b) The diphoton invariant mass distribution in the $\ttH, \Hyy$  analysis~\cite{HIGG-2018-13}. (c) Event yields as a function of $\log_{10}(S/B)$ for data, background and a Higgs boson signal for all decay channels considered, using up to 79.8~\ifb of data, leading to the observation of \ttH production~\cite{HIGG-2018-13}. (d) Measured \ttH production cross-sections normalised to the SM expectation as function of \ptH in the \Hbb final state~\cite{HIGG-2020-23}.
\label{fig:ferm_top}}
\end{figure}

Exploiting the full Run~2 data sample, it is  possible to make cross-section measurements as a function of \ptH in \ttH production, both in the \Hyy \cite{HIGG-2020-16} and \Hbb\cite{HIGG-2020-23} final states. Figure~\ref{fig:ferm_top}~(d) shows the results obtained in the \Hbb final state.

\textbf{The observation of Higgs boson decays into bottom-quark pairs} in 2018 was another significant milestone in the understanding of Higgs boson interactions~\cite{HIGG-2018-04}. This observation provides direct evidence of the Yukawa coupling of the Higgs boson with down-type quarks.

The \Hbb decay has the largest branching ratio in the SM at about 58\%. However, inclusive \Hbb searches are very challenging at the LHC due to backgrounds arising from multijet production that are orders of magnitudes larger than Higgs boson production. The most sensitive analysis is based on $VH$  production, where events are categorised according to the number of charged leptons (muons or electrons) from the vector boson decays. Events containing zero or two leptons target $ZH$ production, while events with one lepton target $WH$ production. All events are required to have at least two jets, of which exactly two must have a high probability to originate from bottom quarks. The primary background sources are $V+$jets and \ttbar production, and their normalisation is constrained using dedicated control regions. The separation between the signal and background is achieved using the invariant mass of the $b\bar{b}$ system  $m_{b\bar{b}}$ and several other kinematic observables, such as the transverse momentum of the vector boson. These observables are used as input to BDTs in the different signal regions. Based on the first 79.8~\ifb of Run~2 data, the analysis yields an observed significance for $VH$, \Hbb production of 4.9$\sigma$ (4.2$\sigma$ expected). Combining this result with the searches for \Hbb in other Higgs boson production modes, a significance of $5.4\sigma$ is observed, with an expectation of 5.5$\sigma$. Furthermore, the combination of this result with the other searches for $VH$ production provides the first observation of this process with a significance of 5.3$\sigma$ (4.8$\sigma$ expected).

In addition, an analysis without multivariate discriminants (cut-based analysis) is performed as a cross-check. In this case, $m_{b\bar{b}}$ is used as the discriminating variable and additional selection criteria are employed, e.g., on the separation between the two $b$-tagged jets, to increase the  purity of the signal regions.
The $VZ$ process with the $Z$ boson decaying into $b\bar{b}$ offers a powerful validation of the cut-based analysis, as this process closely resembles the Higgs boson signal except for the lower invariant mass of the $b\bar{b}$ system. Figure~\ref{fig:ferm_bb}~(a) shows the invariant mass distribution of the $b\bar{b}$ system after subtracting all  backgrounds except for the $VZ$ process, using the first 79.8~\ifb of data in Run 2. The $VZ$ contribution is clearly visible, along with a significant excess at higher $m_{b\bar{b}}$ values due to $VH$ production.

The full Run~2 data sample allows further improvements in the $VH$, \Hbb measurements~\cite{HIGG-2018-51}, yielding a measured signal strength relative to the SM prediction of $\mu_{VH}^{bb}=1.02^{+0.12}_{-0.11}\mathrm{(stat.)}^{+0.14}_{-0.13}\mathrm{(syst.)}$. Figure~\ref{fig:ferm_bb}~(b)  shows the observed and predicted discriminant distribution used to separate the Higgs boson signal from the backgrounds in one of the signal regions. In addition, cross-sections of associated production of a Higgs boson decaying into bottom quark pairs and a vector boson decaying into leptons are measured as a function of the vector boson transverse momentum in kinematic fiducial volumes, with uncertainties ranging from 30\% in the highest $p_{\mathrm{T}}^V$ region ($p_{\mathrm{T}}^V > 250$~\GeV) to 85\% in the lowest ($75~\GeV < p_{\mathrm{T}}^Z < 150$~\GeV).

For transverse momenta of the Higgs boson above about 300~\GeV, the decay products of the Higgs boson in the $VH, \Hbb$ process cannot always be reconstructed through two jets with a small radius parameter $R$ ($R=0.4$). The collimated Higgs boson decay products may however be reconstructed as a single jet with a large radius parameter ($R=1$). This boosted region  is particularly interesting due to its sensitivity to BSM physics at high energy scales. The measured signal strength for the $VH, \Hbb$ events reconstructed as a single large-radius jet  with $p_{\mathrm{T}}^V > 250$~\GeV  with the full Run 2 data is $\mu_{VH}^{bb}=0.72^{+0.39}_{-0.36}$~\cite{HIGG-2018-52}.

A combination of the $VH, \Hbb$ measurements targeting different
transverse momentum regions by reconstructing the Higgs boson as two individual jets~\cite{HIGG-2018-51}  or a single large-radius jet~\cite{HIGG-2018-52} is also performed~\cite{ATLAS-CONF-2021-051}. The overlap between the two analyses is addressed by switching from the former to the latter at a $p_{\mathrm{T}}^V$ of $400$~\GeV.  Cross-sections are measured as function of the transverse momentum of the $W$ and  $Z$ bosons and found to be compatible with the SM expectations, as shown in Figure~\ref{fig:ferm_bb} (c).

Higgs bosons decaying into bottom-quark pairs are also studied in the VBF production mode~\cite{HIGG-2019-04}, resulting in a measured yield of $0.95^{+0.38}_{-0.36}$ times the one predicted by the SM with the full Run~2 data sample, corresponding to an observed (expected) significance of $2.6\sigma$ ($2.8\sigma$).

Run~2 data are also exploited to study the inclusive production of the Higgs boson at high transverse momentum in the \Hbb decay channel~\cite{HIGG-2021-08}. The observed (expected) 95\% CL upper limit on the Higgs
boson production cross-section for $p_{\mathrm{T}}^{H}>1$~\TeV is 9.6 (7.4)~fb, while the SM cross-section in the same kinematic region is 0.13~fb.

\begin{figure}
\centering
\subfigure[]{\includegraphics[width=0.4\textwidth]{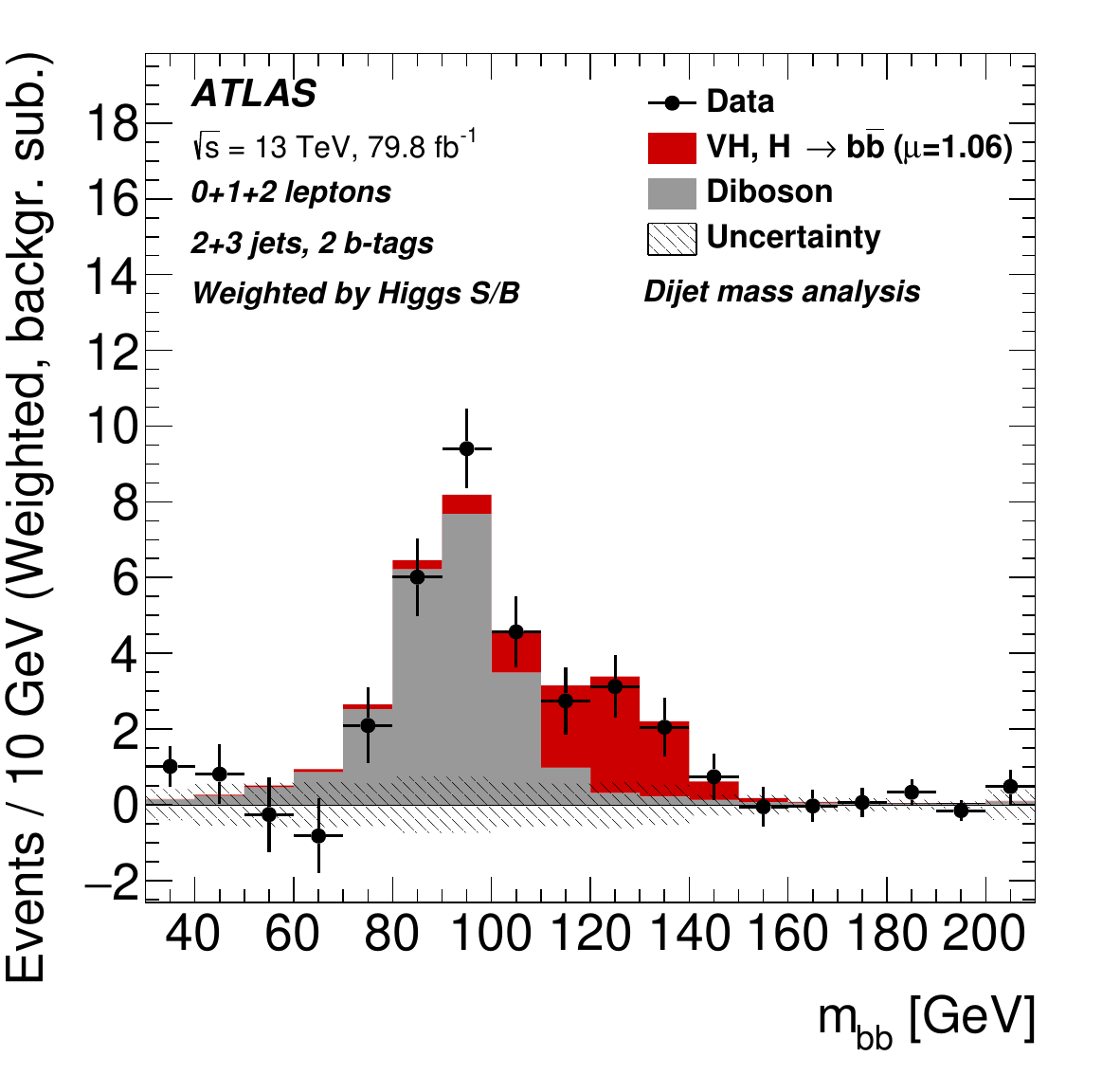}}
\subfigure[]{\includegraphics[width=0.4\textwidth]{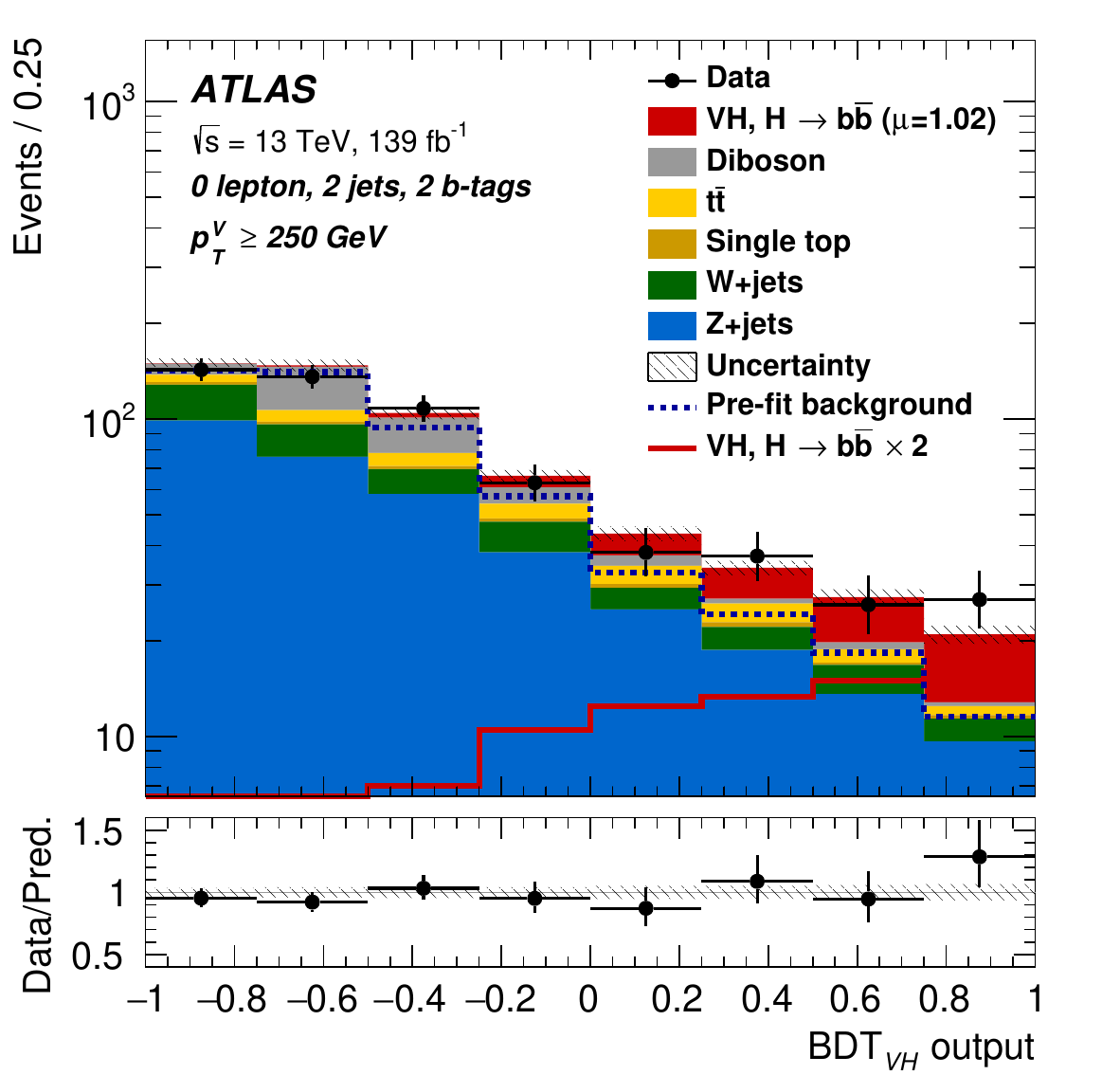}}
\subfigure[]{\includegraphics[width=0.7\textwidth]{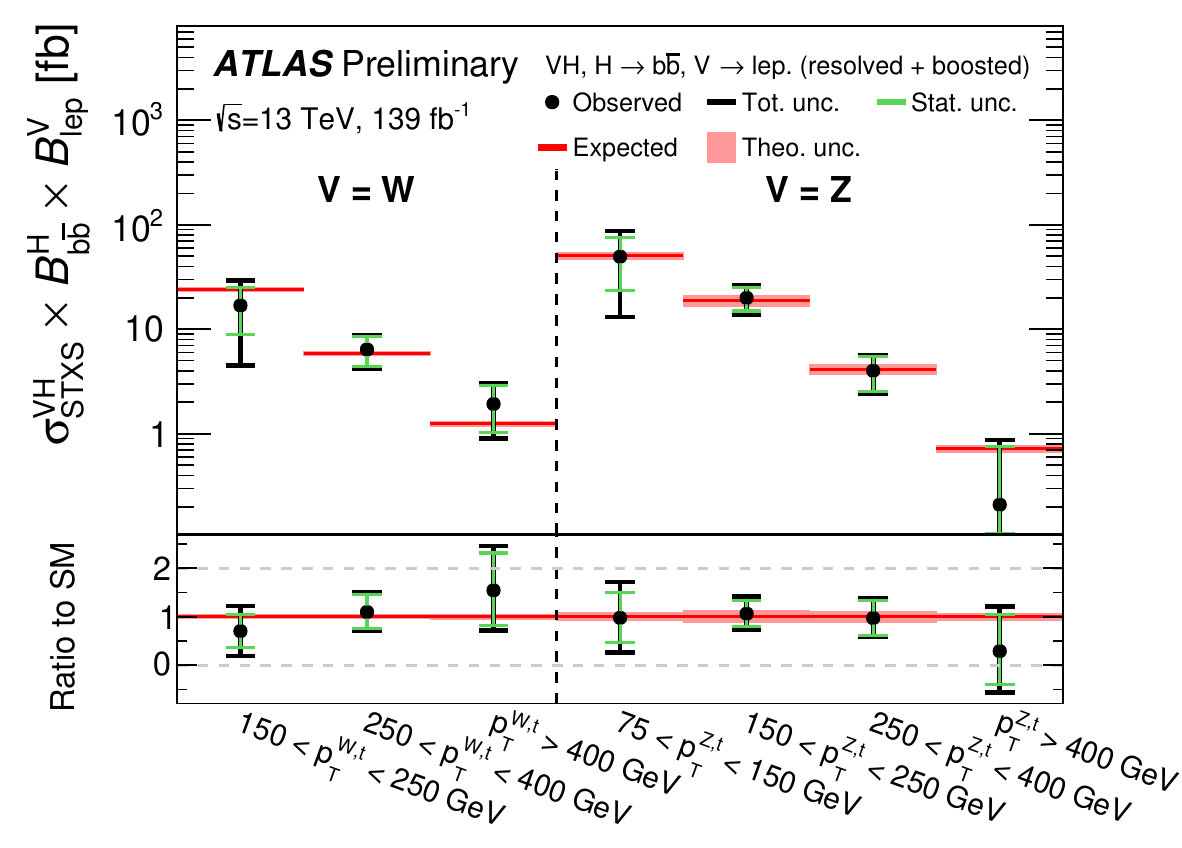}}
\caption{(a) The $S/B$ weighted $b\bar{b}$ invariant mass distribution for the cut-based $VH, \Hbb$ analysis after the subtraction of all backgrounds but the $WZ$ and $ZZ$ diboson processes, based on Run~2 data collected up to 2017~\cite{HIGG-2018-04}. (b) Observed and predicted distribution for one of the 14 BDTs used to separate the Higgs boson signal from the background processes in the $VH, \Hbb$ analysis performed with the full Run~2 data sample~\cite{HIGG-2018-51}. (c) The production cross-section  of the Higgs boson times the \Hbb and $V\to\mathrm{leptons}$ branching ratios as a function of the transverse momentum of the vector boson $p_{\mathrm{T}}^V$~\cite{ATLAS-CONF-2021-051}.
\label{fig:ferm_bb}}
\end{figure}

\subsection{Higgs boson couplings to the second generation fermions: \Hcc and $H\to\mu\mu$}

One of the challenges tackled during Run~2 was the investigation of the Yukawa interactions of the Higgs boson with the fermions of the second generation. As the Yukawa interaction is the only known fundamental interaction that distinguishes between fermion generations, the study of the couplings beyond the third generation is crucial.

\textbf{Higgs boson decays into charm quark--antiquark pairs} are usually searched for in events compatible with $VH$ production~\cite{HIGG-2021-12}. As in the \Hbb case, this production mode provides the possibility for efficient triggering and background rejection through the leptonic decays of the associated $W$ or $Z$ bosons.
Nevertheless, the analysis of \Hcc encounters several additional challenges. The branching ratio for Higgs boson decays into charm quark--antiquark pairs is $\mathcal{B}(\Hcc)$ = 2.89\%,  approximately 20 times smaller than the branching ratio for Higgs boson decays into bottom quark--antiquark pairs. In addition, charm-hadron decays have less distinct signatures compared to those of bottom-hadron decays due to the shorter lifetime of charm hadrons.
A dedicated algorithm requires jets to have a high probability to originate from charm quarks while vetoing jets likely to originate from bottom quarks, allowing the possibility to statistically combine the \Hcc and \Hbb
analyses. This configuration gives an average efficiency of 27\% to tag $c$-jets in simulated $t\bar{t}$ events,
with $b$- and light-jet misidentification rates of 8\% and 1.6\%, respectively.  Similarly to the \Hbb analysis, events are classified according to the number of charged leptons in the events. Additional cuts are applied in the various categories to suppress the $V+$jet, \ttbar and diboson backgrounds. The $VH$, \Hcc signal and background yields are obtained through a simultaneous fit to the invariant mass of the charm-tagged jets ($m_{c\bar{c}}$). The analysis strategy is validated by the simultaneous
measurement of diboson processes with one of the bosons decaying into at least one charm quark. Figure~\ref{fig:ferm_cc} (a) shows the observed  $m_{c\bar{c}}$ distribution in data compared with the SM prediction for one of the analysis categories. The  upper limit  on the $VH$, \Hcc signal strength for each of the vector boson decay channels considered in the analysis is shown in Figure~\ref{fig:ferm_cc} (b),  and their combination yields  an observed (expected) upper limit on the $VH$, \Hcc signal strength of 26 (31) at  95\% CL.

\begin{figure}
\centering
\subfigure[]{\includegraphics[width=0.49\textwidth]{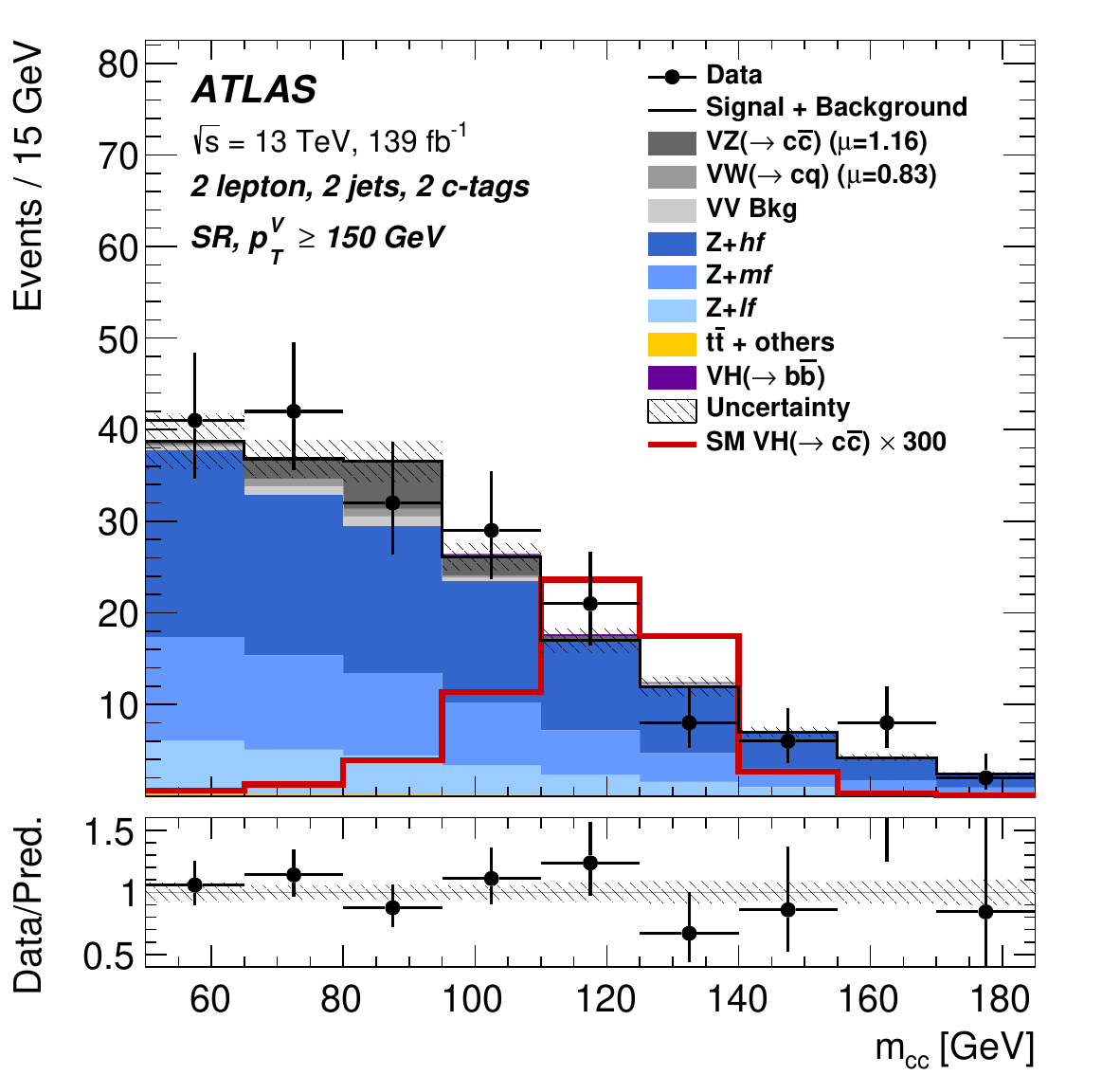}}
\subfigure[]{\includegraphics[width=0.49\textwidth]{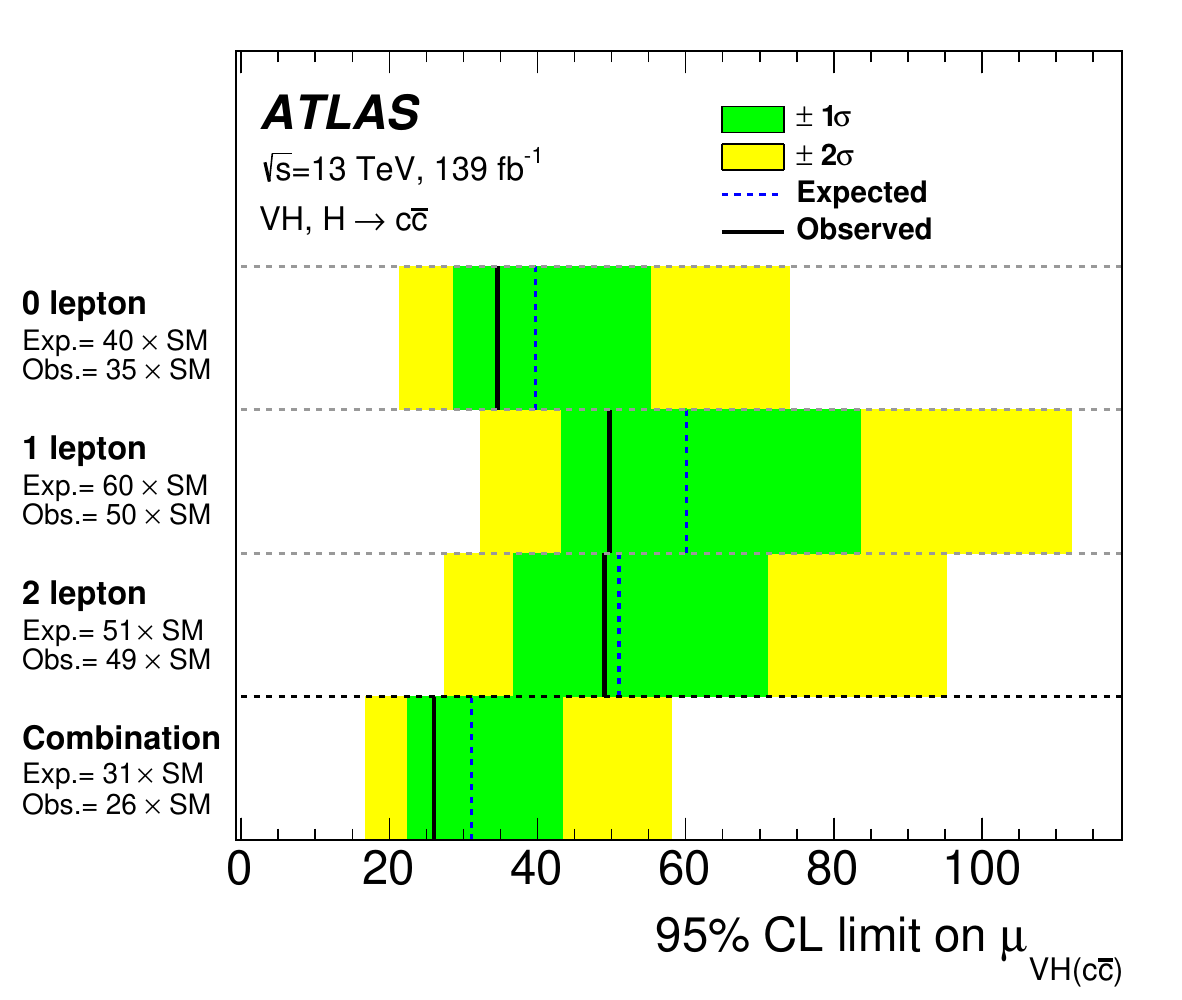}}
\caption{(a) Invariant mass distribution of the two charm-tagged jets in a selected signal region of the \Hcc analysis~\cite{HIGG-2021-12}. The \Hcc signal is also shown, scaled by 300 times the SM prediction. (b) The observed and expected 95\% CL upper limits on the $VH$, \Hcc signal strength  in the different analysis categories and for the combined fit~\cite{HIGG-2021-12}.
\label{fig:ferm_cc}}
\end{figure}

A statistical combination of the \Hcc and \Hbb
analyses is performed  to simultaneously  constrain the coupling-strength modifiers  $\kappa_b$ and $\kappa_c$, which are defined as the ratios of the measured
Yukawa couplings to their SM values~\cite{HIGG-2021-12}. The combination ensures a reduced model dependency of the result, as, even after the \Hcc analysis selection, the $VH$, \Hbb process is expected to be two to eight times larger than the $VH$, \Hcc contribution, depending on the signal region. The
observed bounds in the $\kappa_c$--$\kappa_b$ plane are shown in Figure~\ref{fig:ferm_kb_kc} (a). The observed (expected) upper limit on the ratio yields $|\kappa_c/ \kappa_b| < 4.5$ (5.1) at 95\% CL\@. This bound is slightly
smaller than the $m_b/m_c$ ratio~\cite{FermilabLattice:2018est}, indicating that the Higgs boson to charm-quark coupling
is weaker than the Higgs boson to bottom-quark coupling.
The direct limits on the $\kappa_b$ and $\kappa_c$ coupling-strength modifiers are combined with the ones from the fiducial differential cross-section measurements of \ptH  in the \Hyy\ and \HZZfourl\ final states (Section~\ref{sec:higgs_diffxs}). The resulting constraints are shown in Figure~\ref{fig:ferm_kb_kc}~(b)~\cite{HIGG-2022-04}.

\begin{figure}
\centering
\subfigure[]{\includegraphics[width=0.49\textwidth]{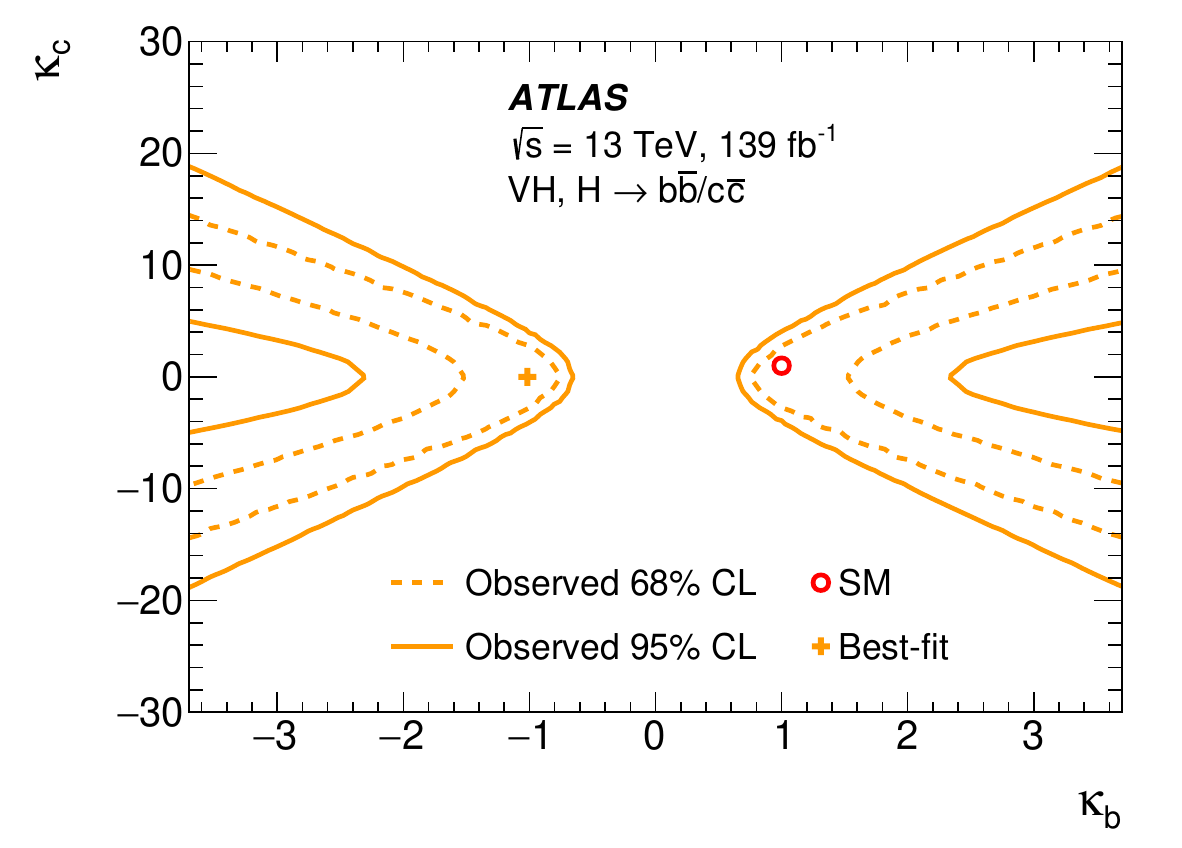}}
\subfigure[]{\includegraphics[width=0.45\textwidth]{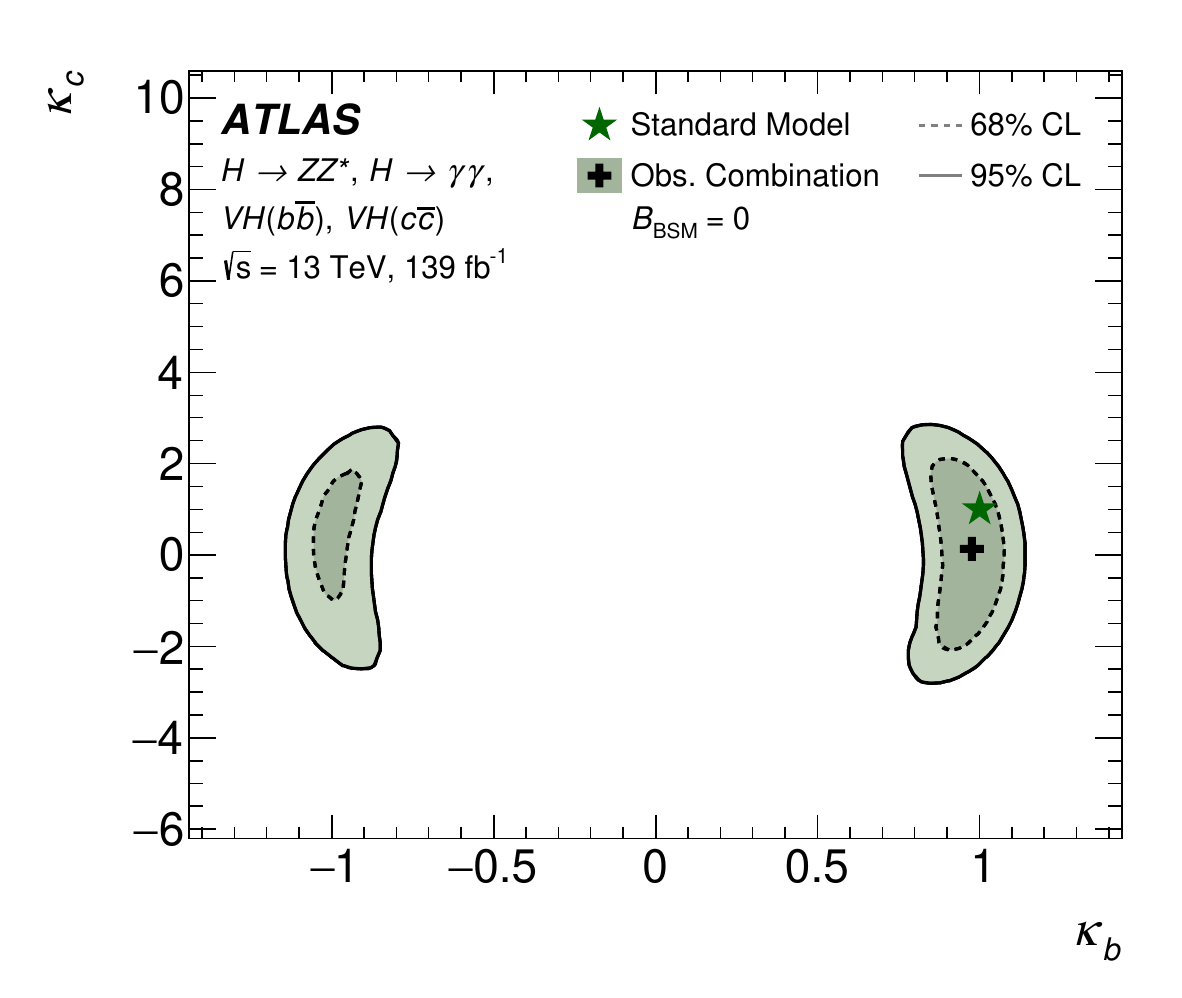}}
\caption{(a) The observed constraints on $\kappa_b$ and $\kappa_c$ at 68\% and 95\% CL from a fit to multivariate discriminants used to identify $VH$ events with Higgs bosons decaying into $b\bar{b}$ or $c\bar{c}$, assuming no decays into BSM particles~\cite{HIGG-2021-12}.
(b) Observed contours for the $\kappa_b$ and $\kappa_c$ parameters from a simultaneous fit to the \ptH distributions measured in \Hyy\ and \HZZfourl decays and to the multivariate discriminants in (a)~\cite{HIGG-2022-04}.
\label{fig:ferm_kb_kc}}
\end{figure}

\textbf{The $H\to\mu\mu$ decay channel} presents a unique opportunity to investigate the interactions of the Higgs boson with second-generation fermions. Despite its low SM branching ratio (approximately $2.17 \times 10^{-4}$), the expected sensitivity in this decay channel surpasses the \Hcc decay channel, thanks to the lower background contamination  and the better invariant mass resolution. Considerable effort was dedicated to further enhance the sensitivity of this decay channel during Run 2~\cite{HIGG-2019-14}, capitalising particularly on continuous improvements in analysis techniques and muon reconstruction.

Events are selected that contain two isolated muons with opposite charges and high transverse momenta, reconstructed within the acceptance of the ATLAS muon spectrometer. The dominant background is the Drell--Yan process. The selected  events are then categorised using multivariate techniques that exploit kinematic and topological differences among the various Higgs boson production modes and background processes, typically resulting in a signal-to-background ratio ($S/B$) ranging from  0.1\% to 8\% in the 120~\GeV~$< m_{\mu\mu} < 130$~\GeV window.

Thanks to the excellent muon momentum resolution, the resonant behaviour of the Higgs boson signal in the $m_{\mu\mu}$ distribution can be utilised after categorisation to extract the signal yield and determine background normalisation and shape parameters, as shown in Figure~\ref{fig:fermions_mumu_invmass}. The resolution of the $m_{\mu\mu}$ resonance is  improved by including photons from QED final-state radiation through which muons can lose a significant fraction of their energy.
The signal is modelled using a double-sided Crystal Ball function, with the width of the Gaussian component ranging from 2.6 to 3.2~\GeV\ depending on the specific category.

\begin{figure}
\centering
\includegraphics[width=0.7\textwidth]{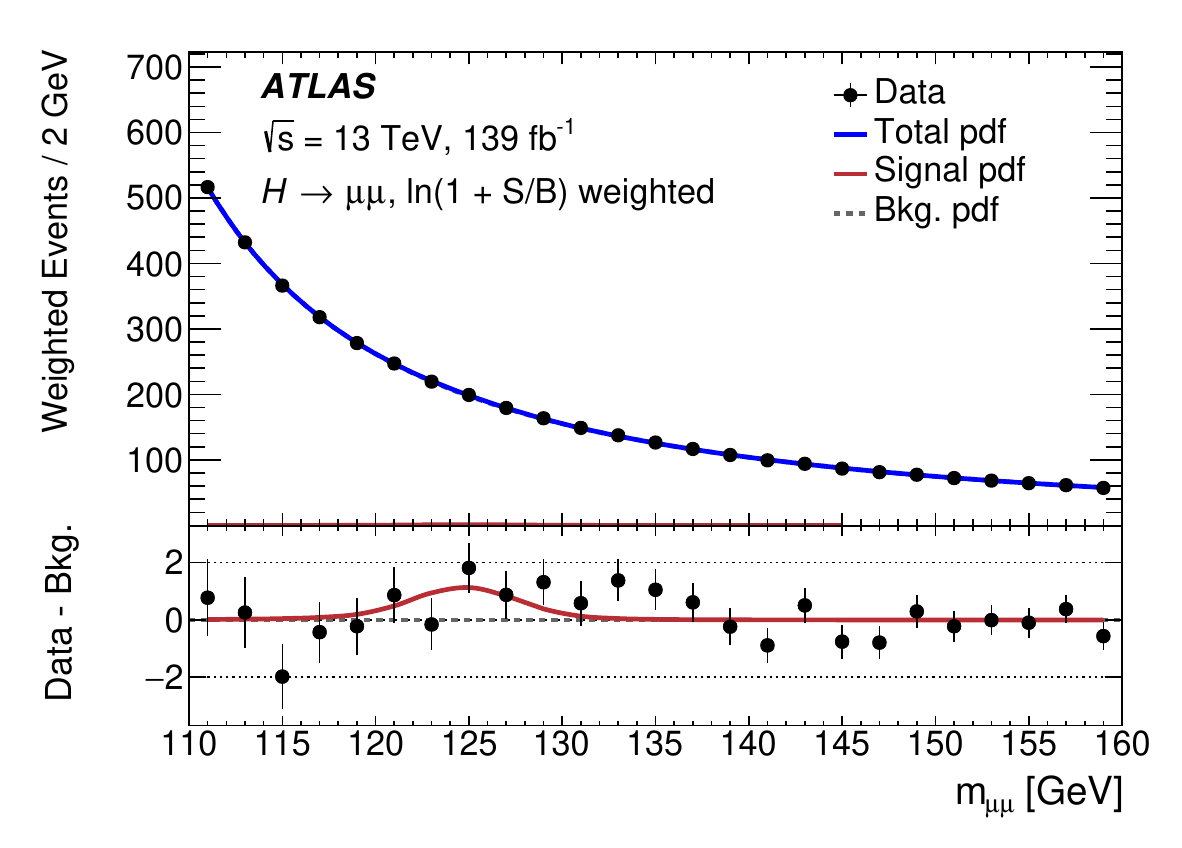}
\caption{  Dimuon invariant mass spectrum in all  analysis categories observed in data, with the events and probability density functions (pdfs) weighted by $\ln(1+S/B)$ ~\cite{HIGG-2019-14} where $S$ corresponds to the observed signal yield in the 120~\GeV~$< m_{\mu\mu} < 130$~\GeV window and $B$ is the background yield in the same window, both derived by the fit to the data.
\label{fig:fermions_mumu_invmass}}
\end{figure}

The observed (expected) significance over the background-only hypothesis for a Higgs boson with a mass of 125.09~\GeV is 2.0$\sigma$ (1.7$\sigma$). The best-fit value of the signal strength parameter, defined as the ratio of the observed signal yield to the one expected in the SM, is $\mu=1.2\pm 0.6$.


\section{Search for rare loop-induced decays of the Higgs boson}
\label{sec:higgs_rare}

Higgs boson decays into two leptons and a photon can occur at tree level through the Yukawa coupling of a Higgs boson to leptons, or through quantum loops of virtual particles. Representative Feynman diagrams are shown in Figure~\ref{fig:rare_decays_feynman}. The  corresponding invariant mass distributions of the electron pair in
\Heey decays are presented in Figure~\ref{fig:rare_decays_mll_theory}. Due to the small Yukawa couplings to electrons, the tree-level diagram in Figure~\ref{fig:rare_decays_feynman} (a) is only relevant for \Hmumuy.
Rare loop processes are interesting to study, as they could be modified by so-far undiscovered particles modifying the branching ratios~\cite{Kachanovich:2020xyg}. Furthermore, the three-particle final state might in the future allow CP studies of the corresponding couplings~\cite{Korchin:2014kha,Chen:2014ona}.

\begin{figure}
\centering
\subfigure[]{\includegraphics[height=2.5cm]{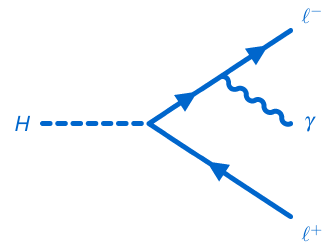}}\hskip1cm
\subfigure[]{\includegraphics[height=2.5cm]{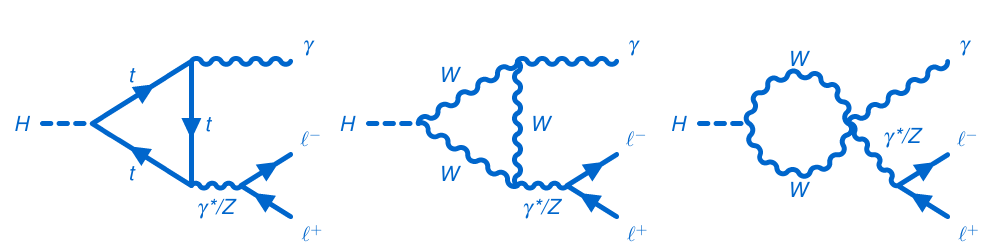}}
\\
\caption{Representative Feynman diagrams of Higgs boson decays into two leptons and a photon.
\label{fig:rare_decays_feynman}}
\end{figure}

\begin{figure}
\centering
\subfigure{\includegraphics[width=0.7\textwidth]{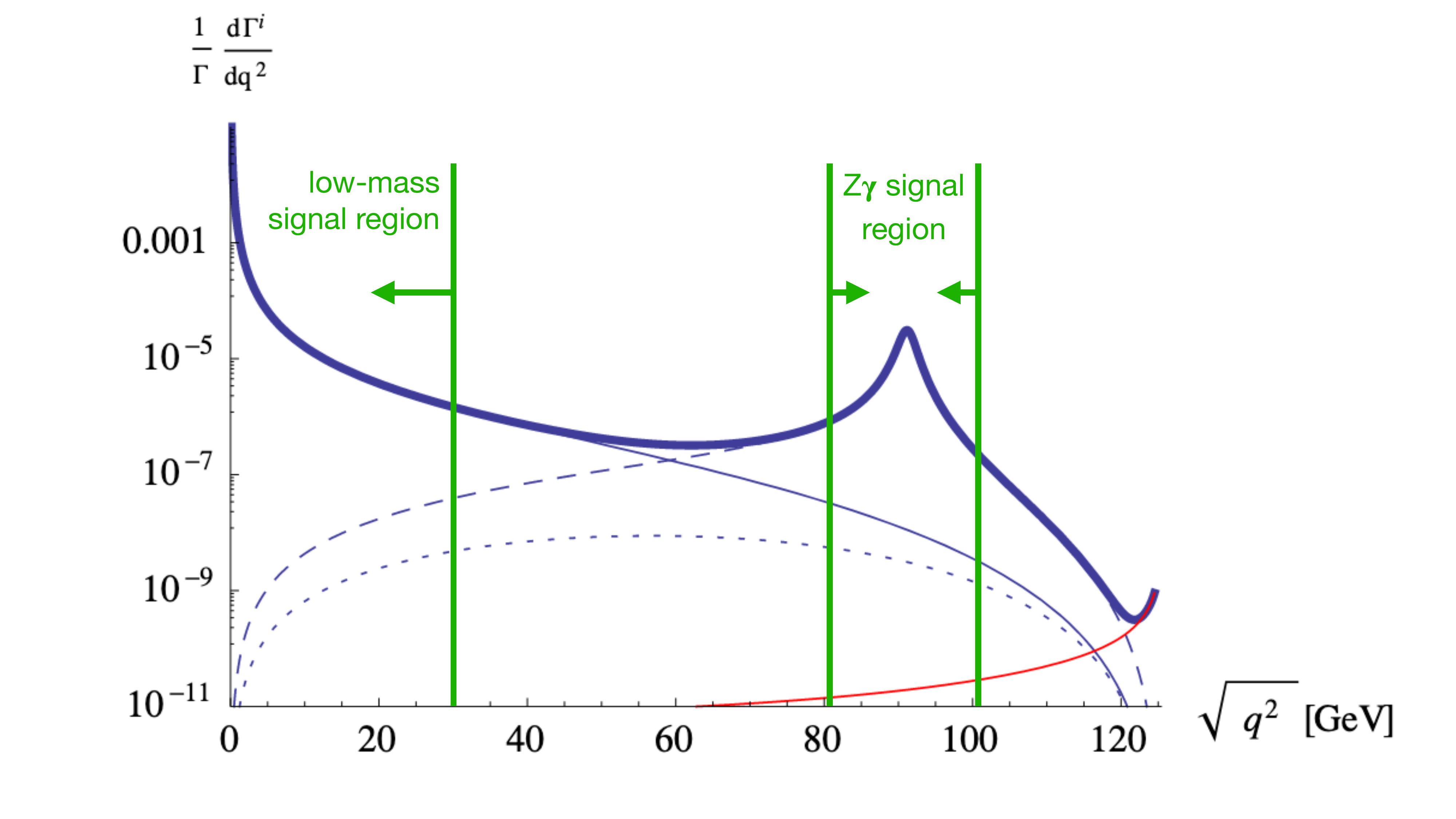}}
\caption{Distribution of the invariant mass of the two electrons in \Heey decays~\cite{Sun:2013rqa}. The red line shows the tree level diagram, the dashed line the decay into a $Z$ boson and a photon, and the thin blue line the decay into a virtual photon and a photon. The dotted line comes from four-point box diagrams, not discussed here. The low-\mll and \Zy signal regions used in the ATLAS analyses are indicated by vertical lines.
\label{fig:rare_decays_mll_theory}}
\end{figure}

ATLAS has performed searches for Higgs boson decays into two leptons and a photon, with \mll below 30~\GeV, mostly targeting the decay into a virtual photon and a photon~\cite{HIGG-2018-43}, as well for decays with \mll close to the $Z$ pole~\cite{HIGG-2018-42}, dominated by decays into a $Z$ boson and a photon. Searches for Higgs bosons decaying into a meson and a photon~\cite{ATLAS:2022rej} are not discussed here, as the SM branching ratios are too small to be detectable at the LHC (see instead Ref.~\cite{HDBSreport}).

Events with a photon and two electrons or two muons are selected. For the low-\mll analysis, the invariant mass of the two leptons \mll is required to be smaller than 30~\GeV. For the \Zy analysis,
the dilepton invariant mass  must be within 10~\GeV\ of the $Z$ boson mass. In the \eey final state,  for low invariant masses and high transverse momenta of the dielectron system, the two electromagnetic showers can overlap in the calorimeter. Therefore, the search for low-\mll final states also considers events if they contain a photon and an object that is compatible with two very close electrons.
In addition to the categorisation into lepton final states, further categorisations are introduced, guided by the different signal-to-background ratios of the various Higgs boson production modes: categories targeting VBF production, \lly final states with high transverse momenta, and the remainder.
In each of the two dilepton mass ranges, a simultaneous parameterised fit is performed over the \mlly mass distributions in all categories, to extract the number of signal events. As in the \yy final state, uncertainties in the choice of fit function for the background are included.
Overall, statistical uncertainties are dominant.
Figure~\ref{fig:rare_decays_results_mlly} shows the \mlly distributions, with each data event weighted by a category-dependent weight that depends on the signal-to-background ratio in a window around the resonance peak.

\begin{figure}
\centering
\subfigure[]{\includegraphics[width=0.43\textwidth]{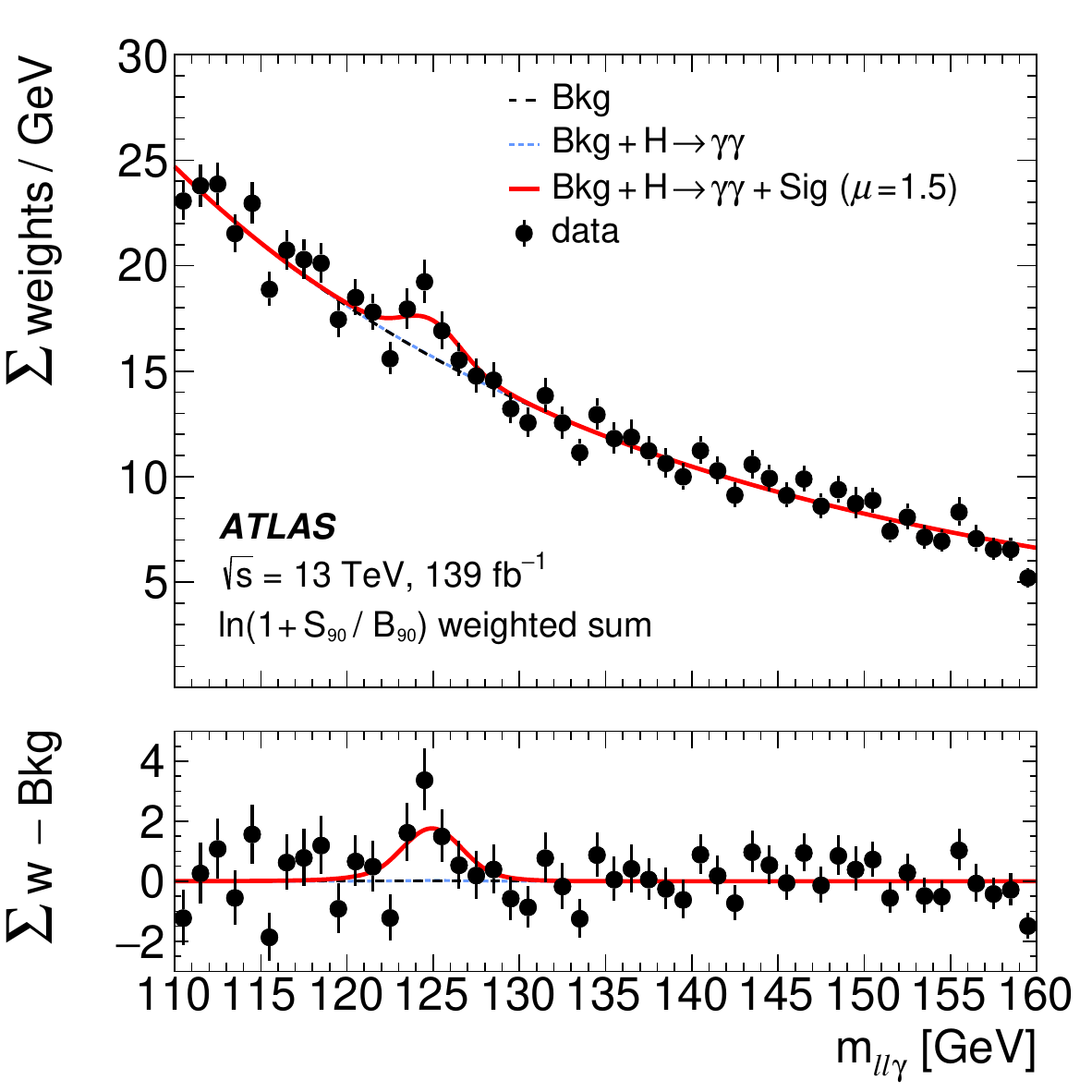}}
\subfigure[]{ \includegraphics[width=0.44\textwidth]{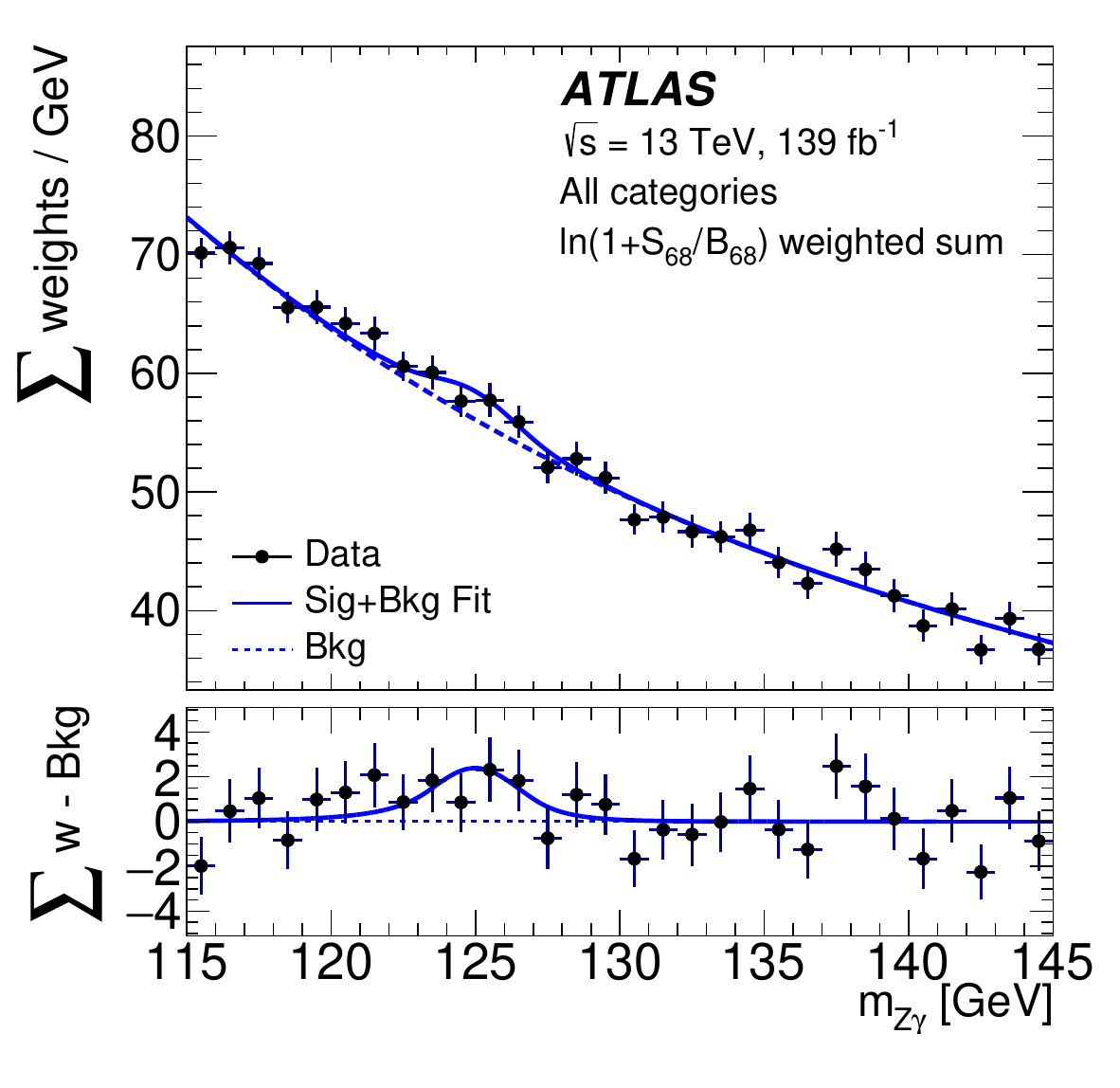}}
\caption{(a) The \mlly distributions for the low-\mll analysis~\cite{HIGG-2018-43}, and (b) the \Zy  analysis~\cite{HIGG-2018-42}, where every data event is reweighted by a category-dependent weight, $\text{ln}(1+S_X/B_X)$, with $X$ = 90\% (68\%) for the low-\mll (\Zy) analysis. Here, $S_X$ is the number of signal events in the smallest window containing $X$\% of the expected signal, and $B_X$ is the expected number of background events in the same window.
\label{fig:rare_decays_results_mlly}}
\end{figure}

Testing the low-\mll regime, evidence for the \Hlly process is found with a
significance of 3.2$\sigma$ over the background-only hypothesis, compared with an expected significance of 2.1$\sigma$ for the SM prediction. The best-fit value of the signal-strength parameter is $\mu  = 1.5 \pm 0.5$. The Higgs boson production cross-section times the \Hlly branching ratio
for \mlly $<$ 30~\GeV\ is determined to be $8.7^{+2.8}_{-2.7}$~fb.

In the search for the \Zy process, the observed (expected) significance is 2.2$\sigma$ (1.2$\sigma$). The best-fit value for the \HZy signal strength is found to be $\mu$ =  2.0~$^{+1.0}_{-0.9}$. As the analysis is limited by the small expected number of  signal events, a combination is performed with the CMS analysis~\cite{CMS-HIG-19-014}, effectively doubling the data sample, which pushes the significance above the evidence threshold: 3.4$\sigma$ with an expectation of 1.6$\sigma$. The corresponding best-fit signal strength is $\mu$ = 2.2 $\pm$ 0.7~\cite{CMS:2023mku}.

\FloatBarrier


\section{Cross-sections and couplings from combined fits}
\label{sec:higgs_combination}

Higgs boson measurements with different decay signatures can be combined to extract as complete a portrait of the Higgs boson as possible~\cite{ATLAS:2022vkf}.
The combination harnesses the statistical power of an unprecedented number of Higgs boson measurements to extract inclusive cross-sections and cross-sections for different production processes and kinematic regions. Each production--decay process is sensitive to different combinations of Higgs boson couplings to other fundamental particles, and the combination allows all accessible couplings to be constrained simultaneously.

Analyses in the following Higgs boson decay channels are considered in the combination: \Hyy~\cite{HIGG-2020-16}, \HZZfourl~\cite{HIGG-2018-28}, \HWWevmuv~\cite{HIGG-2021-20,HIGG-2017-14}, \HZylly~\cite{HIGG-2018-42}, \Hbb~\cite{HIGG-2018-51,HIGG-2018-52,HIGG-2019-04,HIGG-2020-23,HIGG-2021-08}, \Htautau~\cite{HIGG-2019-09}, \Hmumu~\cite{HIGG-2019-14}, \Hcc~\cite{HIGG-2021-12}. They provide measurements in different Higgs boson production modes and phase space regions. A complementary analysis of the \ttH process, sensitive to $WW^*$, $ZZ^*$ and $\tau\tau$ final states is included, as well~\cite{HIGG-2017-02}. %
Furthermore, searches for Higgs bosons decaying into invisible particles, such as dark matter candidates, are considered, most notably those targeting the VBF and \ZH production mechanisms~\cite{EXOT-2020-11,HIGG-2018-26}. These are included in a subset of the fits.

The statistical analysis of the data is based on a combined likelihood function, calculated by forming the product of the likelihood functions describing each of the input measurements~\cite{ATL-PHYS-PUB-2011-011}.
A multitude of results can be extracted from the combination of the different analyses. Each result is obtained under certain assumptions, as detailed below. %

\subsection{Global signal strength, production cross-sections and decay branching ratios}

Assuming each production--decay process scales with the same global signal strength $\mu$, the measured number of produced Higgs bosons relative to the SM prediction is
$\mu  = 1.05 \pm 0.06 = 1.05 \pm 0.03 \mathrm{(stat.)} \pm 0.03 \mathrm{(exp.)} \pm 0.04 \mathrm{(sig. th.)} \pm 0.02 \mathrm{(bkg. th.)}$, where the total uncertainty is split into statistical, experimental and theoretical uncertainties (signal and background).

The cross-sections of the different Higgs boson production processes are measured assuming SM values for the decay branching
fractions and vice versa, as shown in Figure~\ref{fig:combi_XSBR}. All major production modes are now observed, including the $WH$ and $ZH$ processes individually. An unprecedented upper limit at 95\% CL was observed for \tH production: 15 times the SM cross-section (7 times expected), with a strong correlation between the \ttH and \tH measurements. Figure~\ref{fig:combi_XStimesBR} shows the observed event rate divided by the predicted SM event rate for different combinations of Higgs boson production and decay processes. This measurement does not require any assumptions on cross-sections or branching ratios and nicely illustrates the sensitivity of the analyses in the different decay channels. Excellent overall agreement is observed with the SM predictions over multiple orders of magnitude.

\begin{figure}
\centering
\subfigure[]{\includegraphics[width=0.45\textwidth]{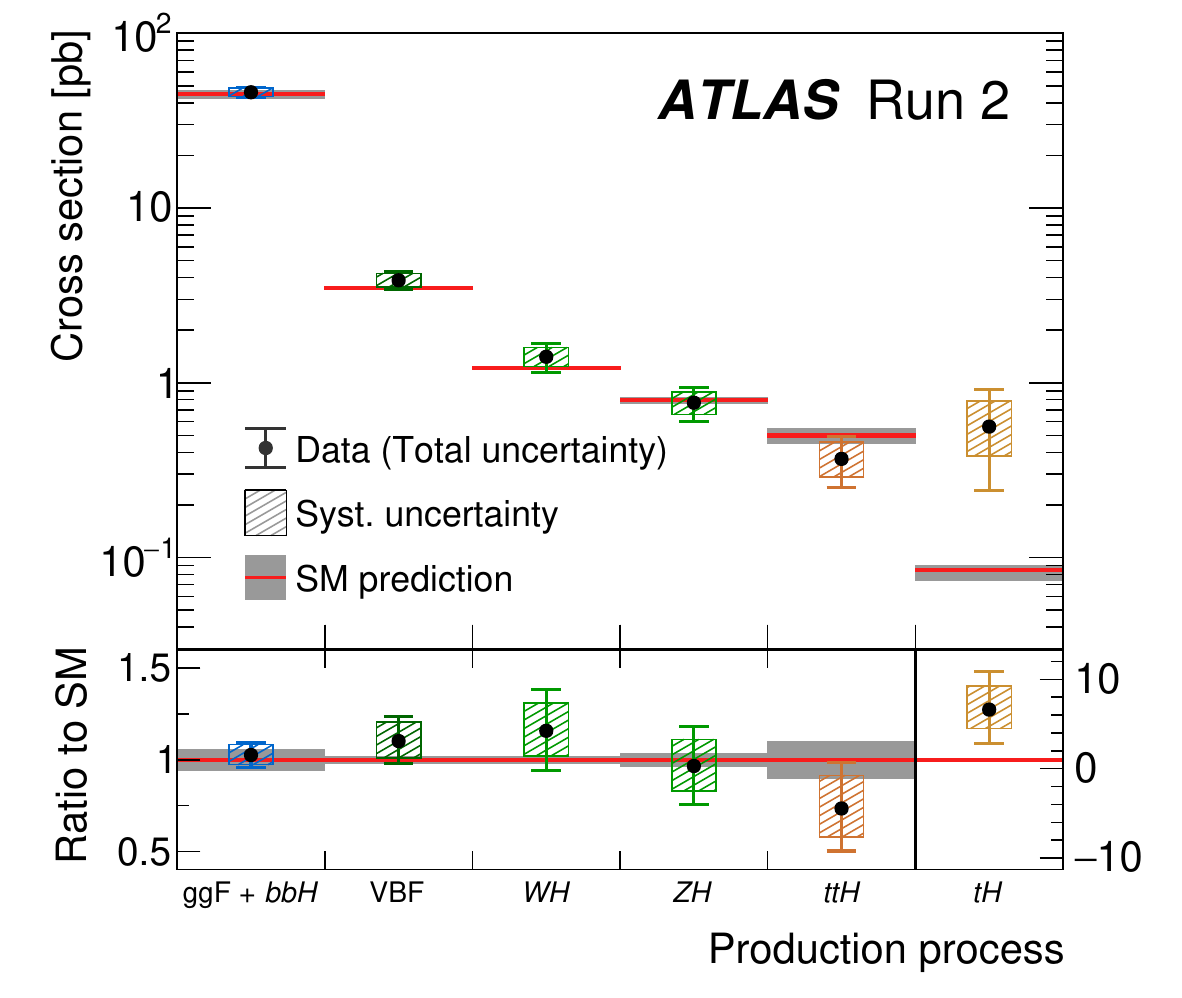}}
\subfigure[]{\includegraphics[width=0.45\textwidth]{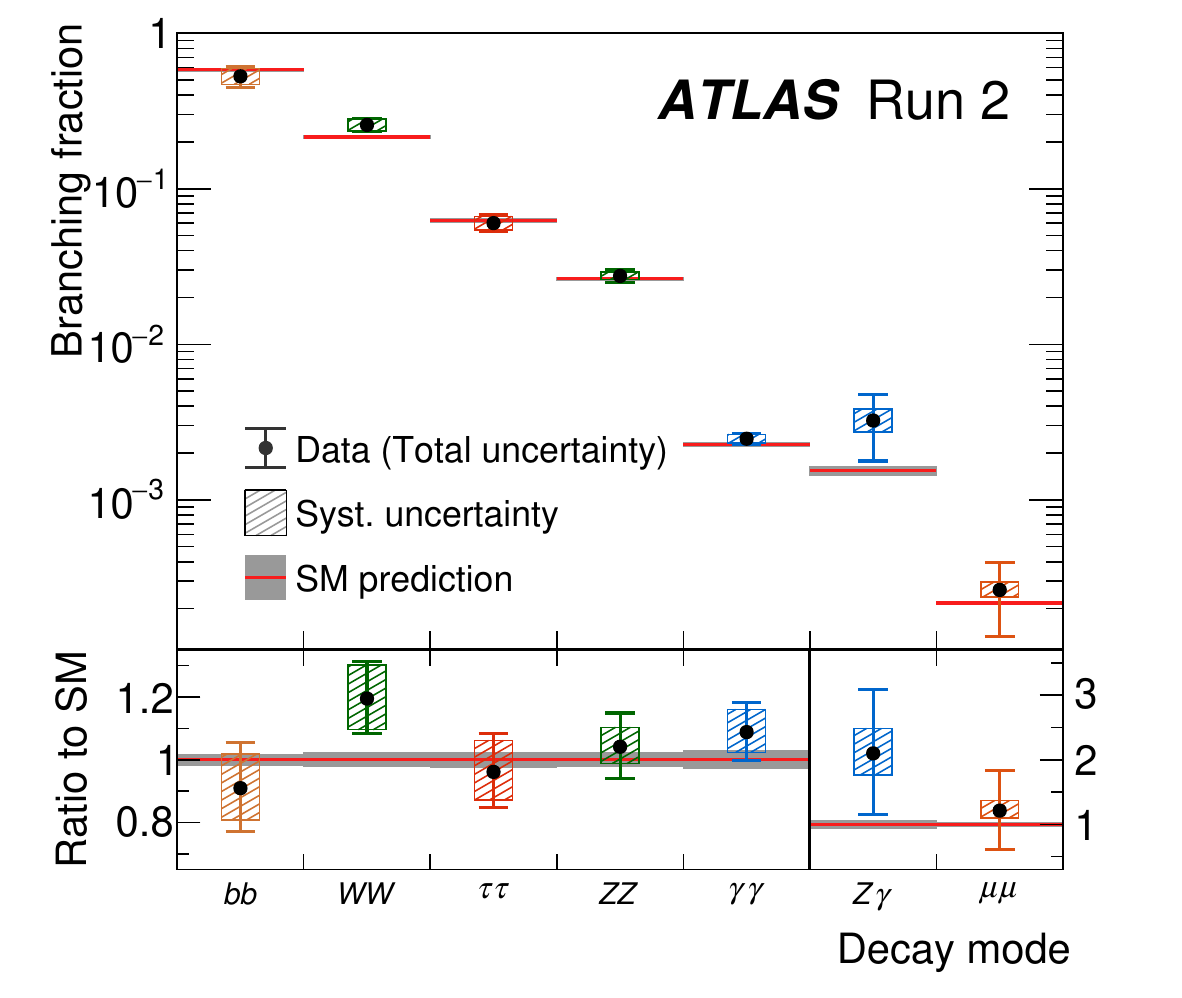}}
\caption{(a) Observed and predicted Higgs boson production cross-sections, assuming SM values for the branching ratios. (b) Observed and expected branching fractions assuming SM values for the cross-sections~\cite{ATLAS:2022vkf}.
\label{fig:combi_XSBR}}
\end{figure}

\begin{figure}
\centering
\subfigure{\includegraphics[width=0.9\textwidth]{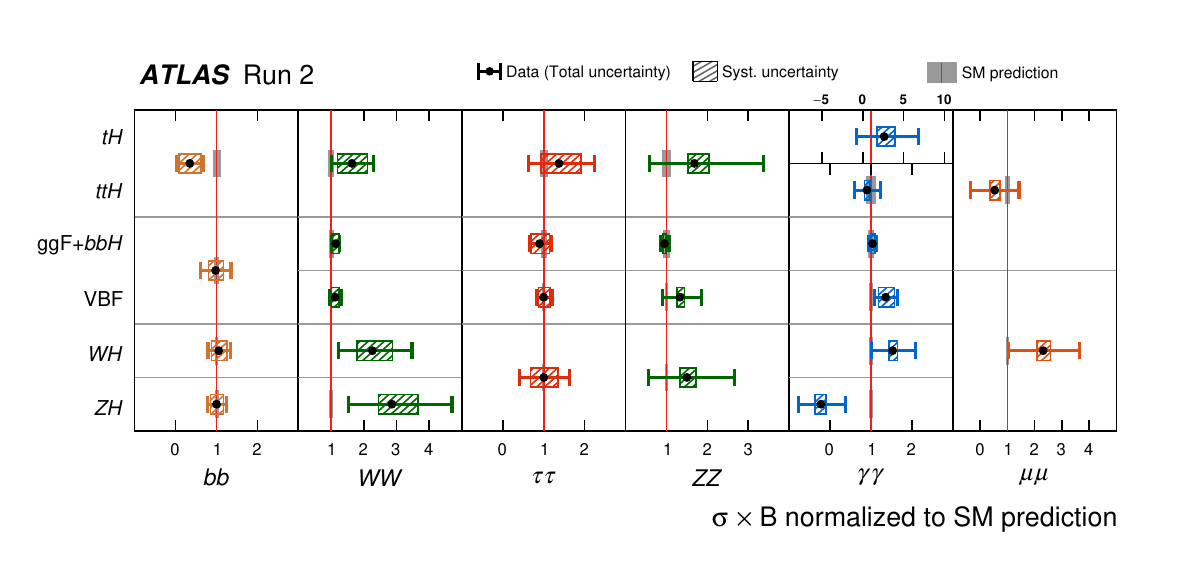}}
\caption{Observed event rate divided by predicted SM event rate for different combinations of Higgs boson production and decay processes~\cite{ATLAS:2022vkf}.
\label{fig:combi_XStimesBR}}
\end{figure}

\subsection{Higgs boson couplings to other particles}

To extract the couplings of the Higgs boson to other particles, a formalism has to be introduced. In the \kap-framework~\cite{Heinemeyer:2013tqa}, the cross-section times the branching fraction for an individual measurement is parameterised in terms of multiplicative coupling-strength modifiers that do not change kinematic distributions but only the strength of the coupling to a given particle compared to that expected in the SM\@. For example, the partial width of the decay into two bottom quarks is parameterised as $\Gamma_b / \Gamma_b^\mathrm{SM} = \kappa_b^2$. Care is taken to adjust the total width of the Higgs boson accordingly, and in some fits BSM contributions to undetected or invisible decays are included, where undetected decays are hidden by large backgrounds. For production and decays through quantum loops (e.g., ggF and \Hyy), either effective coupling-strength modifiers are introduced ($\kappa_g$ and $\kappa_\gamma$), or the loops are resolved into the different particle contributions according to the SM\@.

Assuming all coupling-strength modifiers to fermions ($\kappa_F$) scale the same way, as do all modifiers to vector bosons ($\kappa_V$), one can perform a two-dimensional fit in $\kappa_F$ and $\kappa_V$, finding the best fit value and the 68\% and 95\% CL regions, as shown in Figure~\ref{fig:combi_kappaFkappaV}. Here, the loops are assumed to be as in the SM with no BSM contributions to undetected or invisible decays allowed. The SM value lies within the 95\% CL region.

\begin{figure}
\centering
\subfigure{\includegraphics[width=0.45\textwidth]{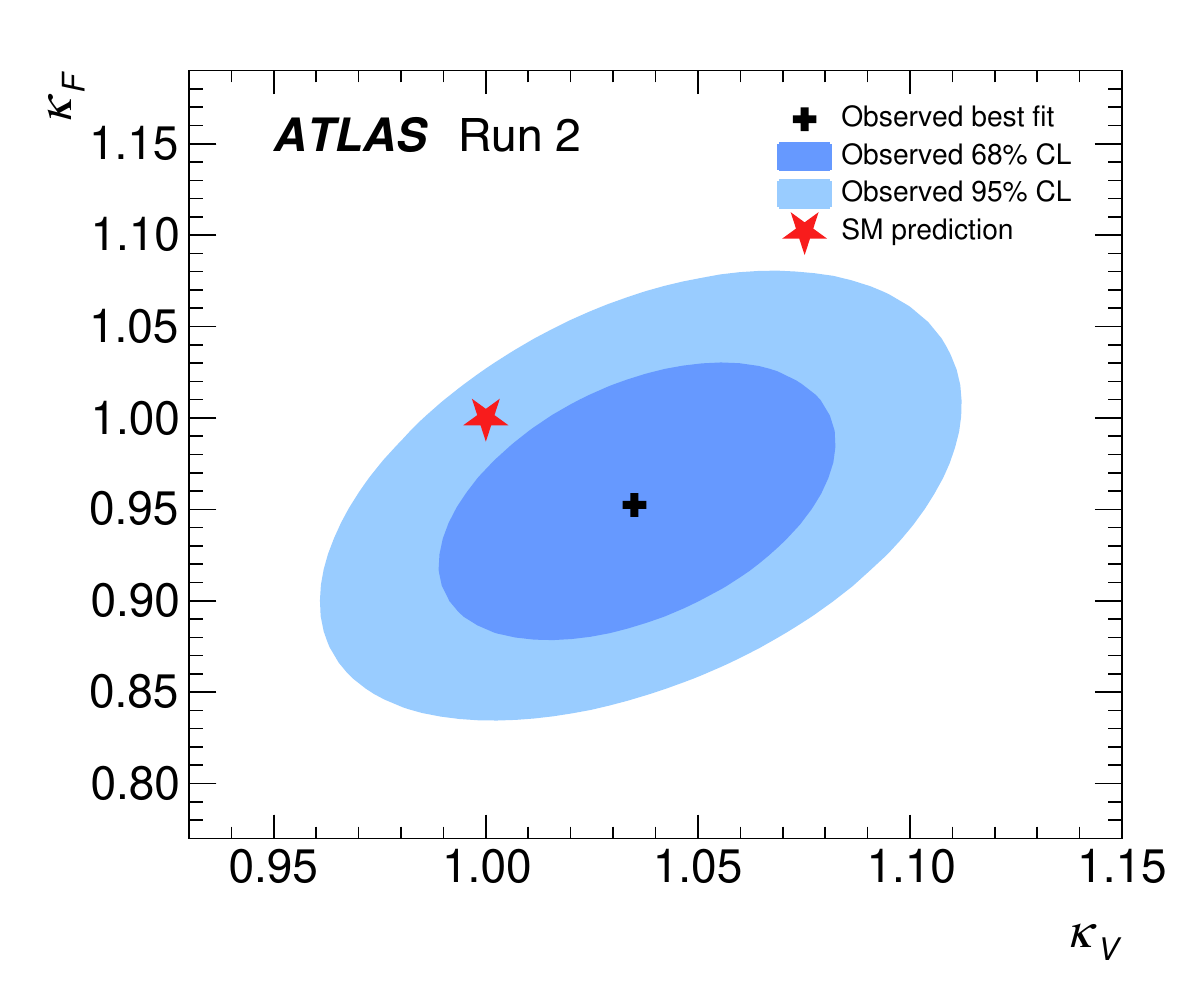}}
\caption{Fit in the $\kappa_F$--$\kappa_V$ plane, showing the best fit value and the 68\% and 95\% CL regions~\cite{ATLAS:2022vkf}. This fit is performed assuming that there are no BSM contributions to invisible or undetected Higgs boson decays.
\label{fig:combi_kappaFkappaV}}
\end{figure}

Still resolving the loop-induced processes according to the SM and not allowing for additional decays, the coupling-strength modifiers for individual particles can be fit separately.  The reduced coupling-strength modifiers are defined as $\sqrt{\kappa_V g_V / 2v} = \sqrt{\kappa_V}(m_V/v)$ for weak bosons with a mass $m_V$, and  $\kappa_F g_F = \kappa_F m_F/v$ for fermions with a mass $m_F$, where $g_V$ and $g_F$ are the corresponding absolute coupling strengths and $v = (\sqrt{2}G_\text{F})^{-\frac{1}{2}} \sim 246$~\GeV is the vacuum expectation value of the Higgs field, which can be derived from the measurement of the Fermi constant $G_\text{F}$~\cite{PDG}. Figure~\ref{fig:combi_diagonal} shows these reduced coupling-strength modifiers as a function of the particle masses. The SM predicts a striking diagonal line, encapsulating the essence of the SM Higgs Lagrangian, which agrees very well with the measured values over about three orders of magnitude. Two versions of the fit are performed, as the sensitivity of the analyses to the charm quark coupling is still rather low. In one, the modification to the charm quark coupling is fixed to be the same as for the top quark coupling, and in the other, the charm quark coupling is a free parameter.

\begin{figure}
\centering
\subfigure{\includegraphics[width=0.8\textwidth]{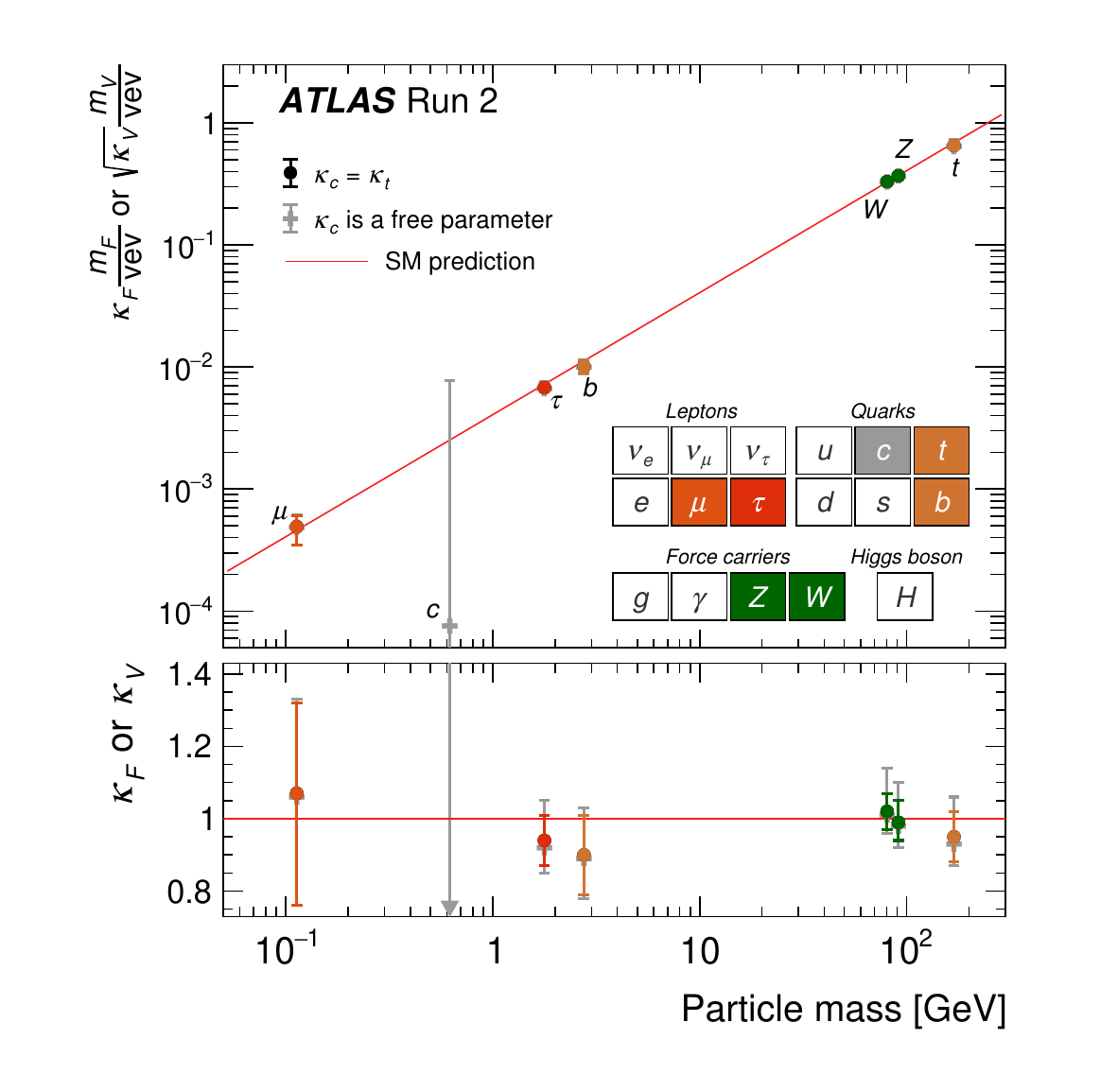}}
\caption{Higgs boson coupling-strength modifiers and their uncertainties versus the particle masses, compared with the SM prediction~\cite{ATLAS:2022vkf}. For the bottom and the top quark, the masses are evaluated at a scale corresponding to that of the Higgs boson mass. Loop-induced processes are resolved according to the SM predictions, and no decays into non-SM particles are allowed.
\label{fig:combi_diagonal}}
\end{figure}

Finally, a fit is performed allowing for new contributions to the loop-induced processes by using effective coupling-strength modifiers for the gluon, photon and \Zy couplings. Figure~\ref{fig:combi_kappas} shows the results of two fits: allowing (and not allowing) for undetected and invisible non-SM contributions to the branching ratios. When these contributions are allowed, the searches for invisibly decaying Higgs bosons produced via VBF or \ZH production are included in the fit,
$\kappa_V \leq 1$ is imposed to avoid degeneracies, and upper limits are set at 95\% CL on the undetected and invisible branching fractions to non-SM particles. It can be seen that allowing for undetected and invisible BSM contributions does not change the fitted coupling-strength modifiers significantly. The modifiers for the coupling strength to top-quark, bottom-quark and \taul are measured with uncertainties ranging from about 7\% to 12\%, to $W$ and $Z$ bosons with uncertainties of about 6\%, and for the effective loop-induced coupling strength to photons and gluons with 6\% and 7\%, respectively. The upper limit on the invisible (undetected) BSM branching ratio is 0.13 (0.12), improving on the direct searches alone (see Ref.~\cite{HDBSreport}).

\begin{figure}
\centering
\subfigure{\includegraphics[width=0.8\textwidth]{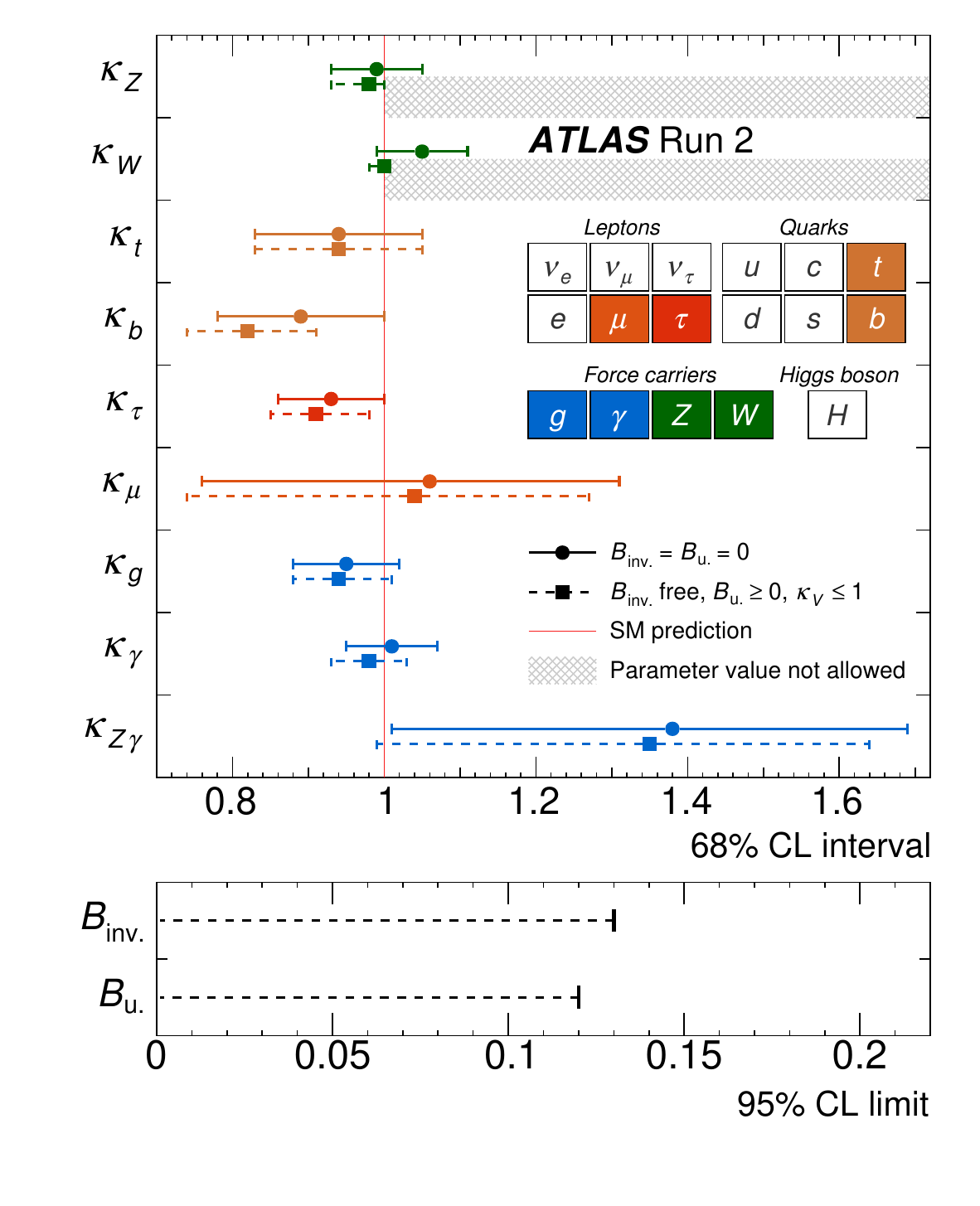}}
\caption{Coupling-strength modifiers and their uncertainties per particle type with effective gluon, photon and \Zy couplings, assuming the BSM contribution to undetected and invisible branching fractions to be zero (solid lines), and allowing them to contribute to the width of the Higgs boson (dashed lines). For the second case, the undetected branching ratio is assumed to be larger or equal to zero, and $\kappa_V$ is assumed to be smaller or equal to one. Limits on the undetected and invisible branching fractions to non-SM particles are shown in the bottom panel~\cite{ATLAS:2022vkf}.
\label{fig:combi_kappas}}
\end{figure}

\subsection{Production cross-sections in different kinematic regions and their interpretation}
\label{sec:higgs_combination_stxt}
A comprehensive study is performed  measuring cross-sections separately per production mechanism in different phase space regions, referred to as simplified template cross-sections (STXS)~\cite{deFlorian:2016spz,Andersen:2016qtm,Berger:2019wnu,Amoroso:2020lgh}. The STXS framework has been designed to maximize the sensitivity of
Higgs boson cross-section measurements to BSM effects, while minimizing their
theory dependence and allowing for the combination of analyses in different decay channels.
The measurements and the SM predictions are shown in Figure~\ref{fig:combi_stxs}. Different  kinematic regimes are targeted, for example Higgs bosons produced with high transverse momenta, which could be modified by BSM phenomena at high energies. Selections are also made based on the properties of the hadronic jets in the events, for example separating VBF- and $VH$-enriched $qq \rightarrow Hqq$ production. The measured cross-sections could reveal phenomena beyond the SM that are not observable through measurements of the coupling strengths. Furthermore they can be used as a basis for  electroweak fits and BSM interpretations.

\begin{figure}
\centering
\subfigure{\includegraphics[width=0.98\textwidth]{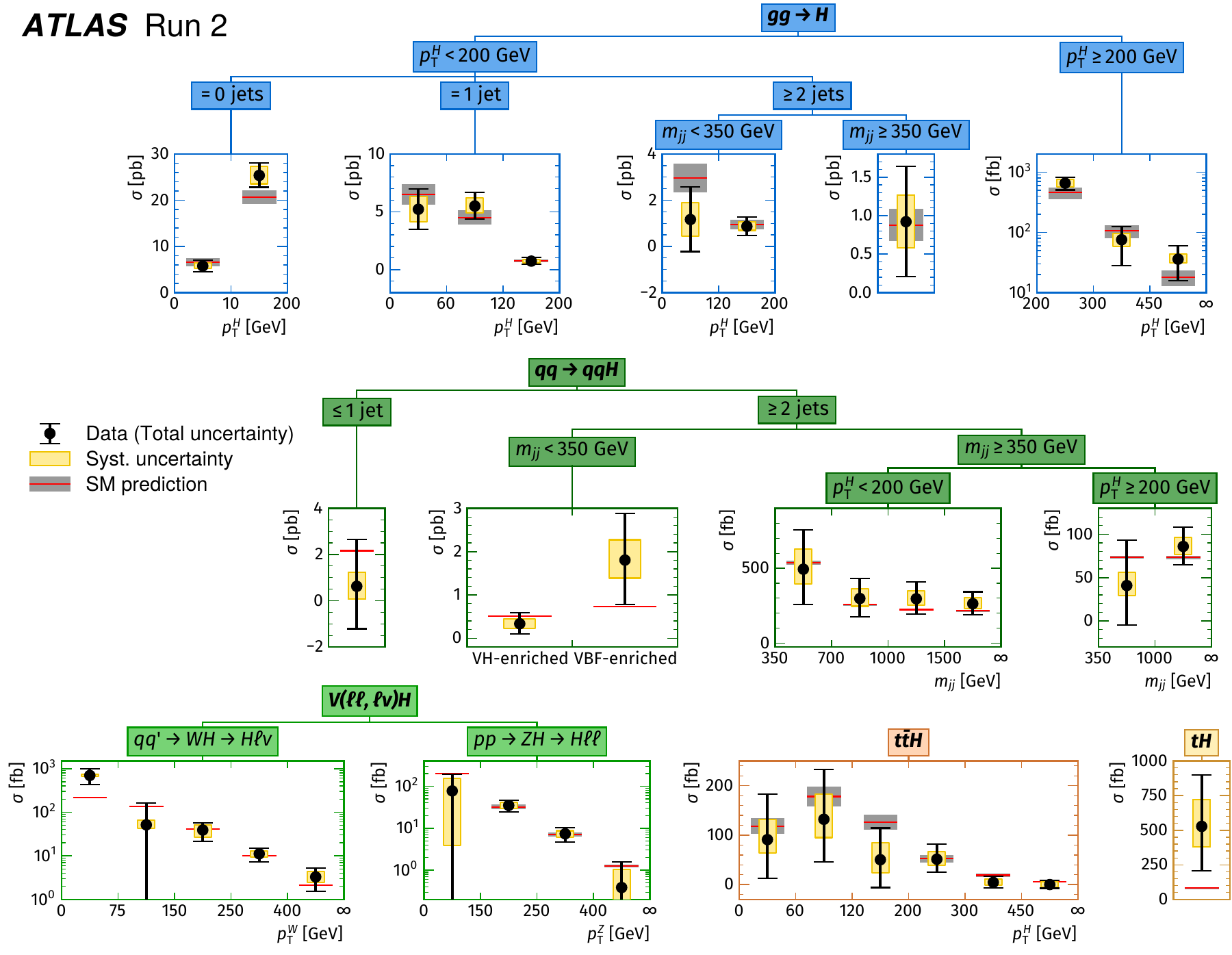}}
\caption{Observed and predicted Higgs boson production cross-sections in different kinematic regions~\cite{ATLAS:2022vkf}.
\label{fig:combi_stxs}}
\end{figure}

The measurement of the production cross-sections in different kinematic regions, separated by decay mode, is interpreted in the SM effective field theory (SMEFT)~\cite{HIGG-2024-EFT}. This allows effects of unknown phenomena at  energy scales $\Lambda$ that are large compared to the vacuum expectation value of the Higgs field, and that reduce to the SM at lower energies, to be searched for without having to rely on a specific model.
A complete set of higher-order-in-mass operators invariant under the SM gauge group is built from the SM fields~\cite{Brivio:2017vri}. They are scaled by Wilson-coefficients, which are zero in the SM\@.
If dimension-five operators are ignored, because they violate lepton number conservation, the leading contributions to physical observables are expected to come from dimension-six operators, which can describe CP-even or CP-odd couplings.  There are several complete sets (bases), of these dimension-six operators e.g., the Warsaw~\cite{Grzadkowski:2010es,Brivio:2017btx,Brivio:2020onw}, Higgs~\cite{deFlorian:2016spz,Azatov:2801789} and HISZ~\cite{Hagiwara:1993ck} bases.

Here, the Warsaw basis is used, and only CP-even operators are considered (see Section~\ref{sec:higgs_cp} for searches for CP-odd effects). When parameterising the cross-section times branching ratios through SMEFT, the series needs to be truncated by construction.
Two different choices are considered: keeping only the linear terms of the new operators (proportional to $\Lambda^{-2}$), or including quadratic terms (proportional to $\Lambda^{-4}$) as well, along with products of different Wilson coefficients.
The second choice is more complete, but also less consistent as it does not include terms from dimension-eight operators with the same suppression factor.
The input measurements do not allow constraints to be put on all relevant Wilson coefficients simultaneously. The solution chosen in this analysis  is a rotated basis, which allows linear combinations that combine operators with similar predicted effects on the measurements~\cite{ATLAS-CONF-2023-052}.

Figure~\ref{fig:combi_eft} shows the fitted values of the rotated coefficients for the linear and the linear+quadratic fit, including their uncertainties. All coefficients are compatible with the SM expectation of zero.
The strongest constraints can be found on coefficients which affect SM processes that are suppressed by small Yukawa couplings or include quantum loops. The operators corresponding to $c_{eH,22}$, $c_{eH,33}$, $c_{bH}$ are effectively modifiers of the Higgs Yukawa coupling to muons, \tauls and bottom quarks, respectively, while non-zero values of $c_\mathrm{Hq}$ would modify $VH$ production. The eigenvectors $e^{[i]}_\mathrm{ggF}$ encapsulate changes to the ggF production and $e^{[i]}_{H\gamma\gamma,Z\gamma}$could affect the \Hyy and \HZy decays.

\begin{figure}
\centering
\subfigure{\includegraphics[width=0.99\textwidth]{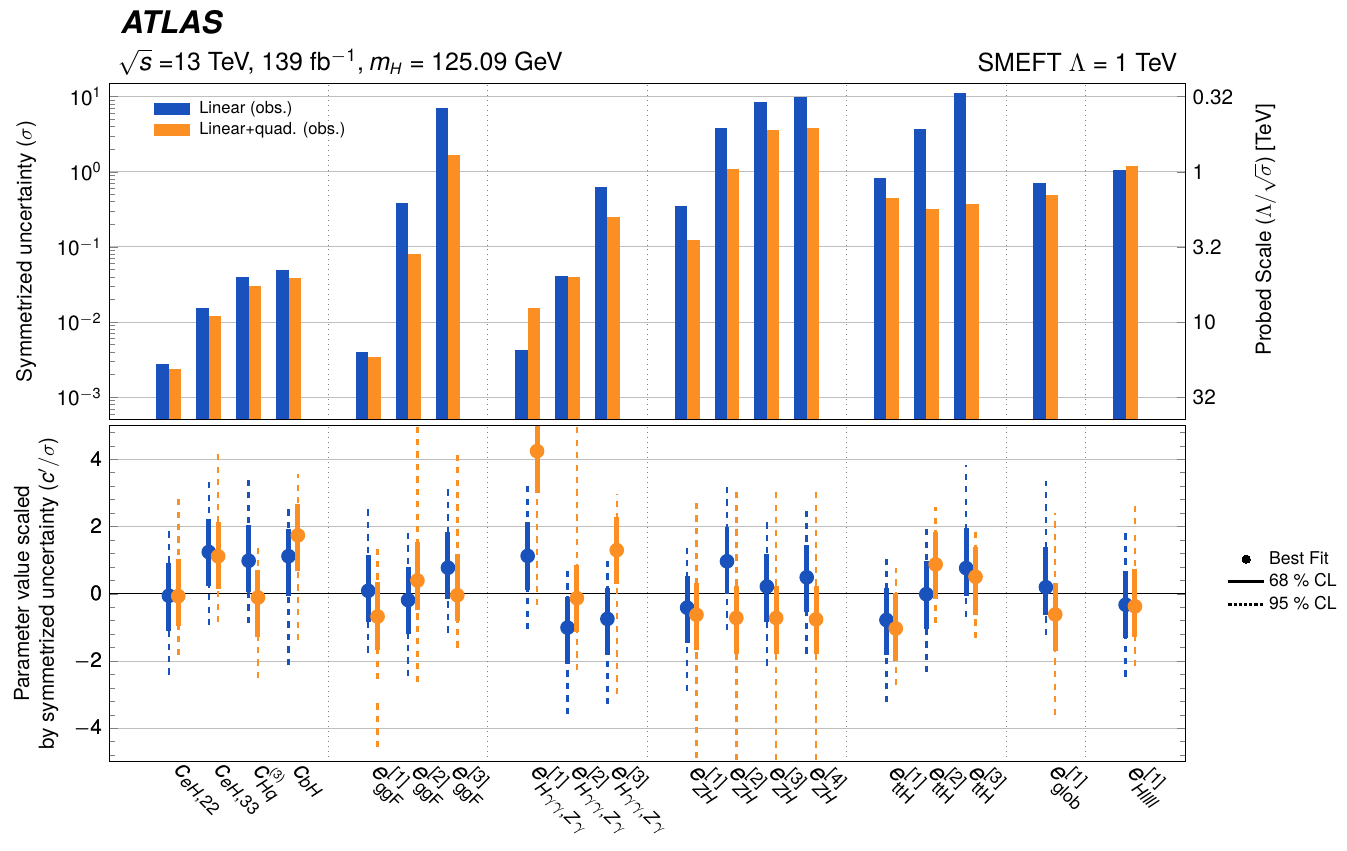}}
\caption{Comparison of the observed parameters of the rotated basis with the SMEFT model that keeps only terms linear in the new operators (blue) and the model including quadratic terms (orange), where all other coefficients and nuisance parameters are profiled, i.e., set to the values that maximize the likelihood~\cite{HIGG-2024-EFT}. The top panel shows the symmetrized 68\% CL uncertainty $\sigma$ of each parameter measurement (left vertical axis) and the corresponding energy scale $\Lambda / \sqrt{\sigma}$, $\Lambda$ = 1~\TeV that is probed (right vertical axis). The bottom panel shows the measured parameter value, 68\% (solid) and 95\% (dashed) CL intervals, divided by the symmetrised uncertainty shown in the top panel. \label{fig:combi_eft}}
\end{figure}

\FloatBarrier


\section{Study of CP properties and search  for anomalous interactions}
\label{sec:higgs_cp}

In the SM, the Higgs boson is a CP-even scalar with the following quantum numbers: $J^{CP} = 0^{++}$. Any sign of CP-odd couplings in the production or decay of the Higgs boson would be an unambiguous indication of BSM phenomena. Moreover, it would provide a so-far unknown  source of CP violation that could play a
fundamental role in explaining the matter–antimatter asymmetry in the universe.

CP-odd Higgs boson couplings to other particles can be described by adding corresponding terms to the SM Lagrangian~\cite{Brivio:2017vri,Berge:2014sra,Berge:2014wta}.
Couplings to bosons and fermions are investigated separately, also probing for mixtures between pure CP-odd and CP-even couplings. While a CP-odd component of the vector-boson couplings to the Higgs boson would be naturally suppressed by the scale at which BSM physics becomes relevant~\cite{Zhang:2010dr,Brivio:2017vri}, this suppression does not happen for Yukawa couplings, where CP-odd Higgs$-$fermion couplings may be significant already at tree level~\cite{Berge:2014sra,Berge:2014wta}.

\subsection{CP properties in interactions with vector bosons}

Higgs boson couplings to vector bosons are usually parameterised in the SMEFT framework (see Section~\ref{sec:higgs_combination}). In the Warsaw basis, there are three relevant dimension-six  operators describing CP-odd interactions at the $HVV$ vertex, scaled by the following Wilson coefficients: $c_{H\tilde{B}}$, $c_{H\tildewide{W}B}$, and  $c_{H\tildewide{W}}$.
Adding these CP-odd operators to the Lagrangian, and assuming the coefficients of the CP-even dimension-six operators to be zero, the new matrix element can be written as the sum of the SM matrix element $\mathcal{M}_{SM}$ and the CP-odd BSM Matrix element $\mathcal{M}_{BSM}$. The HISZ basis also contains three coefficients to describe CP-odd interactions.
By setting $\tilde{d} = \tilde{d}_B$ and the third coefficient to zero, the HISZ basis allows the strength of CP violation to be described  by a single parameter $\tilde{d}$, through
\begin{align*}
\mathcal{M} = \mathcal{M}_{SM} + \tilde{d} \cdot \mathcal{M}_{BSM},
\end{align*}
where the dependence on $\tilde{d}$ has explicitely been factored out~\cite{HIGG-2015-06,ATLAS:2023mqy}.

Experimentally, Higgs boson couplings to vector bosons can be probed in Higgs boson production (most prominently the VBF production mode) and Higgs boson decays into $Z$ or $W$ bosons. BSM effects on  couplings can be probed by measuring cross-sections in different kinematic regions (see Section~\ref{sec:higgs_combination}) and with CP-sensitive observables (this section), which, compared to the cross-section measurements specifically target CP-odd BSM effects.

The CP-sensitive observable of choice is the matrix-element-based `Optimal Observable'~\cite{Atwood:1991ka,Davier:1992nw,Diehl:1993br}, defined as
\begin{align*}
\OO = \frac{2 Re(\mathcal{M}^*_{SM} \mathcal{M}_{BSM})}{|\mathcal{M}_{SM}|^2}.
\end{align*}
This distribution is symmetric with a mean value of zero for a CP-even Higgs boson and becomes asymmetric when contributions from CP-odd BSM couplings are present. In the measurements discussed below, only the shape of the distribution is considered, such that exclusively CP-odd BSM effects are probed.

The Optimal Observable for a given VBF event is calculated based on matrix elements extracted from MC simulations at LO in QCD\@. From the data, the four-momenta of the Higgs boson and the two VBF jets are used as input to \OO. The momentum fractions of the initial-state partons are extracted by making use of the energy--momentum conservation in the collision, taking into account all possible quark flavour combinations as predicted by the parton distribution function~\cite{HIGG-2018-14}.

The CP properties are probed in VBF production for Higgs bosons decaying into two \tauls~\cite{HIGG-2018-14}, two photons~\cite{HIGG-2020-08}, four leptons~\cite{ATLAS:2023mqy}, or two $W$ bosons~\cite{HIGG-2020-25}. The \HZZfourl analysis also makes use of the decay characteristics, which are accessed through the reconstructed four-momenta of the decay leptons, to extract CP information.

In the \Htautau decay channel, the variable \OO is measured in bins of a BDT that is trained to distinguish the VBF signal from the background of non-Higgs and non-VBF processes. A simultaneous fit is performed over these signal regions and a number of control regions, enriched in the most important background processes, to extract the strength of a possible CP-odd parameter $\tilde{d}$.

In the \Hyy decay channel, two BDTs are trained to suppress ggF and non-Higgs contributions creating a sample enriched in VBF events. To estimate the remaining background, the signal yield is extracted from a combined unbinned maximum-likelihood fit applied to the \myy distribution of the observed data in each bin of \OO (see Figure~\ref{fig:CP_yyVBF_OO}). Using the \Hyy channel, constraints are extracted for $\tilde{d}$ and, separately, for the Wilson coefficient of the relevant operator $c_{H\tildewide{W}}$ in the Warsaw basis.

\begin{figure}[t]
\centering
\subfigure{\includegraphics[width=0.8\textwidth]{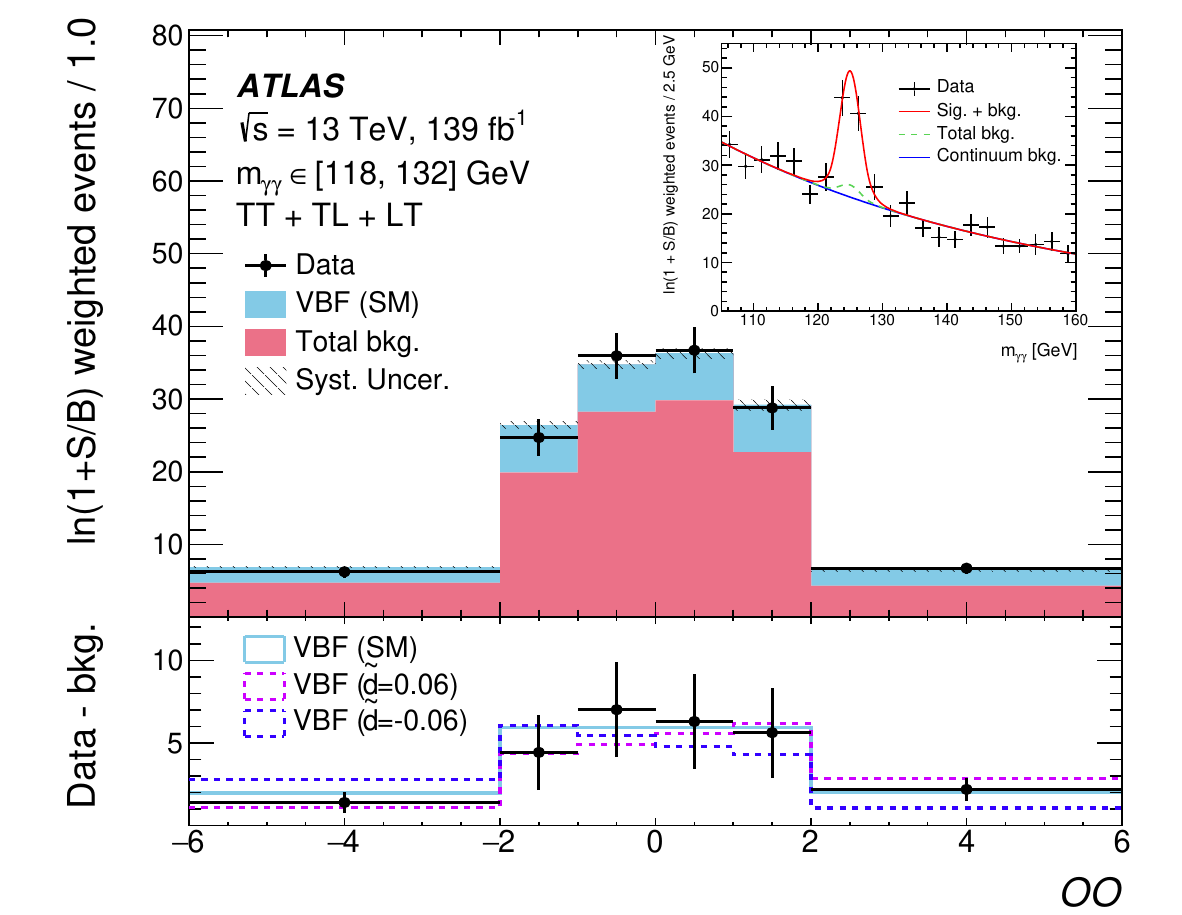}}
\caption{Distribution of the Optimal Observable in the \Hyy decay channel, for events enriched in the VBF production mode, with 118 $<$ \myy $<$ 132~\GeV~\cite{HIGG-2020-08}. The total background includes events from non-VBF Higgs production. The bottom panel contains the background-subtracted distribution for data as well as the prediction for the SM and for two non-zero values of $\tilde{d}$. The invariant mass distribution is shown in the inset on the top right.  Category-dependent weights are applied.
\label{fig:CP_yyVBF_OO}
}
\end{figure}

In order to extract a constraint on $c_{H\tildewide{W}}$ in the \HZZfourl decay channel, both the VBF production and properties of the \HZZfourl decay are used. A NN is trained by creating samples depleted or enriched in VBF, and a fit is performed simultaneously over the different regions. The Wilson coefficients $c_{H\tilde{B}}$ and $c_{H\tildewide{W}B}$, and $\tilde{d}$ are extracted from the decays. Each coefficient is fitted separately, based on a dedicated Optimal Observable.

The measurements of differential VBF cross-sections in the \HWWevmuv decay channel, discussed in Section~\ref{sec:higgs_diffxs}, are used to constrain  the Wilson coefficients $c_{H\tilde{W}}$,  $c_{H\tildewide{B}}$, and $c_{H\tildewide{W}B}$~\cite{HIGG-2020-25}. The distribution of the azimuthal angle difference between the two jets produced together with the Higgs boson is used as discriminant observable.

No deviation from a pure CP-even coupling is found in any of the studies. Figure~\ref{fig:CP_4l_comp} summarises some of the resulting constraints on the Wilson coefficients in both the Warsaw and HISZ bases.

\begin{figure}[t]
\centering
\subfigure{\includegraphics[width=0.8\textwidth]{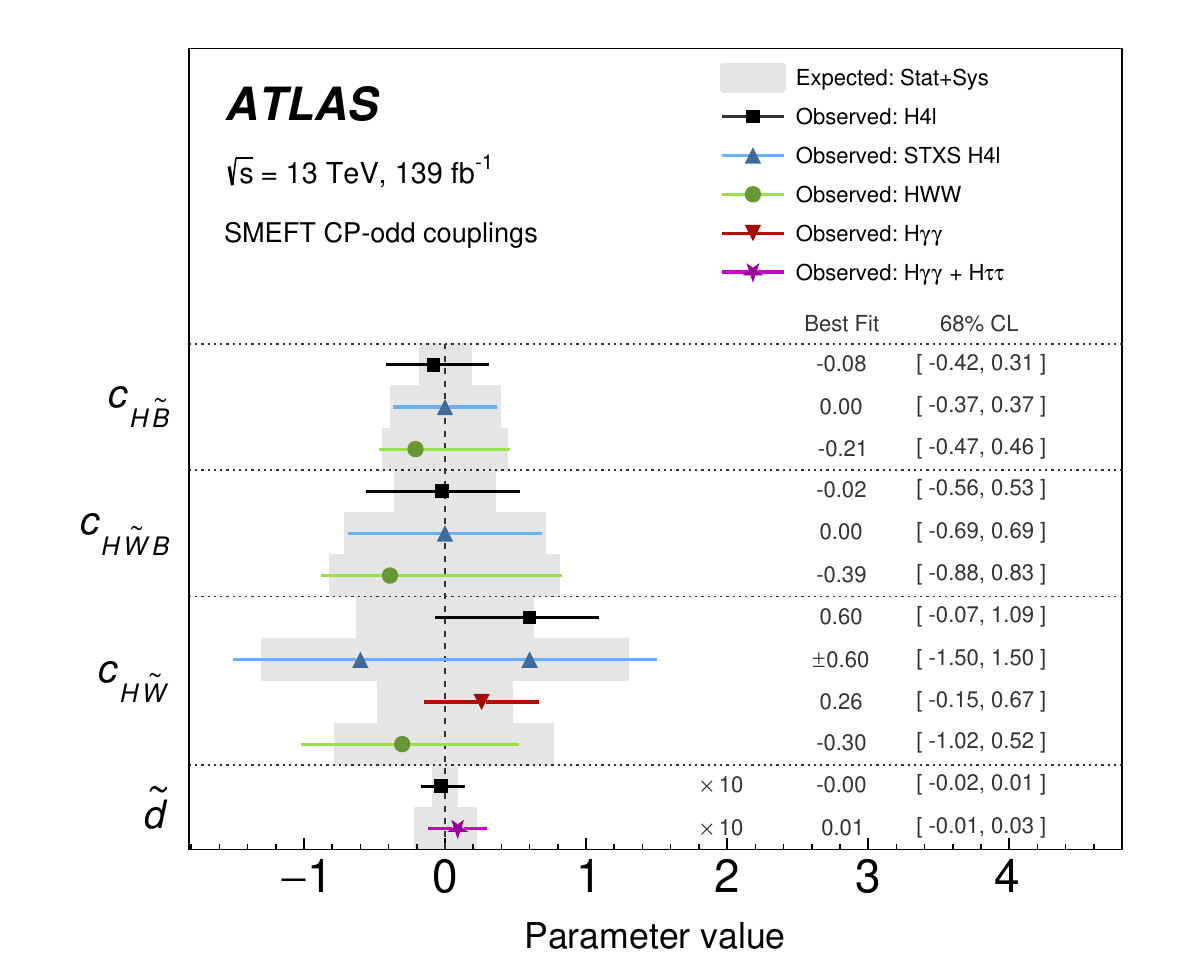}}
\caption{Comparison of results extracted using the CP sensitive observables in the \HZZfourl channel, the \HWWevmuv differential cross-section measurement in the VBF production mode, the \Hyy VBF measurement of $c_{H\tildewide{W}}$ in the Warsaw basis, and the combined VBF measurement of $\tilde{d}$ in the \Hyy
and \Htautau decay channels. Furthermore, the results based on the rate measurements in the \HZZfourl cross-section measurement of Ref.~\cite{HIGG-2018-28} (`STXS H4l')
are added, which cannot distinguish between CP-odd and CP-even effects.
The data points and 68\% CL uncertainty bars show the observed values with statistical and systematic uncertainties. The $\tilde{d}$ best fit points and uncertainty bars are scaled by a factor of ten. In the \HZZfourl channel, only the linear parameterisation is considered (the effects of adding the quadratic terms are found to be negligible), while for the other analyses, the linear plus quadratic parameterisation is applied.
\label{fig:CP_4l_comp}
}
\end{figure}

The CP properties of the effective Higgs--gluon coupling are studied in the \HWWevmuv decay channel~\cite{HIGG-2017-13}. The Lagrangian can be extended by a CP-odd term as
\begin{align}
\mathcal{L}_{Hgg} = -\frac{g_{Hgg}}{4} \kappa_{gg} (\cos{\alpha}\, G^a_{\mu\nu}G^{a,\mu\nu} + \sin{\alpha}\, G^a_{\mu\nu}\tilde{G}^{a,\mu\nu} ) H,
\end{align}
where $G^a_{\mu\nu}$ is the gluon field strength tensor, $\tilde{G}^{a,\mu\nu}
= G^a_{\rho\sigma} \epsilon^{\mu\nu\rho\sigma}/2 $ is the dual tensor, $g_{Hgg}$ is the effective coupling for the SM CP-even ggF interaction, $\kappa_{gg}$ is the coupling-strength modifier for the
effective Higgs$-$gluon interaction and $\alpha$ is the CP-mixing angle, which is zero in the SM\@. The azimuthal angle difference between the two jets produced together with the Higgs boson is used as CP-sensitive observable. Events are selected with two jets and kinematics compatible with the ggF production mode. A BDT is applied to reduce the background contamination and control regions are defined for the most important backgrounds.  A maximum-likelihood fit over all considered regions is performed. Using the angular distribution and the overall rate, a value of $\tan\alpha=0.0 \pm 0.4 \mathrm{(stat.)}  \pm 0.3 \mathrm{(syst.)}$ is obtained for the tangent of the mixing angle for CP-even and CP-odd contributions.

The \HWWevmuv decay channel also allows the Higgs boson coupling to longitudinally and transversely polarised $W$ and $Z$ bosons to be studied separately~\cite{HIGG-2017-13}. The azimuthal angle difference between the jets and the production rates in the signal regions are used. To probe both production and decay, events are selected that are compatible with the VBF production mode.  The polarisation-dependent coupling-strength modifiers are defined as the ratios of the measured polarisation-dependent coupling strengths to
those predicted by the SM, and are determined
to be $a_L = 0.91^{+0.10}_{-0.18} \mathrm{(stat.)}^{+0.09}_{-0.17} \mathrm{(syst.)}$ and $a_T = 1.2 \pm 0.4 \mathrm{(stat.)}^{+0.2}_{-0.3} \mathrm{(syst.)}$, compatible with the SM value of 1.

\subsection{CP properties in interactions with fermions}

The general effective Yukawa interaction between the Higgs boson and fermions, including a CP-odd term at tree level, can be parameterised as
\begin{align}
\mathcal{L}_{\mathrm{\textit{HFF}}} = -\frac{m_F}v \kappa_F (\cos{\alpha}\, \bar{\psi}\psi + \sin{\alpha}\, \bar{\psi} i \gamma_5 \psi   ) H.
\end{align}

Here, $m_F$ is the fermion mass, $v$ is the  vacuum expectation value of the Higgs field, $\kappa_F$ is the scaling factor for the Yukawa coupling, $\psi$ are the fermion fields, and $\alpha$ is the CP-mixing angle. The SM corresponds to a CP-even coupling with $\alpha = 0$ and $\kappa_F = 1$ while a pure CP-odd coupling would be realized for $\alpha = 90^\circ$. CP properties of the Higgs--\taul and Higgs--top-quark couplings are probed.

\textbf{The transverse spin components of the \tauls in the \Htautau decays} are affected by the CP-mixing angle $\alpha$~\cite{ATLAS:2022akr} and can be probed through the directions of the $\tau$-lepton decay products. At least one of the \tauls is required to decay hadronically and decay planes are formed for each \taul from the different visible decay products, i.e., leptons or pions, and, where necessary, the impact parameters of these objects. The angle $\varphi^*_\mathrm{CP}$ between these planes is calculated, which can be analytically related to $\alpha$ \cite{Kramer:1993jn,Berge:2014sra,Berge:2014wta}. The event selection and background estimates follow the analysis discussed in Section~\ref{sec:higgs_fermions}. Categories are created to target VBF production and ggF production with Higgs bosons at high transverse momenta. A BDT classifier is used to  further increase the purity of VBF events. In addition, $Z \rightarrow \tau \tau$ control regions are also formed. Figure~\ref{fig:CP_tautau_phi_ttHyy} (a) shows the measured angle between the decay planes, which is used to extract $\phi_\tau = \alpha$.

Allowing for a free normalisation of the \Htautau branching fraction, $\alpha$ is measured to be $9^\circ \pm 16^\circ$, with an expected value of $0^\circ \pm 28^\circ$
under the SM assumption. The pure CP-odd hypothesis is disfavoured at a level of 3.4$\sigma$.

\begin{figure}
\centering
\subfigure[]{\includegraphics[width=0.4\textwidth]{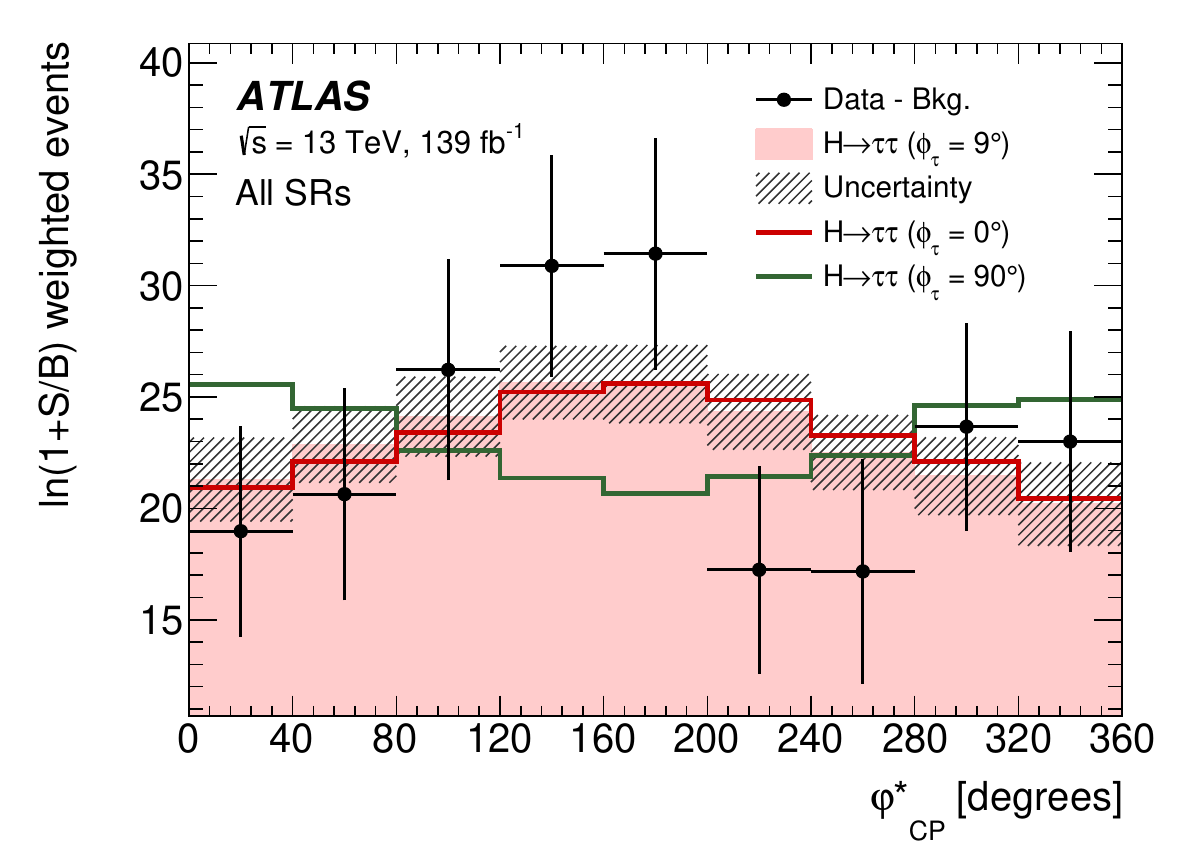}}
\subfigure[]{\includegraphics[width=0.4\textwidth]{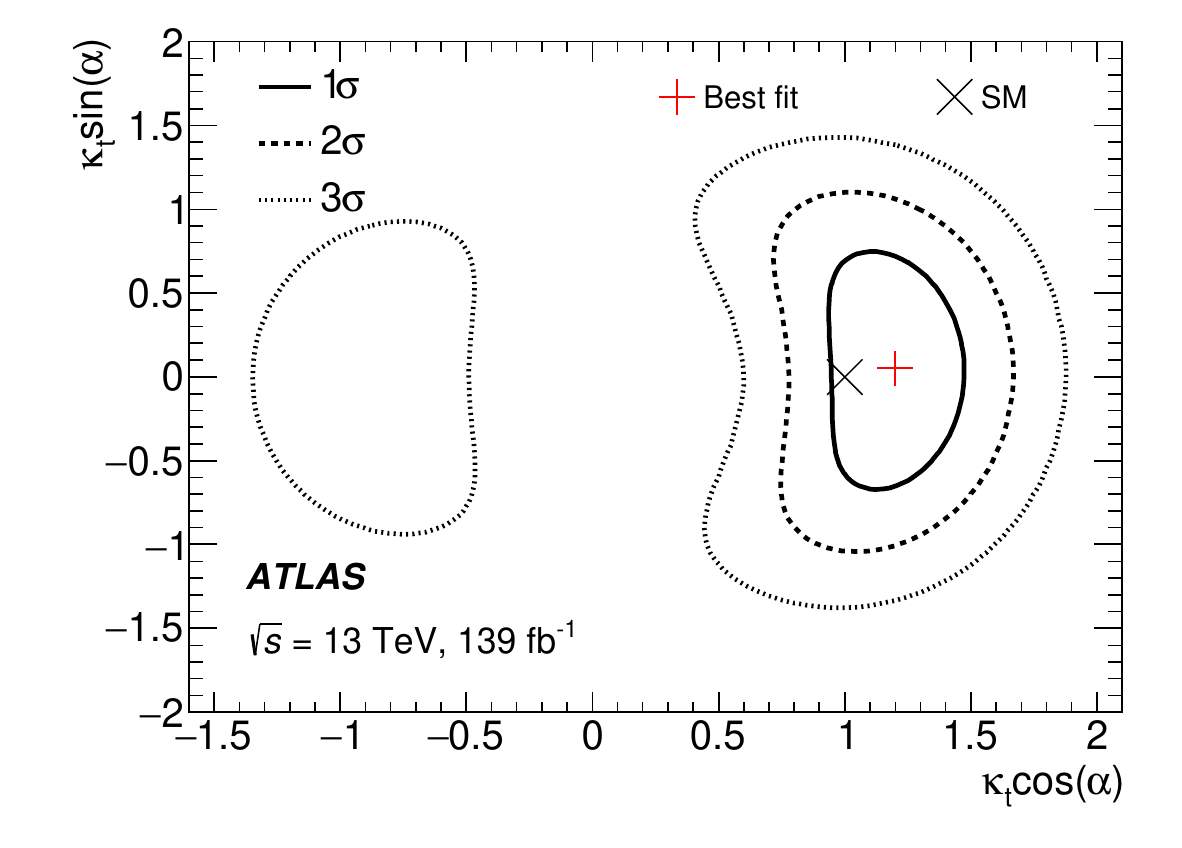}}
\caption{(a) Angle between the \taul decay planes~\cite{ATLAS:2022akr}. Events are weighted by ln{(1+$S$/$B$)}, where $S$ is the signal and $B$ the background yield in the respective category. The best fit is shown in solid pink, while the red and green lines correspond to the predictions for the pure CP-even (scalar, SM) and pure CP-odd (pseudoscalar) hypotheses, respectively, all scaled to the best-fit \Htautau yield. (b) Two-dimensional likelihood contours for $\kappa_t \cos{\alpha}$ and $\kappa_t \sin{\alpha}$, measured in the \Hyy decay channel, with ggF and \Hyy constrained by the Higgs boson coupling combination~\cite{HIGG-2019-01}.
\label{fig:CP_tautau_phi_ttHyy}}

\end{figure}

\textbf{The CP properties of the Higgs--top-quark couplings} are probed in the \hyy~\cite{HIGG-2019-01} and \hbb decay channels~\cite{HIGG-2020-03}, using the \ttH and \tH production mechanisms. Both production rates and kinematic distributions depend on the CP properties, with the \tH rate being particularly sensitive due to interference effects of diagrams. Furthermore,  the ggF production and the \hyy decay rates are also affected, as there is a top-quark contribution in the loop diagrams.

For the \hyy channel, BDTs are used to identify the hadronic decay products for the reconstruction of  the two top quarks, to reduce the background, and to separate CP-even and CP-odd contributions. The latter relies on kinematic distributions of the photons and top quarks and properties of the diphoton and $t\bar{t}$ systems, as well as various azimuthal angles. Events with at least one top quark candidate decaying semileptonically are separated from the fully hadronic events. A combined fit is performed over the \myy distributions in all signal regions and a pure CP-odd coupling of the Higgs boson to top quarks is excluded at 3.9$\sigma$, with $|\alpha| > 43^\circ$ excluded at 95\% CL\@. The likelihood contours as a function of $\kappa_t \cos{\alpha}$ and $\kappa_t \sin{\alpha}$ are shown in Figure~\ref{fig:CP_tautau_phi_ttHyy}.

In the \hbb case, either one or both of the top quarks are required to decay semileptonically. Signal candidate and control regions are formed based on the number of light- and heavy-flavour jets. The signal candidate regions are further split into signal and control regions with different \ttH sensitivity. CP-sensitive observables are formed using the directions and azimuthal separation of the top quarks. A combined likelihood fit allows the mixing angle to be measured, resulting in $\alpha = (11~^{+ 52}_{- 73})^\circ$.

\textbf{In summary}, no deviations from the CP-even hypothesis predicted by the SM are observed in the couplings of the Higgs boson to either bosons or fermions. The Run~2 data excludes CP-odd couplings to \tauls and top quarks at more than 3$\sigma$. Constraints are set on maximum mixing angles for the fermion couplings and on Wilson coefficients and matrix element mixing parameters for the boson couplings.


\section{The Higgs boson self-coupling}
\label{sec:higgs_self}

To completely understand the electroweak symmetry breaking mechanism, not only the properties of the Higgs boson, but also the Higgs field potential have to be probed experimentally.
The Higgs field potential is of great interest from a cosmological perspective, as it determines the structure of the vacuum and influences the electroweak phase transition in the early universe~\cite{Degrassi:2012ry,Reichert:2017puo}. A deviation of the Higgs boson self-coupling from the SM expectation might point to a first-order phase transition, a necessary condition for the electroweak baryogenesis mechanism, which in turn may help solve the matter--antimatter asymmetry puzzle.
The Higgs field potential can be expressed as:
\begin{equation}
V(H)=\frac{1}{2}m_H^2H^2+\lambda_{HHH} \, v \, H^3+ \frac{1}{4}\lambda_{HHHH}H^4 + O(H^5) \, ,
\end{equation}
where $m_H$ is the Higgs boson mass, $v$ is the vacuum expectation value, and $\lambda_{HHH}$ and  $\lambda_{HHHH}$ are the trilinear and quadrilinear Higgs boson self-couplings, respectively.
In the SM, once $m_H$ and the vacuum expectation value are known, the Higgs field potential and the Higgs boson self-couplings are fully determined:
\begin{equation}
O(H^5)=0 \quad \text{and} \quad \lambda_{HHH}^\mathrm{SM}=\lambda_{HHHH}^\mathrm{SM}=\lambda\, , \quad \text{where} \quad \lambda =m_H^2/(2v^2)\sim 0.13 \, .
\end{equation}
To probe deviations from the Higgs boson self-coupling values within the SM Lagrangian structure, the self-coupling-modifier \kl\ is introduced as $\kl = \lambda_{HHH} / \lambda_{HHH}^\mathrm{SM}$.
Many BSM models predict changes in the trilinear and quadrilinear self-couplings and add additional higher-order terms to the Higgs field potential.

At the LHC, the Higgs boson trilinear self-interaction is directly accessible via the production of Higgs boson
pairs (referred to as double-Higgs). The Higgs boson self-interaction also contributes to other processes via NLO EW corrections. In particular, it was shown~\cite{Degrassi:2016wml,Maltoni:2017ims,DiVita:2017eyz,Gorbahn:2016uoy,Bizon:2016wgr,McCullough:2013rea}  that single Higgs boson production cross-sections and branching ratios (referred to as single-Higgs) are  modified if the Higgs boson self-coupling deviates from the SM prediction. Figure~\ref{fig:h-hh-klambda-xsec} shows the expected effect of varying \kl on both the double- and single-Higgs boson production cross-sections. Additionally, BSM physics can give rise to resonant production of Higgs boson pairs mediated by a heavy  scalar. More details of resonant double-Higgs searches can be found in Ref.~\cite{HDBSreport}.

\begin{figure}[tbp]
\centering
\includegraphics[width=0.48\textwidth]{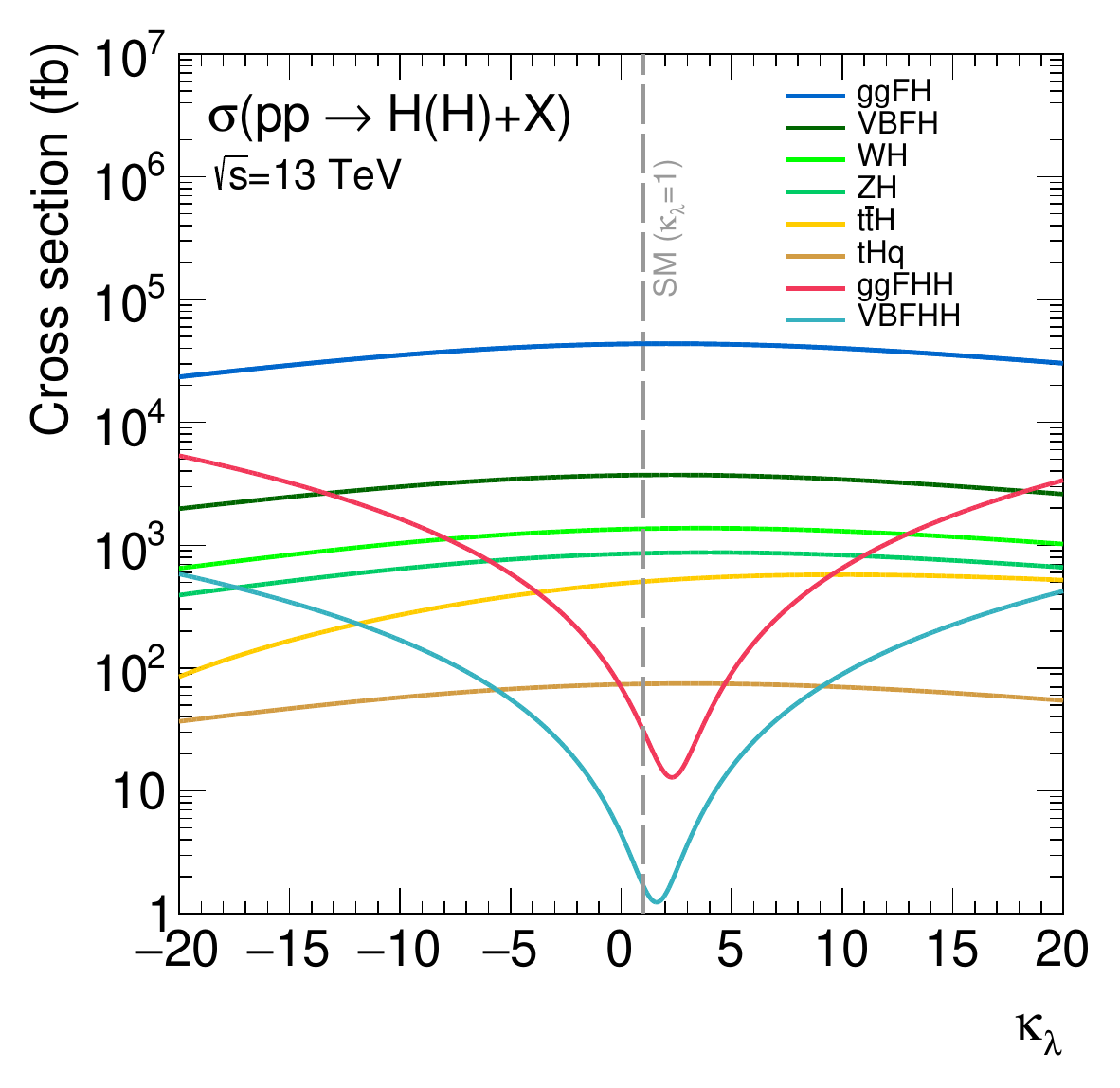}
\caption{Single-Higgs and double-Higgs boson production cross-sections as a function of \kl\ for $\sqrt{s}=13$~\TeV. The dashed line intercepts the values corresponding to the SM hypothesis ($\kl=1$)~\cite{LHCHXSWG-Pubnote}.}\label{fig:h-hh-klambda-xsec}
\end{figure}

\subsection{Double Higgs boson production}

Double Higgs boson production is sensitive to the Higgs boson trilinear self-coupling at LO\@.
In the SM, the \ggFHH\ process accounts for more than 90\% of the Higgs boson pair-production
at the LHC\@. The process with the second-highest cross-section at the LHC is \VBFHH\ production.
At LO, the  \ggFHH\ process proceeds through the destructive interference of two amplitudes: the box diagram in
Figure~\ref{fig:hh-diagrams}~(a), which is proportional to the square of \ktop, and the triangle diagram in Figure~\ref{fig:hh-diagrams}~(b), which is proportional to the product of \ktop\  and  \kl.
The \VBFHH\ process is described at LO by three Feynman diagrams: the two vector bosons radiated by the scattering
quarks fuse into two Higgs bosons via a $HHVV$ vertex (Figure~\ref{fig:hh-diagrams}~(c)), with the corresponding amplitude proportional to $\ktwoV=\lambda_{HHVV}/\lambda_{HHVV}^{\text{SM}}$; the vector bosons fuse into a virtual Higgs boson  that decays into two Higgs bosons via a trilinear vertex  (Figure~\ref{fig:hh-diagrams}~(d)), with the amplitude proportional to the product of \kV\ and \kl; the vector bosons couple with two single Higgs bosons via two $HVV$ interactions (Figure~\ref{fig:hh-diagrams}~(e)), with an amplitude proportional to the square of \kV.
Due to negative interference, the SM predicts double Higgs boson production to be a rare process, with a total cross-section that is more than a thousand times smaller than for single Higgs boson production (see Section~\ref{subsec:simulation} for the numerical values).

The various decays of the two Higgs bosons result in many different final states. However, the most sensitive  decay channels for double Higgs boson production and its properties are those that contain at least one Higgs boson decaying into a bottom-quark pair~\cite{HDBS-2019-29,HDBS-2022-02,HDBS-2018-40,HDBS-2019-27,HDBS-2018-34,HDBS-2021-10}, thanks to the large branching fraction. In particular, the most sensitive ones are the three $HH$ decay channels
\HHbbbb, \HHbbtt, and \HHbbyy.

\textbf{The \HHbbbb\ decay mode} has the advantage of the largest SM $HH$ decay branching fraction of 33.9\%. However, the double-Higgs search in this final state is also affected by a larger SM background compared to the other $HH$ analyses, due to the abundance of QCD multijet processes. The search for \HHbbbb~\cite{HDBS-2019-29} selects
events that contain at least four $b$-tagged jets, which are split between \VBFHH\ and \ggFHH\ categories according to the properties of the additional jets in the event. A further selection aims to reduce the contamination from $t\bar{t}$ production in the two categories through a discriminant that quantifies the probability that an event contains a hadronic top-quark decay. Finally, in each category, the signal region is defined in a two-dimensional plane of the mass of the two Higgs boson candidates $m_{H1}$ and $m_{H2}$, defined respectively as the one with the highest and lowest transverse momentum, by requiring that $X_{HH}<1.6$ with $X_{HH}$ defined as:
\begin{equation}
X_{HH} = \sqrt{ \left ( \frac{m_{H1} - 124~\text{\GeV}}{0.1 \ m_{H1}}\right )^2 + \left ( \frac{m_{H2} - 117~\text{\GeV}}{0.1 \ m_{H2}}\right )^2 } \, ,
\end{equation}
where $124~\text{\GeV}$ and $117~\text{\GeV}$ are determined as the peak of the $m_{H1}$ and $m_{H2}$ distributions for signal events in MC simulation.
The $HH$ invariant mass ($m_{HH}$) is used as a discriminating variable in the maximum-likelihood fit to extract the final results. The background, which is composed mostly of QCD multijet and $t\bar{t}$ processes, is evaluated with a fully data-driven method based on the extrapolation of the $m_{HH}$ distribution from control regions, defined as the signal regions except that exactly two $b$-tagged jets are required in the events. To account for differences in kinematic properties between events with four and two $b$-tagged jets, a NN trained on data is used to reweight the $m_{HH}$ distribution.
The 95\%~CL upper limit on the observed (expected) double-Higgs production signal strength is $\mu_{HH} < 5.4$ (8.1).

ATLAS has also performed a dedicated search for \HHbbbb\ production via \VBFHH\ in the boosted regime~\cite{HDBS-2022-02}, which is particularly sensitive to BSM \ktwoV variations. It results in 95\%~CL observed (expected) constraints on \ktwoV\ equal to $0.52 < \ktwoV < 1.52$ $(0.32 < \ktwoV < 1.71)$. Here, and in the following, unless explicitly stated otherwise, the Higgs boson couplings to the other particles are fixed to their SM expectations.
The 95\%~CL upper limit on $\mu_{HH}$ and the constraints on \kl\ and \ktwoV, obtained by the combination of the two \HHbbbb\ searches, are shown in Figure~\ref{fig:hh-xs-exclusion-channels} and in Table~\ref{tab:results_HH_kl_k2v}.

\textbf{The \HHbbtt\ decay mode} has one of the largest branching fractions (7.3\%) among the possible $HH$ decay channels. The \HHbbtt\ search~\cite{HDBS-2019-27}
includes different signal regions targeting fully hadronic and semileptonic $\tau\tau$ final states.
Events in the semileptonic channel
are required to have exactly one electron or muon and one hadronically decaying \taul with opposite charges. Events in the fully hadronic channel are
required to have exactly two reconstructed hadronically decaying \tauls with opposite charge. In both channels, exactly two $b$-tagged jets are required. For each final state, the events are further split in different signal regions: one targeting \VBFHH\ production, and two regions, dominated by \ggFHH\ and defined by requiring $m_{HH}<350$~\GeV and $m_{HH}\geq350 $~\GeV, respectively. The high mass region targets SM double-Higgs production, while the low-mass region is more sensitive to BSM values of \kl, as the double Higgs boson production cross-section  would increase in this region if the value of \kl is different from one.
The dominant backgrounds arise from $Z$~+~heavy~flavour and $t\bar{t}$ production. Other relevant background contributions are due to events with a single Higgs boson decaying into $\tau^+\tau^-$: $t\bar{t}H$, \ggH$+b\bar{b}$, $Z(\rightarrow b\bar{b})H$, and also $Z(\rightarrow \tau^+\tau^-)H(\rightarrow b\bar{b})$. A multivariate discriminant is trained separately in each signal region.
The fit is performed simultaneously in the signal regions plus an additional control region that is used to constrain the normalisation of the $Z$~+~heavy~flavour background component.
It results in the best expected sensitivity for SM double-Higgs production in a single channel. Figure~\ref{fig:hh-xs-exclusion-channels} shows the final 95\%~CL upper limit on $\mu_{HH}$ and
Table~\ref{tab:results_HH_kl_k2v} presents the constraints on \kl\ and \ktwoV.

\textbf{The \HHbbyy\  decay mode} has a very small branching ratio (0.26\%), but due to its experimentally clean signature allows for  one of the most sensitive $HH$ searches. The \HHbbyy\ search~\cite{HDBS-2021-10}
selects events with two photons and two $b$-tagged jets, each consistent with a SM Higgs boson decay. The dominant background arises from the SM continuum production of photon pairs and jets and from events with a single Higgs boson decaying into a pair of photons. The selected events are split into a low- and a high-mass region according to the value of the four-body mass $m_{b\bar{b}\gamma\gamma}^*=m_{b\bar{b}\gamma\gamma}-m_{b\bar{b}}-m_{\gamma\gamma}+250$~\GeV, where 250~\GeV\ is about twice the Higgs boson mass. Compared to $m_{b\bar{b}\gamma\gamma}$, $m_{b\bar{b}\gamma\gamma}^*$ is chosen because it improves the four-object mass resolution.
In each of the two mass regions, defined by  $m_{b\bar{b}\gamma\gamma}^*<350$~\GeV and $m_{b\bar{b}\gamma\gamma}^*\geq350$~\GeV, respectively, a multivariate discriminant is used to separate the signal, which includes \ggFHH\ and \VBFHH\ production, from the background, and to define the categories which maximise the double-Higgs significance. The final results are obtained from a simultaneous fit to the diphoton invariant mass distribution in these categories, three in the high-mass region and four in the low-mass region.
The 95\%~CL upper limit on $\mu_{HH}$ is shown in Figure~\ref{fig:hh-xs-exclusion-channels} and the constraints on \kl\ and \ktwoV\ are summarised in Table~\ref{tab:results_HH_kl_k2v}.

The analyses in the $HH\rightarrow b\bar{b}b\bar{b}$, $HH\rightarrow b\bar{b}\tau^+\tau^-$ and $HH\rightarrow b\bar{b} \gamma \gamma$ decay
channels are combined~\cite{ATLAS:2024ish} to obtain more stringent constraints on the production cross-section, the Higgs boson self-coupling, and the $HHVV$ coupling. This combination includes also two other searches for double-Higgs production which target rarer or less sensitive final states: the search for $HH$ events in \bbZZ, \VVVV ($V=W\text{ or }Z$), \VVtt, \tttt, \yyVV and \yytt decay channels with leptons in the final states (referred to as  "multilepton")~\cite{HDBS-2019-04}; the search for \HHbbll, which probes the $HH$ decay channels where one of the Higgs bosons decays to a $b$-quark pair and the other to either a boson pair ($ZZ^{*}$, $WW^{*}$) or a $\tau$-lepton pair, which then decays to a pair of opposite-sign leptons and neutrinos~\cite{HDBS-2019-02}. The inclusion of these additional searches maximises the coverage over the combinations of $HH$ decays and improves the sensitivity to SM double-Higgs production by $4\%$.

The combination yields an observed 95\% CL upper limit on $\mu_{HH}$ of 2.9, and an expected upper limit of
2.4 in the absence of $HH$  production.
The individual limits on the signal strength obtained from each channel and their combination are shown in Figure~\ref{fig:hh-xs-exclusion-channels}. The best-fit value of the signal strength  is $\mu_{HH} = 0.5^{+1.2}_{-1.0}$, which is compatible with the SM expectation with a $p$-value of 0.64. The observed (expected) significance for double-Higgs production is $0.4$ ($1.0$) standard deviations, with respect to the hypothesis of no $HH$ production.
The 95\% CL intervals for \kl\  and \ktwoV\ are summarised in Table~\ref{tab:results_HH_kl_k2v}. The  combination and the individual double-Higgs measurements are also interpreted in the context of effective field theory, as shown in Refs.~\cite{ATLAS:2024ish,HDBS-2021-10,HDBS-2019-27,ATL-PHYS-PUB-2022-019}.

\begin{table}[!tbp]
\caption{Summary of the observed and expected constraints
on \kl\ and  \ktwoV\ for the \HHbbyy, \HHbbtt, \HHbbbb, multilepton and \HHbbll\ analyses, and their combination~\cite{ATLAS:2024ish}. In the fit, the Higgs boson couplings to the other particles are fixed to their SM expectations. The expected constraints are derived under the SM hypothesis.}
\begin{center}{\def\arraystretch{1.2}\begin{tabular}{lccc}
\toprule
Analysis &  Obs. 95\% CL  & Exp. 95\% CL & Obs. 68\% CL  \\
\midrule
\HHbbyy & $-1.4 < \kl < 6.9 $    & $-2.9 < \kl < 7.8 $  &  $\kl = 3.0^{+2.2}_{-2.5}$ \\
\HHbbtt & $-3.2 < \kl < 9.0 $   & $-2.5 < \kl < 9.2 $ &  $\kl = 0.0^{+3.3}_{-1.7}$ \\
\HHbbbb & $-3 < \kl < 11 $    & $-5 < \kl < 11 $  &  $\kl = 5.8^{+3.0}_{-5.2}$ \\
\HHml & $-6< \kl < 12 $    & $-4.5 < \kl < 9.6$  &  $\kl \in [-4.4, 0.6] \cup [4.3,9.8] $ \\
\HHbbll & $-6< \kl < 13$    & $-10 < \kl < 17 $  &  $\kl = 3 \pm 6$ \\
\midrule
$HH$ combination & $-1.2 < \kl < 7.2$    & $-1.6< \kl < 7.2$  &  $\kl = 3.8^{+2.1}_{-3.6}$ \\
\midrule
\midrule
\HHbbyy & $-0.5 < \ktwoV < 2.7 $    & $-1.1 < \ktwoV < 3.3 $  &  $\ktwoV = 1.1\pm 0.8$ \\
\HHbbtt & $-0.5 < \ktwoV < 2.7$   & $-0.2 < \ktwoV < 2.4 $ &  $\ktwoV = 0.4^{+1.9}_{-0.5}$ \\
\HHbbbb & $0.5 < \ktwoV < 1.5$    & $0.4 < \ktwoV < 1.7 $  &  $\ktwoV = 1.01\pm 0.23$ \\
\HHml & $-2.5 < \ktwoV < 4.6 $    & $-1.9 < \ktwoV < 4.1 $  &  $\ktwoV = -0.4^{+4.1}_{-1.2}$ \\
\HHbbll & $-0.2 < \ktwoV < 2.4 $    & $-0.5 < \ktwoV < 2.7 $  &  $\ktwoV = 1.1\pm 0.7$ \\
\midrule
$HH$ combination &  $0.6< \ktwoV < 1.5$    & $0.4 < \ktwoV < 1.6$  &  $\ktwoV = 1.02^{+0.22}_{-0.23}$ \\
\bottomrule
\end{tabular}}
\end{center}
\label{tab:results_HH_kl_k2v}
\end{table}

\begin{figure}[htbp]
\centering
\includegraphics[width=0.55\textwidth]{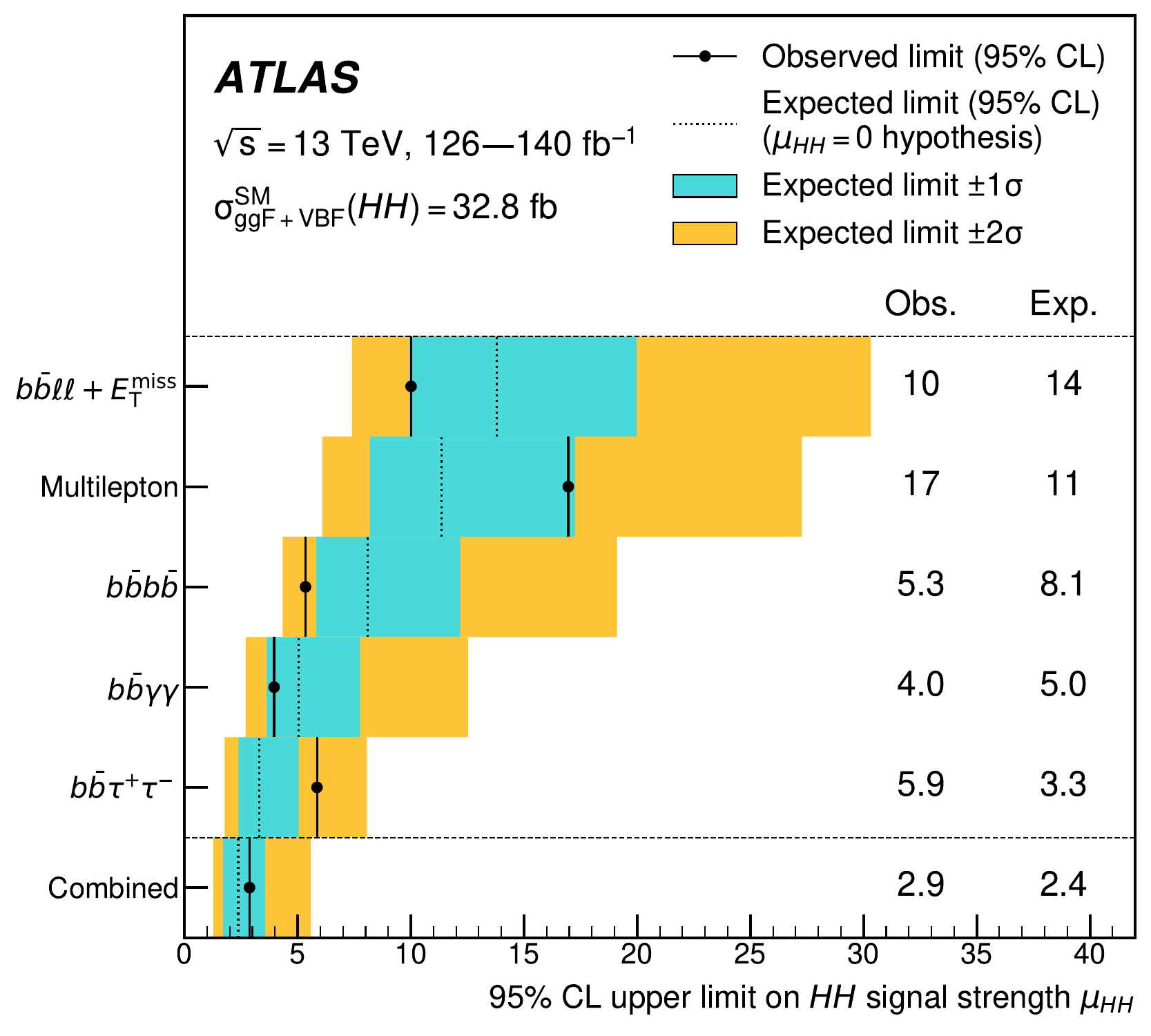}
\caption{Observed and expected 95\% CL upper limits on $\mu_{HH}$
for double Higgs boson production in the \HHbbll, multilepton, \HHbbyy, \HHbbtt\ and \HHbbbb\ decay
channels and their statistical combination~\cite{ATLAS:2024ish}. The expected limit is derived assuming $\mu_{HH}=0$.}
\label{fig:hh-xs-exclusion-channels}
\end{figure}

\FloatBarrier
\subsection{Self-coupling constraints from single Higgs boson measurements and combination with double Higgs boson searches}
An alternative and complementary approach to study the Higgs boson self-coupling was proposed in  Refs.~\cite{Degrassi:2016wml,Maltoni:2017ims,DiVita:2017eyz,Gorbahn:2016uoy,Bizon:2016wgr,McCullough:2013rea}. Processes involving a single Higgs boson do not depend on $\lambda_{HHH}$  at LO, but the Higgs trilinear self-coupling has to be taken into account in the calculation of the complete NLO EW corrections. In particular, $\lambda_{HHH}$ contributes via Higgs boson self-energy loop corrections and additional diagrams, as illustrated for different production modes by the examples in Figure~\ref{fig:self_coupling_diagram}.
\begin{figure}[!htbp]
\centering
\subfigure[]{\includegraphics[height=2.2cm]{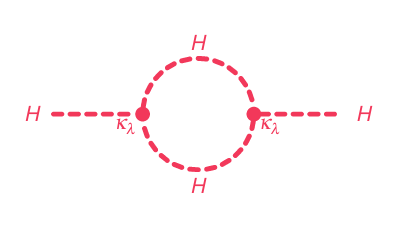}}
\\
\subfigure[]{\includegraphics[height=2.2cm]{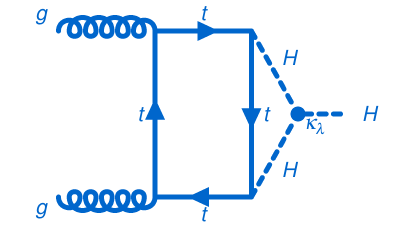}}
\subfigure[]{\includegraphics[height=2.2cm]{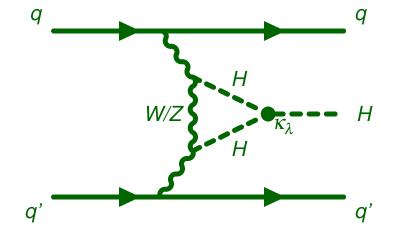}}
\subfigure[]{\includegraphics[height=2.2cm]{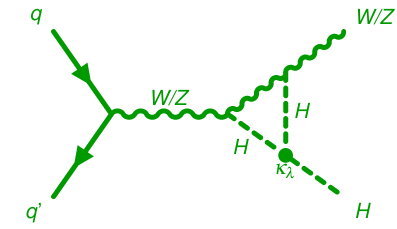}}
\subfigure[]{\includegraphics[height=2.2cm]{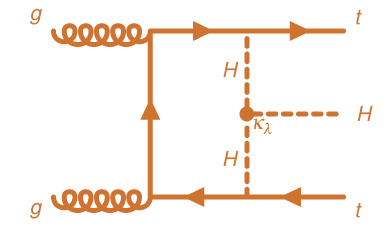}}\\
\caption{
Examples of one-loop $\lambda_{HHH}$-dependent diagrams for (a) the Higgs boson self-energy, and for
single Higgs boson production in the (b) \ggH, (c) \VBF,  (d) \VH,  and (e) \ttH\ modes. The self-coupling vertex is indicated by the filled circle~\cite{ATLAS:2022jtk}.
}
\label{fig:self_coupling_diagram}
\end{figure}

A framework for a global fit to constrain the Higgs boson self-coupling is proposed in Refs.~\cite{Degrassi:2016wml,Maltoni:2017ims}. The SM predictions of the Higgs boson production and branching ratios are modified as:
\begin{equation}
\mu_{if}(\kl)=\mu_i(\kl)\times\mu_f(\kl) \equiv \frac{\sigma_i(\kl)}{\sigma_{\text{SM},i}} \times \frac{\mathcal{B}_f(\kl)}{\mathcal{B}_{\text{SM},f}} \ ,
\label{eq:mu_klambda}
\end{equation}
where $\mu_i$
and $\mu_f$
describe multiplicative corrections to the expected SM Higgs boson production cross-sections ($\sigma_{\text{SM},i}$) and branching fractions ($\mathcal{B}_{\text{SM},f}$) as a function of $\kl$.
The functional dependence of $\mu_i(\kl)$ and $\mu_f(\kl)$ on $\kl$ varies according to the production mode and the decay channel. Moreover, these functions depend on the kinematic region considered within each process, especially for the \VH\ and \ttH\ production modes.
The cross-sections measured in different kinematic regions from the combined analysis in Section~\ref{sec:higgs_combination_stxt} are used  to constrain \kl. While potential variations across kinematic regions are not considered for \ggH, the differential effects of \kl for \VBF, \WH, \ZH  and \ttH\ production modes are parameterised as reported in Ref.~\cite{LHCHXSWG-Pubnote}.

The observed (expected) 95\% CL intervals for \kl\ from single Higgs boson measurements are $-4.0 < \kl < 10.3$~($-5.2 < \kl < 11.5$)~\cite{HDBS-2022-03}, which are not much looser than those obtained from the search for double Higgs boson production, despite the milder dependence of the single Higgs boson cross-sections on \kl. Moreover, since the two approaches are independent, they can be combined to provide the current most stringent 95\% CL interval for $\kl$, equal to $-0.4 < \kl < 6.3$~\cite{HDBS-2022-03}. This result and the following shown in this section, are based on the combination of a first version of the $HH\rightarrow b\bar{b}b\bar{b}$, $HH\rightarrow b\bar{b}\tau^+\tau^-$ and $HH\rightarrow b\bar{b} \gamma \gamma$ searches in the 140~\ifb dataset~\cite{HDBS-2019-29,HDBS-2018-40,HDBS-2018-34}, differently from the combination presented in the previous section which benefits from improved input analyses.

Double Higgs boson production does not only depend on \kl, but also has a significant dependence on $\kappa_t$  because the two most important diagrams contributing to the \ggFHH\ amplitude (see Figure~\ref{fig:hh-diagrams}) contain the coupling of the Higgs boson to the top quark. Using only double Higgs boson analyses, it is not possible to simultaneously constrain \kl\ and $\kappa_t$  without spoiling the sensitivity of the measurements of the two coupling-strength modifiers, as shown in Figure~\ref{fig:h-hh-klambda-kt-scan}. The inclusion of single Higgs boson measurements makes it possible to eliminate the degeneracy between \kl\ and $\kappa_t$ and leads to a limit almost identical to that obtained by determining only \kl. The combination of single and double Higgs boson analyses allows a further generalisation of the fit model
by simultaneously profiling
$\kappa_t$, $\kappa_b$,
$\kappa_{\tau}$, and $\kappa_V$.  The constraints on \kl\ in the different configurations are summarised in Table~\ref{tab:results_kl}.
\begin{figure}[!htbp]
\centering
{\includegraphics[width=0.48\textwidth]{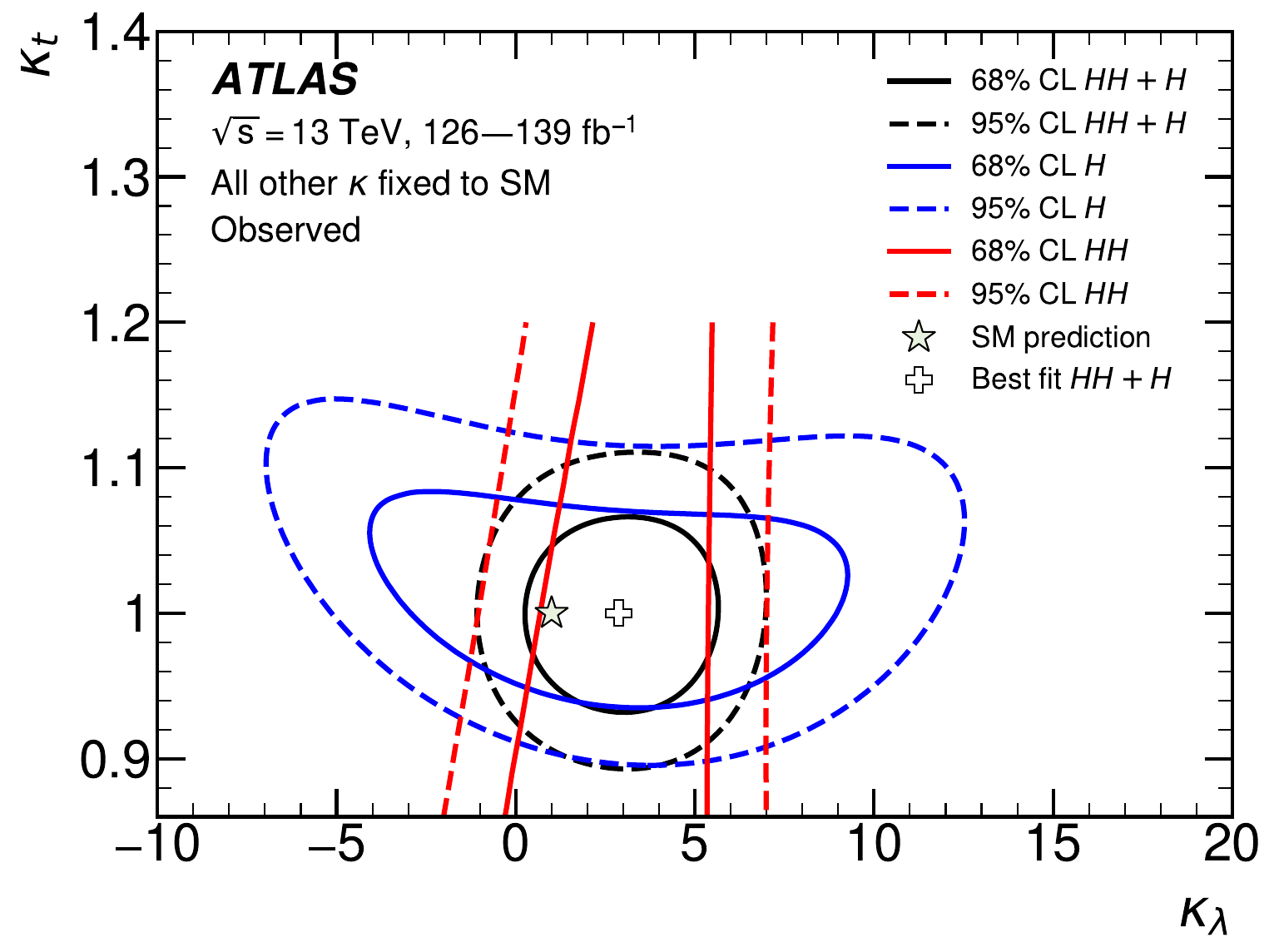}}
\caption{Observed 68\% (95\%) CL contours in the \kl--\ktop\ plane from single (blue) and  double (red) Higgs boson analyses and
their combination (black)~\cite{HDBS-2022-03}. The Higgs boson couplings to the other particles are fixed to their SM expectations.}
\label{fig:h-hh-klambda-kt-scan}
\end{figure}

\begin{table}[!htbp]
\caption{Summary of observed and expected constraints on \kl and corresponding observed best-fit values with their uncertainties~\cite{HDBS-2022-03}. The $HH$ combination is based on the earlier full Run 2 analyses performed by ATLAS. }
\begin{center}{\def\arraystretch{1.2}\begin{tabular}{llccc}
\toprule
&Floating parameters & Obs.\ 95\% CL  & Exp.\ 95\% CL & Obs.\ value$^{+1\sigma}_{-1\sigma}$ \\
\midrule
$HH$ combination     & \kl &     $-0.6 < \kl < 6.6$    & $-2.1 < \kl < 7.8$  &  $\kl = 3.1^{+1.9}_{-2.0}$ \\
Single-$H$ combination & \kl &   ~$-4.0 < \kl < 10.3$   & ~~$-5.2 < \kl < 11.5$ &  $\kl = 2.5^{+4.6}_{-3.9}$ \\
$HH$+$H$ combination & \kl &       $-0.4 < \kl < 6.3$    & $-1.9 < \kl < 7.6$  &  $\kl = 3.0^{+1.8}_{-1.9}$ \\
$HH$+$H$ combination & \kl, \ktop                   & $-0.4 < \kl < 6.3$  & $-1.9 < \kl < 7.6$  & $\kl = 3.0^{+1.8}_{-1.9}$\\
$HH$+$H$ combination & \kl, \ktop, \kV, \kb and \ktau   & $-1.4 < \kl < 6.1$  & $-2.2 < \kl < 7.7$  & $\kl = 2.3^{+2.1}_{-2.0}$ \\
\bottomrule
\end{tabular}}
\end{center}
\label{tab:results_kl}
\end{table}

Although still not achieving the sensitivity required to probe the SM, the constraints of the trilinear Higgs boson self-coupling are starting to enter ranges predicted by BSM physics. They also have exceeded the expectations based on the Run~2 intermediate analyses~\cite{HIGG-2016-15,HIGG-2016-16,EXOT-2016-31,HIGG-2016-27}. This success can be attributed to new analysis techniques and improvements in the object identification, such as new algorithms for tagging heavy-flavour jets~\cite{ATLAS:2022qxm}. Future data taking will be critical to get closer to the first evidence of double Higgs boson production and increase the sensitivity to  possible BSM effects on \kl.
\FloatBarrier


\FloatBarrier

\section{Conclusion and outlook}
\label{sec:conclusion}

In Run~2 of the LHC (2015--2018), the ATLAS detector recorded an unprecedented amount  of proton--proton collision data: 140~fb$^{-1}$, produced at a record centre-of-mass energy of 13~\TeV. The analysis of these data significantly advances our understanding of the Higgs boson properties and its interactions with other particles.

The mass of the Higgs boson is not predicted by the SM and is a fundamental parameter in the calculation of its expected properties. It is measured to be $125.11\pm 0.11$~\GeV, which corresponds to a precision better than a permille, leading to negligible variations in the SM predictions for other Higgs boson properties compared to the current experimental uncertainties of the measurements. All the measurements, in particular the production cross-sections and decay branching fractions, show very good agreement with the SM predictions within the uncertainties.

Higgs boson production in association with a vector boson and its production in association with two top quarks is observed for the first time; together with the Run~1 discoveries, all major Higgs boson production modes at the LHC have now been observed.  With the observation of \Htautau and \Hbb decays, couplings to third generation fermions are now established, and the search for \Hmumu, probing the Yukawa couplings to second generation fermions, yields a significance of 2$\sigma$. Evidence is found for \Hlly decays, both for \HZy (in a combination of ATLAS and CMS analyses) and lepton pairs with low invariant masses.

Assuming the SM coupling structure, the Higgs boson couplings to the $W$ and $Z$ bosons are measured with uncertainties of about 6\%, and the effective loop-induced couplings to photons and gluons with 6\% and 7\%, respectively. Couplings to  \tauls are measured with uncertainties of 7\%, and to the top and bottom quarks with about 12\%. The measured uncertainty in the coupling to muons, which are the first second-generation fermions to be probed, is about 30\%. Potential deviations from the SM predictions are also constrained within the framework of effective field theories. Searches for Higgs boson pair production achieve uncertainties in the $HH$ cross-section of close to 100\%, and reduce the  possible range of the Higgs boson self-coupling ($\lambda_{H H H}$) to the interval $-0.4 < (\lambda_{H H H}/\lambda_{SM}) < 6.3$ at 95\% CL.

Analyses of the data from Run~3 (2022--2025) and the High-Luminosity Runs of the LHC will build on the achieved results. Observation of the \Hlly and \Hmumu decays are expected, as well as Higgs boson pair production, and the precision of the measured couplings and their structures will be significantly enhanced~\cite{ATLAS:2019mfr, ATL-PHYS-PUB-2022-018, ATL-PHYS-PUB-2022-053}, further improving the characterisation of this key particle.


%
\section*{Acknowledgements}
%

%
%

%
%

We thank CERN for the very successful operation of the LHC and its injectors, as well as the support staff at
CERN and at our institutions worldwide without whom ATLAS could not be operated efficiently.

The crucial computing support from all WLCG partners is acknowledged gratefully, in particular from CERN, the ATLAS Tier-1 facilities at TRIUMF/SFU (Canada), NDGF (Denmark, Norway, Sweden), CC-IN2P3 (France), KIT/GridKA (Germany), INFN-CNAF (Italy), NL-T1 (Netherlands), PIC (Spain), RAL (UK) and BNL (USA), the Tier-2 facilities worldwide and large non-WLCG resource providers. Major contributors of computing resources are listed in Ref.~\cite{ATL-SOFT-PUB-2023-001}.

We gratefully acknowledge the support of ANPCyT, Argentina; YerPhI, Armenia; ARC, Australia; BMWFW and FWF, Austria; ANAS, Azerbaijan; CNPq and FAPESP, Brazil; NSERC, NRC and CFI, Canada; CERN; ANID, Chile; CAS, MOST and NSFC, China; Minciencias, Colombia; MEYS CR, Czech Republic; DNRF and DNSRC, Denmark; IN2P3-CNRS and CEA-DRF/IRFU, France; SRNSFG, Georgia; BMBF, HGF and MPG, Germany; GSRI, Greece; RGC and Hong Kong SAR, China; ISF and Benoziyo Center, Israel; INFN, Italy; MEXT and JSPS, Japan; CNRST, Morocco; NWO, Netherlands; RCN, Norway; MNiSW, Poland; FCT, Portugal; MNE/IFA, Romania; MESTD, Serbia; MSSR, Slovakia; ARIS and MVZI, Slovenia; DSI/NRF, South Africa; MICIU/AEI, Spain; SRC and Wallenberg Foundation, Sweden; SERI, SNSF and Cantons of Bern and Geneva, Switzerland; NSTC, Taipei; TENMAK, T\"urkiye; STFC/UKRI, United Kingdom; DOE and NSF, United States of America.

Individual groups and members have received support from BCKDF, CANARIE, CRC and DRAC, Canada; PRIMUS 21/SCI/017, CERN-CZ and FORTE, Czech Republic; COST, ERC, ERDF, Horizon 2020, ICSC-NextGenerationEU and Marie Sk{\l}odowska-Curie Actions, European Union; Investissements d'Avenir Labex, Investissements d'Avenir Idex and ANR, France; DFG and AvH Foundation, Germany; Herakleitos, Thales and Aristeia programmes co-financed by EU-ESF and the Greek NSRF, Greece; BSF-NSF and MINERVA, Israel; Norwegian Financial Mechanism 2014-2021, Norway; NCN and NAWA, Poland; La Caixa Banking Foundation, CERCA Programme Generalitat de Catalunya and PROMETEO and GenT Programmes Generalitat Valenciana, Spain; G\"{o}ran Gustafssons Stiftelse, Sweden; The Royal Society and Leverhulme Trust, United Kingdom.

In addition, individual members wish to acknowledge support from CERN: European Organization for Nuclear Research (CERN PJAS); Chile: Agencia Nacional de Investigaci\'on y Desarrollo (FONDECYT 1190886, FONDECYT 1210400, FONDECYT 1230812, FONDECYT 1230987); China: Chinese Ministry of Science and Technology (MOST-2023YFA1605700), National Natural Science Foundation of China (NSFC - 12175119, NSFC 12275265, NSFC-12075060); Czech Republic: Czech Science Foundation (GACR - 24-11373S), Ministry of Education Youth and Sports (FORTE CZ.02.01.01/00/22\_008/0004632), PRIMUS Research Programme (PRIMUS/21/SCI/017); EU: H2020 European Research Council (ERC - 101002463); European Union: European Research Council (ERC - 948254, ERC 101089007), Horizon 2020 Framework Programme (MUCCA - CHIST-ERA-19-XAI-00), European Union, Future Artificial Intelligence Research (FAIR-NextGenerationEU PE00000013), Italian Center for High Performance Computing, Big Data and Quantum Computing (ICSC, NextGenerationEU); France: Agence Nationale de la Recherche (ANR-20-CE31-0013, ANR-21-CE31-0013, ANR-21-CE31-0022, ANR-22-EDIR-0002), Investissements d'Avenir Labex (ANR-11-LABX-0012); Germany: Baden-Württemberg Stiftung (BW Stiftung-Postdoc Eliteprogramme), Deutsche Forschungsgemeinschaft (DFG - 469666862, DFG - CR 312/5-2); Italy: Istituto Nazionale di Fisica Nucleare (ICSC, NextGenerationEU), Ministero dell'Università e della Ricerca (PRIN - 20223N7F8K - PNRR M4.C2.1.1); Japan: Japan Society for the Promotion of Science (JSPS KAKENHI JP21H05085, JSPS KAKENHI JP22H01227, JSPS KAKENHI JP22H04944, JSPS KAKENHI JP22KK0227); Netherlands: Netherlands Organisation for Scientific Research (NWO Veni 2020 - VI.Veni.202.179); Norway: Research Council of Norway (RCN-314472); Poland: Polish National Agency for Academic Exchange (PPN/PPO/2020/1/00002/U/00001), Polish National Science Centre (NCN 2021/42/E/ST2/00350, NCN OPUS nr 2022/47/B/ST2/03059, NCN UMO-2019/34/E/ST2/00393, UMO-2020/37/B/ST2/01043, UMO-2021/40/C/ST2/00187, UMO-2022/47/O/ST2/00148); Slovenia: Slovenian Research Agency (ARIS grant J1-3010); Spain: Generalitat Valenciana (Artemisa, FEDER, IDIFEDER/2018/048), Ministry of Science and Innovation (MCIN \& NextGenEU PCI2022-135018-2, MICIN \& FEDER PID2021-125273NB, RYC2019-028510-I, RYC2020-030254-I, RYC2021-031273-I, RYC2022-038164-I), PROMETEO and GenT Programmes Generalitat Valenciana (CIDEGENT/2019/023, CIDEGENT/2019/027); Sweden: Swedish Research Council (Swedish Research Council 2023-04654, VR 2018-00482, VR 2022-03845, VR 2022-04683, VR 2023-03403, VR grant 2021-03651), Knut and Alice Wallenberg Foundation (KAW 2018.0157, KAW 2018.0458, KAW 2019.0447, KAW 2022.0358); Switzerland: Swiss National Science Foundation (SNSF - PCEFP2\_194658); United Kingdom: Leverhulme Trust (Leverhulme Trust RPG-2020-004), Royal Society (NIF-R1-231091); United States of America: U.S. Department of Energy (ECA DE-AC02-76SF00515), Neubauer Family Foundation.

%
%


%
%
%
\printbibliography

\clearpage
 
\begin{flushleft}
\hypersetup{urlcolor=black}
{\Large The ATLAS Collaboration}

\bigskip

\AtlasOrcid[0000-0002-6665-4934]{G.~Aad}$^\textrm{\scriptsize 103}$,
\AtlasOrcid[0000-0001-7616-1554]{E.~Aakvaag}$^\textrm{\scriptsize 16}$,
\AtlasOrcid[0000-0002-5888-2734]{B.~Abbott}$^\textrm{\scriptsize 121}$,
\AtlasOrcid[0000-0002-1002-1652]{K.~Abeling}$^\textrm{\scriptsize 55}$,
\AtlasOrcid[0000-0001-5763-2760]{N.J.~Abicht}$^\textrm{\scriptsize 49}$,
\AtlasOrcid[0000-0002-8496-9294]{S.H.~Abidi}$^\textrm{\scriptsize 29}$,
\AtlasOrcid[0009-0003-6578-220X]{M.~Aboelela}$^\textrm{\scriptsize 44}$,
\AtlasOrcid[0000-0002-9987-2292]{A.~Aboulhorma}$^\textrm{\scriptsize 35e}$,
\AtlasOrcid[0000-0001-5329-6640]{H.~Abramowicz}$^\textrm{\scriptsize 153}$,
\AtlasOrcid[0000-0002-1599-2896]{H.~Abreu}$^\textrm{\scriptsize 152}$
% ...
\bigskip
\\

% ...
$^{16}$Department for Physics and Technology, University of Bergen, Bergen; Norway.\\
% ...
$^{29}$Physics Department, Brookhaven National Laboratory, Upton NY; United States of America.\\
% ...
$^{35}$$^{(a)}$Facult\'e des Sciences Ain Chock, Universit\'e Hassan II de Casablanca;$^{(b)}$Facult\'{e} des Sciences, Universit\'{e} Ibn-Tofail, K\'{e}nitra;$^{(c)}$Facult\'e des Sciences Semlalia, Universit\'e Cadi Ayyad, LPHEA-Marrakech;$^{(d)}$LPMR, Facult\'e des Sciences, Universit\'e Mohamed Premier, Oujda;$^{(e)}$Facult\'e des sciences, Universit\'e Mohammed V, Rabat;$^{(f)}$Institute of Applied Physics, Mohammed VI Polytechnic University, Ben Guerir; Morocco.\\
% ...
$^{44}$Physics Department, Southern Methodist University, Dallas TX; United States of America.\\
% ...
$^{49}$Fakult\"{a}t Physik , Technische Universit{\"a}t Dortmund, Dortmund; Germany.\\
% ...
$^{55}$II. Physikalisches Institut, Georg-August-Universit\"{a}t G\"ottingen, G\"ottingen; Germany.\\
% ...
$^{103}$CPPM, Aix-Marseille Universit\'e, CNRS/IN2P3, Marseille; France.\\
% ...
$^{121}$Homer L. Dodge Department of Physics and Astronomy, University of Oklahoma, Norman OK; United States of America.\\
% ...
$^{152}$Department of Physics, Technion, Israel Institute of Technology, Haifa; Israel.\\
$^{153}$Raymond and Beverly Sackler School of Physics and Astronomy, Tel Aviv University, Tel Aviv; Israel.\\
% ...
$^{a}$ Also Affiliated with an institute covered by a cooperation agreement with CERN.\\
$^{b}$ Also at An-Najah National University, Nablus; Palestine.\\
$^{c}$ Also at Borough of Manhattan Community College, City University of New York, New York NY; United States of America.\\
$^{d}$ Also at Center for Interdisciplinary Research and Innovation (CIRI-AUTH), Thessaloniki; Greece.\\
$^{e}$ Also at Centro Studi e Ricerche Enrico Fermi; Italy.\\
$^{f}$ Also at CERN, Geneva; Switzerland.\\
% ...

\end{flushleft}


%

\end{document}